\documentclass[useAMS,usenatbib,aps,floatdix,nofootinbib]{mn2e}

\usepackage{graphics}
\usepackage{graphicx}
\usepackage{caption}
\usepackage{amssymb,amsmath}
\usepackage{amsmath}
\usepackage{psfrag}
\usepackage{graphicx}

\usepackage{color}

\newcommand{\f}{\frac}

\newcommand{\be}{\begin{equation}}      
\newcommand{\ee}{\end{equation}}      
      
\newcommand{\bef}{\begin{figure}}      
\newcommand{\eef}{\end{figure}}      
\newcommand{\bea}{\begin{eqnarray}}    
\newcommand{\eea}{\end{eqnarray}}      
  
\def\spose#1{\hbox to 0pt{#1\hss}}
\def\ltapprox{\mathrel{\spose{\lower 3pt\hbox{$\mathchar"218$}}
\raise 2.0pt\hbox{$\mathchar"13C$}}}
\def\gtapprox{\mathrel{\spose{\lower 3pt\hbox{$\mathchar"218$}}
\raise 2.0pt\hbox{$\mathchar"13E$}}}
\def\inapprox{\mathrel{\spose{\lower 3pt\hbox{$\mathchar"218$}}
\raise 2.0pt\hbox{$\mathchar"232$}}}

\def\bse{\begin{subequations}}
\def\ese{\end{subequations}}

\def\bF{{\mathbf F}}

\def\lsim{\raise 0.4ex\hbox{$<$}\kern -0.8em\lower 0.62ex\hbox{$\sim$}} 
\def\gsim{\raise 0.4ex\hbox{$>$}\kern -0.7em\lower 0.62ex\hbox{$\sim$}}

\def\f0N{f_0^{(N)}}
\def\bec{\begin{center}}
\def\eec{\end{center}}

\title[Non-linear clustering in scale-free 
cosmologies]
{Self-similarity and stable clustering in a family of scale-free cosmologies}
\author[D. Benhaiem, M. Joyce and B. Marcos]
{David Benhaiem${^1}$, Michael Joyce${^1}$ and Bruno Marcos${^{2,3}}$\\
$^1$
Laboratoire de Physique Nucl\'eaire et Hautes \'Energies, \\
Universit\'e Pierre et Marie Curie - Paris 6, CNRS IN2P3 UMR 7585, 4 Place Jussieu, 75752 Paris Cedex 05, France\\
$^{2}$ Laboratoire J.-A. Dieudonn\'e, UMR 7351, \\
Universit\'e de Nice - Sophia Antipolis, Parc Valrose 06108 Nice Cedex 02, France \\
$^{3}$Instituto de F\'{\i}sica, Universidade Federal do Rio Grande do Sul, Caixa Postal 15051, CEP 91501-970,
Porto Alegre, RS, Brazil}

\begin{document}

\date{\today}

\maketitle

\begin{abstract}
We study non-linear gravitational clustering from cold gaussian power-law initial conditions in a family of scale-free EdS models, characterized by a free parameter $\kappa$ fixing the ratio between the mass driving the expansion and the mass which clusters. As in the ``usual'' EdS model, corresponding to $\kappa=1$, self-similarity provides a powerful instrument to delimit the physically relevant clustering resolved by a simulation. Likewise, if stable clustering applies, it implies scale-free non-linear clustering. We derive the corresponding exponent $\gamma_{sc} (n, \kappa)$ of the two point correlation function. We then report the results of extensive N-body simulations, of comparable size to those previously reported in the literature for the case 
$\kappa=1$, and performed with an appropriate modification of the GADGET2 code. We observe in all cases self-similarity in the two point correlations, down to a lower cut-off which decreases monotonically in time in comoving coordinates. The self-similar part of the non-linear correlation function is fitted well in all cases by a single power-law with an exponent in good agreement with $\gamma_{sc} (n, \kappa)$. Our results thus indicate that stable clustering provides an excellent approximation to the non-linear correlation function over the resolved self-similar scales, at least down to $\gamma_{sc} (n, \kappa) \approx 1$, corresponding to the case $n=-2$ for $\kappa=1$. We conclude, in contrast notably with the results of \cite{smith}, that a clear identification of the breakdown of stable clustering in self-similar models - and the possible existence of a ``universal'' region in which non-linear clustering becomes independent of initial conditions - remains an important open problem, which should be addressed further in significantly larger simulations.

\end{abstract}


\section{introduction}


Cosmological $N$ body simulations are the primary instrument used to make theoretical predictions 
for structure formation in current models of the universe. However, the many and rich results which have been derived from such simulations for the non-linear regime remain essentially phenomenological
in nature, and analytical understanding of this regime, crucial for many non-trivial tests of these models, remains poor.  Further, there are open questions concerning their reliability and precision in 
reproducing the relevant continuum limit. To address both of these issues in a systematic manner,  
it is of interest to focus on models which are much simplified with respect to the currently 
favoured ``real" cosmological models. 
The canonical such model is the EdS model with cold 
power-law initial conditions, which has indeed been used as a reference point for controlled 
study of  structure formation in the literature (see e.g. \cite{peebles, efstathiou_88, padmanabhan_etal_1996, colombi_etal_1996, 
jain+bertschinger_1997, jain+bertschinger_1998, valageas_etal_2000, smith}). The 
simplicity of this model lies in its ``scale free" nature: the cosmology has a single characteristic 
time scale (the Hubble time)  
and the initial  conditions a single characteristic length scale which should be relevant in the evolution
(the scale at which fluctuations are of order unity). From this one can infer the 
property of  ``self-similarity": clustering at different times should be invariant when 
the spatial coordinates are appropriately rescaled. This property provides a 
strong test of the extent to which the clustering which develops in 
$N$-body simulations is actually independent of the additional 
length scales they introduce : the initial interparticle spacing (which provides
the required ultraviolet cut-off in the power spectrum of initial fluctuations), 
the size of the periodic simulation box, and also the force smoothing scale.  
Further the model, again because it is scale-free, leads to a very simple analytical 
prediction for the non-linear regime if the hypothesis of stable clustering, 
i.e. ``freezing" of clustering in physical coordinates, is made. The model thus 
provides a very tightly controlled testing ground for cosmological simulations, 
and more specifically of the validity of stable clustering. While stable clustering 
can be at most, in reality,  a good  approximation --- it is clear that structures 
are not strictly frozen in physical coordinates, because of tidal forces and 
even mergers --- it is fundamental to establish how good an approximation 
it actually is:  stable clustering leads to an imprinting of initial conditions on 
the non-linear regime which implies a strong ``non-universality" of
clustering.  Thus the study of stable clustering and it breakdown
is intimately related to the question of the existence of a 
``universal" features in non-linear gravitational clustering, i.e.,
independence of non-linear structures of initial conditions and/or
cosmological background evolution.  

In this paper, we return to these questions which have been previously
studied --- but, as we will discuss, only partially resolved --- in the previous
literature:
\begin{itemize}
\item Do scale-free cosmological models (and specifically the usual EdS model) 
with cold power law initial conditions lead to self-similar clustering? If so, for 
what range of $n$? 
\item How good is the approximation of stable clustering in the strongly 
non-linear regime? Specifically, how well can stable clustering predict
the form of the two point correlation function (or power spectrum)? 
Does it break down completely at some $n$ and lead to a regime which 
is independent of initial conditions? 
\end{itemize}
To address them we study not just this usual EdS model, but a family of such 
models which includes this  model as a special case. This means in practice 
that we have an extra control parameter on which we can study the 
dependence of the crucial behaviours --- the range of self-similarity 
and the validity of stable clustering. This will allow us, as we will show,
to pinpoint clearly the range of scales over which  these behaviours extend.  
Stated another way, with the usual EdS model one has a single free 
parameter, the exponent $n$ of the power spectrum of density fluctuations
characterizing the initial conditions, while we have now an additional
free parameter probing the role of the cosmological expansion. 
We thus have a two dimensional space of ``initial conditions/cosmology"
in which to study non-linear structure formation in a very controlled
manner.

Non-linear structure formation, and more specifically $N$-body simulations,
may be studied using different tools. The most evident set of tools, and those
employed widely in the literature on cosmological simulations since their
beginnings, are those provided by standard correlation analysis of correlated 
point processes,  using real space correlation functions, or the associated 
Fourier space analysis with power spectra (see e.g. \cite{peebles, book}). 
It is this approach we will use in this paper, focussing on two point
correlation analysis. This is both a natural starting point and the
best for comparison with the existing literature. An extension of analysis
of these models using the now very widespread characterization in terms
of the properties of halos and models constructed with them will be
presented in subsequent work. The present study is the sequel of a 
a study \cite{joyce+sicard_2011, benhaiem+joyce+sicard_2013} of an 
even simpler class of  truly ``toy" models:  the one dimensional analogy
of the family of  three dimensional scale-free EdS models we study here. 
As we will discuss further in our conclusions, there are remarkable 
similarities between the two cases, and the one dimensional case, which 
admits very accurate numerical simulations and a very extended spatial range 
of non-linear clustering, appears to be a very useful test-ground\footnote{
The exploration of 1D models as a tool for understanding the physics
of non-linear structure formation in a cosmological context goes back at least as
far as the work of  \cite{melott_prl_1982,melott_1d_1983} for the case of 
hot dark matter models. Different initial conditions and variants of the model 
have been discussed by a number of other authors, notably  
\citep{yano+gouda, miller+rouet_2002, aurell+fanelli_2002a, miller+rouet_2006, valageasOSC_2,
miller_etal_2007, agmjfs_pre2009, miller+rouet_2010a}.}.
Indeed this present study follows very much a path and structure
suggested by the one-dimensional study reported in 
\cite{joyce+sicard_2011, benhaiem+joyce+sicard_2013}.

Non-linear clustering in the EdS model with cold power law
initial conditions have been explored using $N$-body simulations 
in various studies in the literature.  The study of \cite{efstathiou_88}  
explored exponents in the range from  $n=-2$ to $n=1$, and found 
good evidence for self-similarity of the evolution, as well as 
reasonable agreement with the predictions of stable clustering 
for two point correlation functions, albeit over a quite 
limited dynamic range; that of  \cite{padmanabhan_etal_1996}
reported results for the cases $n=-2$ and $n=-1$ again in
line with self-similarity but reported deviations from stable
clustering predictions, albeit again with a very limited 
resolution.  A study by \cite{colombi_etal_1996}
explored the same range of $n$, and, using different statistical
tools, found small but significant deviations from the predictions 
of stable clustering for the cases $n=0$ and $n=1$.
The range of $n$ in which self-similarity applies has been the 
subject  of some discussion in the literature, notably for the range
$n < -1$ (see \cite{jain+bertschinger_1997} and references
therein). The study of  \cite{jain+bertschinger_1998} found
both self-similarity and agreement of two point properties
with the scaling prediction for the case $n=-2$. The
most recent extended studied of these models in
the literature of  \cite{smith}, explored the range $n=-2$ to 
$n=0$, and found again results in line with self-similarity
in all cases, but report, on the other hand, very strong 
deviations from stable clustering as probed through 
exponents of two point correlations properties.  As we will 
discuss in some detail, our conclusions concerning the
stable clustering approximation for the case $\kappa=1$ 
are in fact in disagreement with those of \cite{smith}, but 
in line with those of \cite{jain+bertschinger_1998}.

We note also that the case of a static universe, which corresponds
also to a special case of the class we treat,  has been treated 
in \cite{bottaccio2, sl1, sl2, sl3} for the cases $n=0$ and $n=2$. 
These studies found evolution which is self-similar \cite{sl1}
and evidence  for independence of the non-linear correlation 
properties of $n$ \cite{sl3}.  Indeed, as we will discuss further,
in a non-expanding universe it is clear that the stable clustering 
would never be expected to be a reasonable approximation.

The paper is organised as follows. In  the next section we define the class of 
models we study and establish some notation. We derive results for the
linear regime and use it to specify the scalings for self-similar behaviour.
We then discuss the stable clustering hypothesis, and describe a simple
derivation of the prediction it gives, when combined with self-similarity, 
for the exponent of the associated power-law behaviour of the 
non-linear two point correlation. We explain then the simple physical 
meaning of this exponent:  it controls directly the relative size of structures
when they have virialized compared with their original comoving
sizes. As a consequence (and also on the basis of our previous one
dimensional study)
we expect the stable clustering 
to be an increasingly good approximation as
the value of the exponent increases.
Further the value of the exponent can be related to the range 
over which self-similar stable clustering may be measured in
a finite simulation --- with the accessible range at fixed simulation
size being increasingly large as the exponent increases.
Thus the region in which stable clustering is not valid is
intrinsically more difficult to access numerically. 

In the second part of the paper we report our numerical results.
We first describe how we have implemented a modification to
the GADGET2 code to simulate the family of cosmologies, 
and tests we have performed to check this modification.
We discuss in some detail also control on the code using
the so-called Layzer-Irvine test.  We then report our results 
of our analysis of two point correlation properties
in both real and reciprocal space. In the final section
we summarize our results and compare them 
with previous work, discuss the relation of our
results to the question of universality of non-linear
clustering, and finally outline some directions for future 
work.

\section{A family of 3D scale-free models}  
\label{A family of 3D scale-free models}  
Dissipationless cosmological N-body 
simulations (see e.g. 
\cite{bertschinger_98, bagla_review, dehnen+reed_2011} for reviews) solve 
numerically the equations 

\begin{equation}
	\frac{d^2 {\bf x}_i}{dt^2} +
	2H \frac{d{\bf x}_i}{dt}  = \frac{1}{a^3} \bF_i 
	\label{3d-equations-1}
\end{equation}
where 
\begin{equation}
	\bF_i  =- Gm \sum_{j \neq i}^P
	\frac{{\bf x}_i - {\bf x}_j}{\vert {\bf x}_i - {\bf x}_j \vert^3} 
	W_\varepsilon (\vert {\bf x}_i - {\bf x}_j \vert) \,.
	\label{3d-equations-2}
\end{equation}

In Eq.~\eqref{3d-equations-2}, ${\bf x}_i$ are the comoving particle coordinates of the $i=1...N$ particles
of equal mass $m$, 
enclosed in a cubic box of side $L$, and subject to {\it periodic boundary conditions},
$a(t)$ is the appropriate scale factor for the cosmology considered, and 
$H(t)={\dot a}/{a}$ is the Hubble constant. The function $W_\varepsilon$ is
a regularisation of the divergence of the force at zero separation 
--- below a characteristic scale, $\varepsilon$, which is typically (but
not necessarily) fixed in comoving units. 
The superscript `P' in the sum in (\ref{3d-equations-2})  indicates 
that it runs over the infinite periodic system, i.e., the force on a particle is
that due to the $N-1$ other particles {\it and all their copies}. The
sum, as written, is formally divergent, but it is implicitly regularized
by the subtraction of the contribution of the mean mass density. 
This is physically appropriate 
in an expanding universe, as the 
mean mass density sources the expansion, while the particles
move in a potential sourced only by the fluctuations of mass
about this mean density (see e.g. \cite{peebles}).
Note that the limit in which $a(t)$ is constant, i.e., of a static universe, is 
well defined, provided the same method of calculating the 
force is used. This regularization of the cosmological problem
in the limit of a static universe is known as the ``Jeans swindle''
(see e.g. \cite{binney, kiessling, Assisi}). 

The one parameter family of models we will study is simply 
\begin{equation}
	H^2= \kappa^2 \frac{8\pi G \rho_0}{3a^3}
	\label{kappa-def}
\end{equation}
where $\rho_0$ is the mean mass density of the particles in
the simulation, and  $\kappa$ is a {\it positive constant}. 
Thus we have
\begin{equation}
	a(t) = \left(\frac{\kappa t}{t_0}\right)^{2/3} \qquad {\rm where}¬¨‚Ä†\qquad t_0= \frac{1}{\sqrt{6\pi G \rho_0}}.
\end{equation}
The case $\kappa=1$ is the usual EdS model,
while for any $\kappa > 0$ and $\kappa \neq 1$ the cosmology 
is, formally, an EdS model in which the total {\it matter-like} energy density 
driving the expansion differs by a factor from that of the matter 
which clusters.
Alternatively, and equivalently, it can be considered as
the class of models obtained by ``renormalizing"   
Newton's constant in the expansion
rate, i.e., $H^2={8\pi {\tilde G} \rho_0/3}$,  where  ${\tilde G}= 
\kappa^2 G$, i.e., in which the gravitational ``constant"
which appears in the expansion rate is different to the
one relevant for the scale of cosmological structure
formation. Finally the case $\kappa=0$ corresponds 
to the static universe case.

Our primary interest here is not in the cosmological 
interpretation (or ``realism") of this family of models, but in 
how the parameter $\kappa$, which controls the rate of 
expansion compared to the usual EdS model, affects 
structure formation. We note, however, that for $\kappa >1$ 
the model is equivalent to one in which,  in addition to the 
``ordinary"  clustering matter,  there is an additional 
pressureless component of the energy density which remains
uniform. An example of such a model is a homogeneous scalar 
field with an exponential potential which has an attractor
solution in which it contributes a fixed fraction of the mass
density (see e.g. \cite{ferreira+joyce_1998} and reference 
therein).  For $0<\kappa<1$ there is no such interpretation 
--- the additional mass density is negative --- and we are
not aware of a cosmological model which can realize
such a behaviour. )

There is another very simple way of describing these
models, which makes their choice as a class for study
very natural.
To see this it is sufficient to change time 
variable in Eq.~(\ref{3d-equations-1}) by defining 
\begin{equation}
	\tau=\int \frac{dt}{a^{3/2}}\,.
	\label{tau-definition}
\end{equation}
This gives the equations of motion in the
form 
\begin{equation}
	\frac{d^2 {\bf x}_i}{d\tau^2} +
	\Gamma (\tau) \frac{d{\bf x}_i}{d\tau}  =  \bF_i \,,
	\label{3d-equations-3}
\end{equation}
i.e., in which all effects of the cosmology appear only as a simple
{\it fluid damping term} with
\begin{equation}
	\Gamma =\frac{1}{2} a^{3/2} H =\frac{1}{2} a^{-1} \frac{da}{d\tau}. 
		\label{Gamma-a-relation}
\end{equation}
For our class of models we thus have
\begin{equation} 
	\Gamma=\kappa \sqrt{2 \pi G \rho_0/3} = \frac{\kappa}{3t_0}\,,
	\label{kappa-gamma}
\end{equation} 
i.e., in an appropriate time variables they correspond to the 
class of  {\it infinite non-expanding 
self-gravitating system (with Jeans regulation of the force) subjected to
a simple fluid damping}. 

\begin{figure}
	\centering\includegraphics[scale=0.6]{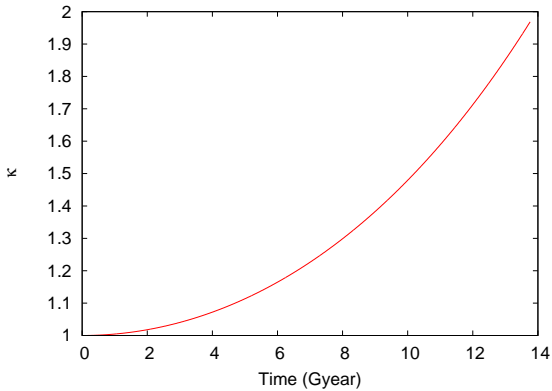} 
	\caption{Evolution of the damping parameter $\kappa$ in a
	$\Lambda$CDM model with $\Omega_{\Lambda}=0.7$,  $H_0=72 km/s/Mpc$}
	\label{fig-kappa-lambdaCDM}
\end{figure}

The change of variable employed is equally valid for any cosmological 
Friedmann-Robertson Walker type model, and thus in general any
such model is equivalent to such a model but in which the fluid
damping coefficient $\Gamma$ may vary as a function of time. 
Indeed the relation (\ref{kappa-gamma}) is always valid when
(\ref{kappa-def}) is simply taken as a definition, which,
in terms of  the standard cosmological parametrisation
is simply
\begin{equation}
	\kappa=\frac{1}{\sqrt{\Omega_m}}\, 
\end{equation}
where $\Omega_m=\rho_0/\rho_c$  (with $\rho_c=3H_0^3/8\pi G$ the 
critical energy density). For a $\Lambda$CDM cosmology  therefore 
$\kappa=\left(1 - \frac{\Omega_{\Lambda,0}}{\Omega_m,0}a^{-3} \right)^{-1/2}$,
which can be calculated easily using the known analytic expression 
for $a(t)$ in  this model. The resulting $\kappa$ is shown as a function
of time in Fig.  ~\ref{fig-kappa-lambdaCDM}, for the case $ \Omega_{\Lambda,0}=0.7$ 
(i.e. $\Omega_{m,0}=0.3$) and $H_0=72 km/s/Mpc$.

Likewise any generic homogeneous dark energy model will correspond
to some slowly varying damping parameter $\kappa$ with $\kappa >1$.
The case of an open universe also corresponds to the same range,
with a slightly gentler increase of $\kappa$, while the case of a
closed universe, in its expanding phase, corresponds to a
decreasing $\kappa <1$, with $\kappa \rightarrow 0$  as turnaround 
is approached. To the extent that any such background cosmology 
may be considered then as a continuous interpolation of the class of
models with $\kappa$ constant, our study may thus help to understand 
in particular  the role of modification of the expansion rate due to a 
cosmological  constant, or more generally dark energy, on structure 
formation  (a question addressed more  directly in many recent works, 
e.g., \cite{rasera_etal_2009}). We will comment further on this in our 
conclusions section below.

In what follows we will often find it convenient to do our analysis in
the time variable $\tau$ and thus consider the equations of motion 
in the form (\ref{3d-equations-3}) where $\Gamma$ is related to
$\kappa$ by (\ref{kappa-gamma}).  This has the advantage of allowing 
a simultaneous treatment of the static limit ($\Gamma=0$), and
will also be the form we will exploit in our numerical simulations.
When necessary or instructive we will change back, for 
$\kappa \neq 0$, to the more familiar cosmological 
time variable $t$ or scale factor $a$, noting that the 
relevant transformations are
given by

\begin{equation}
	t=t_0 e^{\kappa \tau /t_0}=t_0 e^{3\Gamma \tau} 
\end{equation}
 and thus
\begin{equation}
	a=e^{2\kappa \tau /3t_0}=e^{2\Gamma \tau}. 
\end{equation} 

\subsection{Collisionless limit}

The Vlasov-Poisson limit for our class of models,
in the coordinates $(\vec{x}, \tau)$ in which
the equations of motion are given by (\ref{3d-equations-3})
with $\Gamma$ constant and ${\bf F}_i$ the
(regulated) gravitational force, can be written
in the simple form \footnote{The term on the right 
hand side arising from the fluid damping 
can be rewritten on the left hand side as 
 $-\Gamma {\bf v} \cdot \nabla_{\bf v} f - 3 \Gamma f$;
the first term describes the modification of the mean 
field force, the second term the  ``shrinking" of the 
phase space volume. It is straightforward to verify that
this writing is equivalent to the usual form of this 
equation employed in the cosmological literature. }
 \begin{equation}
\partial_\tau f+ {\bf v} \cdot  \nabla_{\bf x} f
- (\nabla_{\bf x} \Phi)\cdot \nabla_{\bf v} f =  \Gamma \nabla_{\bf v} \cdot ( {\bf v} f) 
\end{equation}
where $f=f(\vec{x}, \vec{v}, t)$ is the phase space mass density
with $\vec{v}=\frac{d\vec{x}}{\delta \tau}$, and 
\begin{equation}
\nabla_{\bf x}^2 \Phi = 4 \pi G [\int f d^3 x d^3 v - \rho_0]
\end{equation}

\subsection{Linear perturbation theory in the collisionless limit}

Following the analogous steps to the usual treatment 
(see \cite{peebles, binney}) we obtain, by taking moments
of these equations and neglecting 
velocity dispersion, the continuity and Euler equation.
Linearizing in the velocity field and density perturbations,
and using the Poisson equation (with subtracted mean density),
we obtain 
\begin{equation}
\frac{d^2 {\delta}}{d\tau^2} +
\Gamma \frac{d {\delta}}{d\tau} = 4\pi G \rho_0 \delta
\label{linear theory}
\end{equation}
where 
 \begin{equation}
	\delta \left(x,t\right)=\frac {\rho(x,t) - \rho_0 \left(t\right)}{\rho_0 \left(t\right)}
\end{equation}
is the normalized density fluctuation.

These equations have a decaying and growing mode solution, the
latter being given by 
\begin{equation}
{\delta} \propto D(\tau)=e^{2 \alpha \Gamma \tau} \quad {\rm where} \quad
\alpha = \frac{1}{4}[-1 + \sqrt{1+\frac{24}{\kappa^2}}]
\label{linear theory-growing-general}
\end{equation}
For $\kappa\neq 0$ this can be written simply   
as 
\begin{equation}
D(a) = a^{\alpha} 
\label{linear theory-growing-expanding}
\end{equation}
For $\kappa=1$ we thus recover the growth law as in 
the standard EdS model. Comparatively growth is
``slowed down" for $\kappa >1$, and ``sped up" for 
$\kappa < 1$. Thus, as would be expected, linear theory
retains its scale free nature --- characteristic of Newtonian 
gravity --- but it is modified 
by the addition of damping. Note that in the static 
limit, $\kappa=0$, we have
\begin{equation}
D(\tau)=e^ {\sqrt{4\pi G \rho_0} \tau} 
\label{linear theory-growing-static}
\end{equation}

We will study (as in almost all cosmological simulations)
the case where we start from a time where the theoretical
input perturbations are very small and solely in the growing 
mode. The linear evolution from the same initial conditions 
in any model in our family can thus be trivially mapped onto 
that of any other one. To do so it is very convenient to
work in the dimensionless time variable $t_s (\tau)$ defined 
by 
\begin{equation}
D(\tau)=e^{t_s},
\end{equation}
i.e.,
\begin{equation}
t_s = 2 \alpha \Gamma \tau={\sqrt{\frac{\pi G \rho_0}{6}}} [-\kappa + \sqrt{\kappa^2+24}] \, \tau
\end{equation}
in which the linear evolution in all models will be identical (in the growing mode).  
For $\kappa \neq 0$ we have $t_s = \alpha \ln a$. We will refer to this time
variable as {\it the static time variable}, because in the static limit
it coincides with the physical time,  exactly in ``natural" units 
$\sqrt{4 \pi G \rho_0}$. Note also that for $\kappa \neq 0$ we have 
$a=e^{t_s/\alpha}$.  In our numerical analysis below we will often use 
this time variable in order to isolate the specific non-trivial dependence 
of the non-linear clustering on $\kappa$.

\subsection{Self-similarity}
\label{section-SS}

For a given model with $\kappa$ fixed, self-similarity of
the clustering is obtained in principle under precisely the 
same assumptions as in the usual EdS model \cite{peebles}: 
if (1) the clustering which develops is independent of any
length scale other than the single one defined by the 
non-linearity scale, and (2) the latter scale evolves as
predicted by linear theory, the statistical properties 
should be invariant when expressed in length
units which rescale in the same way. Specifically
the 2-point auto-correlation function of the 
density field
\begin{equation}
	\xi \left(x\right)=\left< \delta({\bf x^\prime}) \delta({\bf x^\prime}+{\bf x}) \right>
\end{equation}
should obey the relation 
\begin{equation}
\xi(x, \tau)=\xi_0 \Big(\frac{x}{R_s(\tau)} \Big)
\label{self-similarity-xi}
\end{equation}
where $\xi_0=\xi(x, 0)$ and $\tau=0$ is some (arbitrary)
reference time. For the power spectrum, which is the
Fourier transform of $\xi(x)$, the corresponding relation 
is\footnote{We assume here as is canonical (see, for example, \cite{peebles, book}) 
that the distribution of points is described as a stochastic point process which is 
statistically homogeneous and isotropic, and thus that $\xi$ 
is a function of the separation $x=|\bf x|$ only. The power spectrum
is then its Fourier transform and a function of $k=|\bf k|$ only.
We use the convention 
$P({\bf k})=\frac{1}{(2\pi)^3} \int d^3 x e^{-i {\bf k} \cdot {\bf x}} \xi ({\bf x})$.}  
 \begin{equation}
P(k, \tau)= R_s^3 (\tau) P_0 \Big( kR_s(\tau) \Big) \,.
\label{self-similarity-Pk}
\end{equation}
It is commonplace also to define the dimensionless quantity
\begin{equation}
\Delta^2 (k) =  \frac{k^3\, P(k)}{2 \pi^2}\,,
\end{equation}
for which the self-similarity relation reads 
\begin{equation}
\Delta^2 (k, \tau)= \Delta_0^2 \Big( kR_s(\tau) \Big) \,
\label{self-similarity-Delta2}\,.
\end{equation}
 
The function $R_s(\tau)$ is derived from linear theory,
which describes a self-similar evolution from power
law initial conditions once the growing mode dominates.
Assuming the validity of linear theory at small $k$, 
so that $P(k, t) \propto D^2$, and taking 
$P(k) \propto k^n$, it follows immediately from 
(\ref{self-similarity-Pk}), or equivalently from 
(\ref{self-similarity-xi}) noting that  
$\xi(x) \propto 1/x^{n+3}$ at large $x$, 
that
\begin{equation}
R_s = e^{\frac{4 \alpha}{3+n} \Gamma \tau}=e^{\frac{2 t_s}{3+n}}=a^{\frac{2 \alpha}{3+n}}
\label{SS-scaling}
\end{equation}
where the last equality holds for $\kappa \neq 0$.

The two central assumptions stated above in deriving this 
result impose constraints on the range of $n$ for which it can
be expected to hold \cite{peebles}. 
For the first condition we require $n>-3$, since if $n\leq  -3$ density
fluctuations diverge unless an infrared cut-off is imposed (and
this cut-off then becomes a relevant length scale).
The second condition requires $n \leq 4$ since for $n >4$
linear theory is no longer valid: non-linear evolution 
produces a $k^4$ tail in the power spectrum at small 
$k$ which will dominate over the initial power\footnote{This can be considered as a special case 
of the bound originally derived by Zeldovich on 
fluctuations generated by local momentum and 
mass conserving physics \cite{zeldovich-k4, peebles}.}.

Whether there are in fact more restrictive bounds on
self-similarity has been discussed quite extensively in the
literature, and it has been one of the aims of numerical
simulation to determine its actual range of validity.
Whether the fact that the variance of the velocity fields, 
proportional in linear theory to that of the gravitational
field, requires an infrared regularization for $-3<n<-1$
may lead to breaking of self-similarity has been 
the subject of some consideration (see \cite{jain+bertschinger_1997}
and references therein). The conclusion of both theoretical 
and numerical study (for the standard EdS model) is that this
is not the case. The essential point is that what
is relevant in the clustering dynamics is the
{\it difference} in gravitational forces on point 
at a finite separation, and the fluctuations in this 
quantity have exactly the same infrared properties as
those in the mass density field (\cite{jain+bertschinger_1997},
see also discussion in \cite{gabrielli_etal_2010a}).
Other authors have placed in question the
validity of self-similarity at larger $n$, on 
the grounds that bluer spectra may require
ultra-violet regulation of physically relevant
quantities. Notably the range $1<n<4$ has 
sometimes been hypothesized to be excluded,
because the mass fluctuation in a top-hat window 
becomes dependent on the ultraviolet cut-off 
\cite{peebles}, or because the fluctuations in 
gravitational fluctuations require such a 
cut-off (see e.g. \cite{smith}). To our knowledge
the only numerical study of this range in three 
dimensions prior to the present one is that 
of \cite{sl1} for $n=2$ in a static  universe, which 
has found that self-similarity does indeed apply
in this case. The one dimensional study 
of \cite{yano+gouda} treats the cases $n=1,2,3$ 
for $\kappa=1$, that of \cite{joyce+sicard_2011} 
both $n=2$ and $n=4$ for $\kappa=1$ and
$\kappa=\sqrt{3}$, while 
\cite{benhaiem+joyce+sicard_2013} treats
$n=2$ for many models with $\kappa$ 
ranging from $0$ to $2$. All find self-similar 
evolution in these cases. Below we will report
 results showing clearly that self-similarity is 
 manifestly valid for $n>1$ in three dimensions 
 not only in the usual EdS model, but also in the 
 full family of models we treat.

\subsection{The stable clustering approximation for subsystems}

The basis of the stable clustering hypothesis in the usual EdS model
is the fact that a finite substructure in an expanding universe, {\it provided tidal
forces exerted by all other mass may be neglected}, evolves
in {\it physical coordinates} as if it were an isolated self-gravitating
system in a non-expanding universe. Formally this can be seen by
transforming the equations of motion in the form (\ref{3d-equations-1}),
using (\ref{3d-equations-2}) with the smoothing set to zero,
to physical coordinates ${\bf r}_i= a(t) \, {\bf x}_i$.  This gives
\begin{equation}
	\frac{d^2 {\bf r}_i}{dt^2}  = - Gm \sum_{j \neq i}^P
	\frac{{\bf r}_i - {\bf r}_j}{\vert {\bf r}_i - {\bf r}_j \vert^3} + \frac{1}{a}\frac{d^2 a}{dt^2}\bf{r}_i.
	\label{3d-equations-4}
\end{equation}
The infinite sum on the right hand (with the implicit regularization given by
subtracting the mean density) can be calculated by summing in a 
sphere with centre at the centre of mass of the chosen 
substructure, and taking its radius to infinity. This sum can then be divided 
into three parts:  (1) the force exerted by the other particles in the 
substructure, (2) the force exerted by the particles outside the substructure, 
and (3) the force associated with the subtraction of the background, 
which by Gauss' theorem, may be written as $+\frac{4\pi G}{3} \rho_0 
[\bf{r}_i - \bf{r}_c] $ where $\bf{r}_c$ is the position of the centre of
the subsystem. Neglecting in the second contribution all but the 
net force exerted at $\bf{r}_c$,  using 
$\frac{1}{a}\frac{d^2 a}{dt^2}=-\frac{4\pi G}{3} \rho_0$,
and setting ${\bf r}_c=0$ (so that ${\bf r}_i$ is the position
relative to the centre of mass), we obtain for particles in the 
substructure ${\cal S}$, 
\begin{equation}
	\frac{d^2 {\bf r}_i}{dt^2}  = - Gm \sum_{j \neq i, j \in {\cal S}}
	\frac{{\bf r}_i - {\bf r}_j}{\vert {\bf r}_i - {\bf r}_j \vert^3}, 
	\label{3d-equations-4b}
\end{equation}
where ${\bf r_i}$ is now the position relative to the centre of
mass of the subsystem,  i.e., the equations of motion of a 
completely isolated substructure\footnote{A more rigorous
derivation  of these equations for an isolated subsystem in a 
cosmological simulation, using a precise standard prescription 
for the calculation of the infinite periodic sum, is given 
in \cite{joyce+syloslabini_2012}.}. 

For the class of models we are studying the same steps may
be followed, the only difference being that we have now 
$\frac{1}{a}\frac{d^2 a}{dt^2}=-\kappa^2 \frac{4\pi G}{3} \rho_0$
and therefore obtain instead 
\begin{equation}
	\frac{d^2 {\bf r}_i}{dt^2}  = - Gm \sum_{j \neq i, j \in S}
	\frac{{\bf r}_i - {\bf r}_j}{\vert {\bf r}_i - {\bf r}_j \vert^3}  + \frac{1}{a^3}	
	(1- \kappa^2)\frac{4\pi G}{3} \rho_0 \bf{r}_i,
	\label{3d-equations-5}
\end{equation}
where ${\bf r}_i$ is, again, the position with respect to the centre of
mass of the substructure.
 
Thus, for $\kappa \ne 1$, the equations in physical coordinates for
a subsystem, in absence of tidal forces from other mass, include
a contribution from the background coming from the component 
of the comoving mass density $(\kappa^2-1)\rho_0$ which
does not cluster. Note that the result applies also for $\kappa=0$,
in which the physical and comoving coordinates are identical, and
(\ref{3d-equations-5}) is then simply a direct rewriting of 
(\ref{3d-equations-1}) in which the background 
subtraction (``Jeans' swindle") is explicited (and the
tidal forces neglected). 

Considering a subsystem of radius $\sim R$ containing $N$
particles,  we note that the first force term ${\bf F}_{\cal S}$
is of order $GmN/R^2$. If  the force term associated with the 
background is denoted  ${\bf F}_{B}$ we then have that
\begin{equation}
\frac{\|{\bf F}_{B}\|}{\|{\bf F}_{\cal S}\|} \sim \frac{|1- \kappa^2|}{a^3} \frac{\rho_0}{\rho_{\cal S}}
\end{equation}
where $\rho_{\cal S}$ is the physical mass density of the subsystem.
The condition that the subsystem is non-linear is simply that its
density be large compared to the mean density, i.e., 
 ${\rho_{\cal S}} \gg  \rho_0/a^3 $. Thus, in the same
approximation, the background term in the equations of 
motion in physical coordinates becomes a small perturbation.
In absence of this term the substructure will virialize and
remain stationary in physical coordinates, i.e., with
$\rho_{\cal S}$ approximately constant. Thus for 
$\kappa >0$ a non-linear system on which
the tidal forces of other mass can be neglected, will 
always asymptotically virialize and remain macroscopically 
stable in physical coordinates. The smaller is $\kappa$, the 
longer time (and higher density) will be needed to reach this regime,
while in the static limit it may never be better than 
a reasonable approximation (as the density of the structure
relative to the background will not, a priori,  grow).
In any case, as we will discuss below, the approximation
that structures may evolve independently of others
in this way is clearly one which will break down for
sufficiently small $\kappa$.

\subsection{Exponent of non-linear two point correlation function}

Let us now make the assumption that, {\it in some range of scale},
the statistical properties of the non-linear structure which develop 
are stable in physical coordinates. Assuming that the 
associated clustering is self-similar, this implies immediately 
that the clustering in this range is strictly scale-free 
in our models,  i.e., it has {\it no characteristic length scale}: 
if there were such a scale it would scale in time in 
comoving units in proportion to $1/a$, which is inconsistent
with the self-similar form of the two point
correlation function (\ref{self-similarity-xi}). In short:
if there is a regime of self-similar stable clustering in
these models, it must be truly scale-free. As underlined
in \cite{peebles}, this implies that the associated 
clustering is fractal in nature, corresponding
to a ``virialized hierarchical" clustering.

Such clustering is characterized by its exponents, and 
we focus here just on the two point correlation function. 
If there is a regime of stable clustering, it must then
have a power law behaviour $\xi(x) \sim x^{-\gamma_{sc}}$.
To derive this exponent it suffices to consider that 
it applies at any time between two scales: a scale
 $x_{min}$ and $x_{max}$ say. The upper cut-off,
marking the breakdown of stable clustering, must
scale as required by self-similarity, i.e.,  
$x_{max} \propto R_s(\tau)$, and correspond
to some {\it constant} value $\xi (x_{max})= \xi_{max}$.
The lower cut-off $x_{min}$, on the other hand,
must scale in the same way only {\it if  the clustering is 
self-similar below this scale}. Alternatively 
this lower cut-off $x_{min}$ can scale 
differently, and in this case it marks also a 
lower cut-off to self-similarity.  In this case
$x_{min}$ is necessarily related to some 
other scale in the system, such as the 
initial interparticle distance, or the force 
smoothing length. As we will discuss,
this appears to be the case actually realized
in all our numerical simulations (which
have finite resolution). 

To calculate the exponent associated
with stable clustering, we consider the correlation 
function between an {\it arbitrary} physical scale 
$r_{sc}$ in the range of stable  clustering, and 
the scale $x_{max}$. In comoving units the former 
corresponds to a scale which depends on time as 
$x_{sc} (\tau)= x_{sc} (0) a_0/a$. In the
non-linear regime, $\xi \gg 1$, and stability 
of clustering in physical coordinates implies 
that $\xi$ varies only because of its normalisation 
to the mean density in physical coordinates, i.e.,  
$\xi( x_{sc} (\tau)) = \xi (x_{sc} (0)) (a/a_0)^3$. 
Assuming a power law behaviour of the
correlation function, $\xi(x) \sim x^{-\gamma_{sc}}$,
and using these scalings, we have
\begin{equation}
	\gamma_{sc}  = - \frac{{\ln \left( \frac{{{\xi _{max}}}}{{{\xi _{sc}}}} \right)}}{{\ln \left( \frac{{{x _{max}}}}{{{x_{sc}}}} \right)}}  =\frac{3{\ln \left( \frac{a}{a_0} \right)}+{\ln \left( \frac{{{\xi _{sc} (a_0)}}}{{{\xi _{max}}}} \right)}}{\left(\frac{2\alpha}{3+n}+1\right){\ln \left( \frac{a}{a_0} \right)}+{\ln \left( \frac{{{x _{max}(a_0)}}}{{{x_{sc}(a_0)}}} \right)}}	\,.
\end{equation}
We can always choose $a_0$ such that $x _{max}=x_{sc}(a_0)$, at which
time $\xi _{sc} (a_0)=\xi _{max}$, and therefore we have
\begin{equation}
	\gamma_{\rm sc} \left( {n, \kappa } \right) = 
		\frac{3(3+n)}{3+n+2\alpha}
	=\frac{6(3+n)}{5+ \sqrt{1+\frac{24}{\kappa^2}} +2n}\,.
\label{sc-prediction}
\end{equation}

\begin{figure}
	\resizebox{!}{5.8cm}{\includegraphics[]{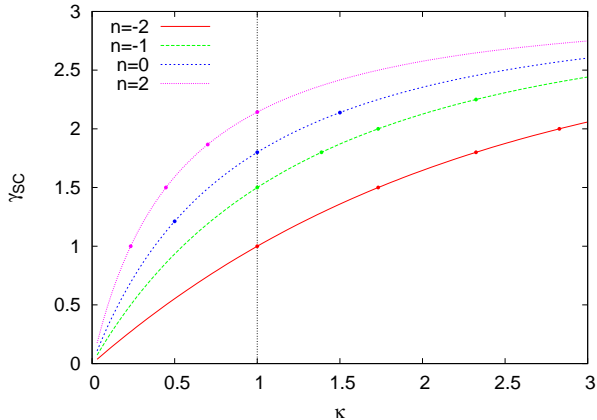}}
	\caption{Exponent $\gamma_{sc}$ of the non-linear two point correlation function 
	predicted using the stable clustering hypothesis, as a function of $\kappa$ for different chosen $n$.
	The usual EdS prediction corresponds to the case $\kappa=1$. The points on the solid line indicate
	the models for which we report results of numerical simulations in the second part of the paper.}
	\label{fig-Gamma_SC}
\end{figure}

This result is plotted in Fig. \ref{fig-Gamma_SC} showing 
the prediction  $\gamma_{\rm sc} \left( {n, \kappa } \right)$
as a function of $\kappa$ for different $n$.
In the case $\kappa=1$ we recover, as expected, the well 
known result of \cite{davis+peebles_1977, peebles} :
\begin{equation}
	\gamma_{\rm sc} \left( {n, \kappa=1 } \right) = 
		\frac{3(3+n)}{5+n}\,.
	\label{sc-prediction-Peebles}
\end{equation}
Note that $\gamma_{\rm sc}\left( {n, \kappa } \right)$ is positive
provided $n > -3$, which is, as we have discussed above,  
precisely the lower bound on $n$ for which self-similarity
can apply.  Further $\gamma_{\rm sc}\left( {n, \kappa } \right)$
is a monotonically increasing function of both $\kappa$ and $n$,
and it is bounded above by the spatial dimension. This is 
precisely the bound which is required if the correlation
function is that of a scale invariant mass distribution 
(as $3 -\gamma$ is the associated fractal dimension).
For $\kappa=0$ 
on the other hand, we obtain 
$\gamma_{\rm sc}=0$. This means that
in this limit the predicted exponent is not 
consistent with a scale invariant distribution,
and indeed, as we will discuss further below,
the stable clustering hypothesis ceases to be 
physically reasonable in this case.

\subsection{Validity of stable clustering}

Let us now consider the 
validity of the stable clustering hypothesis. In models
of the kind we study, with cold initial conditions, 
and with $n$ in the range $-3 < n \leq 4$, 
structure formation is hierarchical in nature: 
fluctuations go non-linear and collapse at 
successively larger scales. In the stable clustering 
hypothesis we suppose that structures  
collapse and virialize at a given 
time and are thereafter essentially undisturbed 
by the subsequent evolution, i.e., they are
subsumed in larger structures but remain
essentially unchanged in physical
coordinates (with respect to their own
centre of mass). Clearly this can be at
best a good approximation for some
time: any given structure will evolve
in physical coordinates because of
interactions with other structures,
and indeed can even merge with 
other ones. The relevant question is 
therefore {\it how good an approximation 
stable clustering provides},  rather than 
whether the stable clustering hypothesis is strictly valid or not.
More specifically the question is how
well the predictions for macroscopic quantities 
furnished by this hypothesis work, and {\it over
what range of scale}. 
In the second part of this paper we will focus, as 
in various other works in the literature, on trying
to answer this question for the two point correlation 
function in the non-linear regime, for which we have just 
derived the prediction.

The class of models we are studying is a two
dimensional family, and, if stable clustering is
a relevant approximation for describing non-linear
clustering, we would expect that the degree 
to which it is valid will depend on the 
parameters $\kappa$ and $n$. In the preceeding 
study \cite{joyce+sicard_2011, benhaiem+joyce+sicard_2013} 
of models in one spatial dimension, 
we have shown that there is in fact a simple 
qualitative answer to this question which 
is suggested by simple theoretical 
considerations, and which turns out to be 
remarkably well born out by numerical study. 
With trivial modifications, as we will now 
explain, the same considerations apply 
in three dimensions: assuming stable 
clustering (and self-similarity) to apply, the 
exponent $\gamma_{sc}$ we have just 
calculated can be shown to control directly 
the {\it relative size} of virialized objects 
of different masses; it is natural, as we
will explain, to consider this as probably
the essential parameter controlling the 
validity  of stable clustering.
More specifically, this reasoning suggests 
that  the criterion for the validity of stable 
clustering can be expected to be that
$\gamma_{sc}$ be sufficiently large.
This expectation has been born out
in the one dimensional models, with
the ``critical value" situated at
$\gamma_{sc} \approx 0.15$.
The goal of our numerical study in 
this paper is to see if an 
analogous result holds
in three dimensions.

Let us consider then two overdensities of mass
$M_1$ and $M_2 > M_1$, corresponding to
initial comoving scales $L_1^0$ and $L_2^0$
respectively. Assuming that their collapse
is self-similar (as, for example, in the spherical
collapse model), the ratio of their sizes when
they virialize is equal to the initial value 
of this ratio. One can infer also that the time-delay 
between their respective virialization, at scale
factors $a_1$ and $a_2$ say, is given by 

\begin{equation}
	R_s (\tau_2-\tau_1)= \left(\frac{a_2}{a_1}\right)^{\frac{2\alpha}{3+n}}= 
	\left(\frac{L_2^0}{L_1^0}\right)
\end{equation}

Now, if we assume that the first structure is stable
from the time it virializes, we can deduce that the 
ratio of the sizes of the two structures decrease 
by a factor of $\frac{a_1}{a_2}$ during the interval
between their virialization. Thus, when the 
second (larger) structure virializes, we have 
that the ratio of the sizes of the two structures
is
\begin{equation}
	\frac{(L_2/L_1)}{(L_2^0/L_1^0)}=\left(\frac{L_2^0}{L_1^0}\right)^{-\frac{3+n}{2\alpha}} =\left(\frac{L_2^0}{L_1^0}\right)^{-\frac{\gamma_{sc}}{3 - \gamma_{sc}}} 
\label{scaling}
\end{equation}

Quite simply, the larger is $\gamma_{sc}$, the more a structure
which has virialized can ``shrink" relative to a larger structure
which virializes later. Or, in other words, the larger $\gamma_{sc}$
is, the more ``concentrated"  are the pre-existing virialized 
substructures inside a larger structure when it collapses. 
This is precisely the property one would expect to be
relevant for the validity of stable clustering: if the
sub-structures inside a structure are smaller 
(and therefore more tightly bound), the process
of their disruption by tidal forces and mergers
will be much slower and less efficient.
Indeed, in the limit that $\gamma_{\rm sc}$
tends to its upper limit, any structure which 
collapses and virializes will see the 
substructures which have collapsed
before it essentially as point particles,
and thus stable clustering should 
become exact in this limit. As $\gamma_{\rm sc}$ 
decreases, on the  other hand, we expect that the 
interaction between structures can lead more 
easily to their disruption, and in particular that 
mergers of substructures become much more probable. 

More precisely we would expect that, in a given
scale-free model (i.e. for given $n$ and $\kappa$), 
there will be a time scale $\tau_s$ characteristic of the 
stability of any structure. Now {\it if self-similarity applies},
i.e., if no other length scales (particle
discreteness scale, force smoothing scale)
have any influence on the macroscopic
evolution of the structure, this time
scale should be the same, for any structure, when 
expressed in terms of its own dynamical time scale.
This means that if a structure virializes at scale 
factor $a_v$, its stability will remain a good 
approximation until some scale factor $a_s$ 
where the ratio $a_v/a_s$ has some fixed
value, $\eta_s$ say. Supposing this to be
the case leaves the derivation given above 
of the exponent $\gamma_{sc}$ unchanged
--- no assumption about the behaviour
of the lower cut-off to stable clustering
was made. However it gives us also 
a prediction that the power law region
in the correlation function should have a
lower cut-off $x_{min} \sim \eta_s x_{max}$.
Adopting the arguments above, we
would expect this ratio $\eta_s$ to depend 
on $n$  and $\kappa$ only through $\gamma_{\rm sc}$,
and to decrease monotonically as 
$\gamma_{\rm sc}$ increases, so that the 
range of scale in which stable clustering
may apply will stretch monotonically
as $\gamma_{\rm sc}$ increases.
Note that, in any case, the scale $x_{min}$, 
if it exists, must then also scale as defined by 
self-similarity, i.e., $x_{min} \propto R_s(\tau)$.

\subsection{Range of stable clustering in a finite simulation}
\label{range of SC}
The above considerations, and the evidence supporting
them in the one dimensional studies of 
\cite{joyce+sicard_2011, benhaiem+joyce+sicard_2013}, 
motivate and structure our numerical study: the goal is
to measure, in our $(n,\kappa)$ model space, as well
as possible the form of the self-similar two point 
correlation function and to determine in particular in 
what range of scale the prediction of stable clustering 
may describe it well. In particular  we would like to 
determine, whether, as anticipated, $\gamma_{sc} (n,\kappa)$ 
is the parameter relevant to answering this
question, and if there a characteristic
or ``critical" value of $\gamma_{sc} (n,\kappa)$ below
which the stable clustering approximation breaks down 
completely, as has been found in the one dimensional 
studies (at a value $\gamma_{sc}$ in the range between 
$0.15 - 0.2$). These studies also illustrate the numerical 
difficulties which arise in addressing these questions, and 
the extent to which they can actually be understood and
anticipated from the considerations above. Indeed,
as we will now explain, we expect $\gamma_{sc}$
not only to be an indicator for the range of scales
in which the ``true" self-similar two point 
correlation function ---- without any limit of
spatial resolution --- may be well described by 
the stable clustering predictions, but also to 
control the range of scale over which it
may potentially be measured in a numerical
simulation of a given finite size.

Given that we set out to detect self-similar clustering,
and to assess the validity in particular of the stable 
clustering approximation, it is  evidently relevant
to estimate the range of scales over which 
such self-similar stable clustering would be
expected to develop in a simulation of 
given particle number $N$, {\it if we assume 
that this approximation apply}. The particle number
$N$ fixes the simply the temporal range over which 
evolution can be simulated, as this
is bounded above by the time at which the scale
of non-linearity approaches the box size. Let us
call $\tau_i$, corresponding to $a=a_i$, 
the time when the {\it first} non-linear structure
--- say of order one hundred particles --- 
virializes, with a comoving size $x_i$.
The simulation can be run (if numerically
feasible) until a time  
$a=a_f$  when the largest approximately 
virialized region is of a size $x_f$reaches 
some small fraction of the box-size $L$.  
Using self-similarity it follows that
\begin{equation}
\left(\frac{a_f}{a_i}\right)^{\frac{2\alpha}{3+n}} \approx 
\left(\frac{x_f}{x_i}\right)
\end{equation}
Assuming further that structures are stable once they
virialize we have that  $x_{min} (a_f)$,  the lower cut-off 
to stable clustering the end of the simulation, obeys
the relation
\begin{equation}
\ln \left(\frac{x_{min} (a_f)}{x_i}\right)  \approx \ln \left(\frac{a_i}{a_f}\right) \approx 
\frac{\gamma_{sc}}{3-\gamma_{sc}} \ln \left(\frac{x_f}{x_i}\right) \,.
\label{SCrange-estimate0}
\end{equation}
Likewise, using again that the upper cut-off to stable clustering $x_{max}$
at the final time is simply $x_f$, we have
\begin{equation}
\ln \left(\frac{x_{min} (a_f)}{x_{max}(a_f)}\right) \approx
\frac{3}{3-\gamma_{sc}} \ln \left(\frac{x_f}{x_i}\right) 
\label{SCrange-estimate}
\end{equation}
These estimates can be modified easily to incorporate
a breakdown of stable clustering as discussed above.
This becomes relevant if it is possible to evolve
sufficiently long so that $a_i/a_f < \eta_s$, in
which case the scale $x_{min} (a_f)$ can then becomes 
smaller than the ``true" $x_{min}$, and this lower cut 
can in principle be resolved. 

For a simulation of given size, the ratio $\frac{x_f}{x_i}$
is, to a first rough approximation, independent 
of both $n$ and $\kappa$: $x_i$ is proportional to the 
initial interparticle distance, and $x_f$ is limited to
be some fraction of the box size\footnote{More 
exactly, both $x_i$ and $x_f$ will in fact 
depend on $n$ and $\kappa$:  $x_i$ depends 
on the density at virialization which is  
expected to increase (slowly) with $\kappa$
(from an analysis of the spherical collapse 
model for this class of models, which we will 
present elsewhere); 
the scale $x_f$  attainable is expected
to be smaller for redder spectra  (i.e. smaller $n$) 
because of their greater sensitivity 
to small $k$ power which is cut-off by the box.
Further, as we will see, considerations of numerical
cost will mean that it is not always feasible in
practice to evolve all models to the same $x_f$.}.
Therefore, if stable clustering applies, it will become 
increasingly difficult to robustly probe
it numerically as $\gamma_{sc}$ decreases.
In short just as the range $[x_{min}, x_{max}]$ 
is expected to shrink as $\gamma_{sc}$ decreases,
the range of numerically accessible scales does too. 
Conversely,  to access as much of the 
range over which we can measure the exponents 
for larger $\gamma_{sc}$, we will need to use a 
small force smoothing (to resolve down to $x_{min} (a_f)$) 
which is more challenging numerically.

\section{Numerical simulations: methods and results}

The aim of our study here is to characterize how non-linear clustering
depends on the initial conditions, parametrized by $n$, and the cosmology,
parametrized by $\kappa$.  We have thus aimed to produce a large library of 
$N$ body simulations (of purely self-gravitating ``dark matter" particles)
covering a significant range of these parameters. In this paper we will focus 
our analysis on two point correlation properties, while other complementary 
analyses using other tools will be reported in future work. In particular,
as discussed, we will focus here on the degree of validity of self-similarity
and the relevance of the stable clustering approximation.

\subsection{Simulation Code}
To do our simulations we have chosen to use the widely used and very
versatile GADGET2 code \citep{gadget}.  As the class of models we have
described are not equivalent to those which can be simulated by the
code --- models specified by the standard cosmological parameters
--- we need to modify it appropriately in order to realize this
possibility.  To do so one possibility is to modify the cosmological
version of the code. Another one, which is the method we have chosen,
is to modify the static universe version of the code
(i.e. non-expanding system in periodic boundary conditions) exploiting
the fact, which we have highlighted, that in our models the equations
of motion may be written in the form (\ref{3d-equations-3}) where
$\Gamma$ is a constant, i.e., the system is equivalent to a static
universe with a constant fluid damping.  We have thus modified the
time-integration scheme of GADGET2 keeping the original
``Kick-Drift-Kick'' structure of leap-frog algorithm and modifying
appropriately  the ``Kick'' and ``Drift'' operation. The structure of the code is
otherwise unchanged. Details can be found in Appendix ~\ref{app-gadget2}.

\begin{figure*}
	\resizebox{!}{5.8cm}{\includegraphics[]{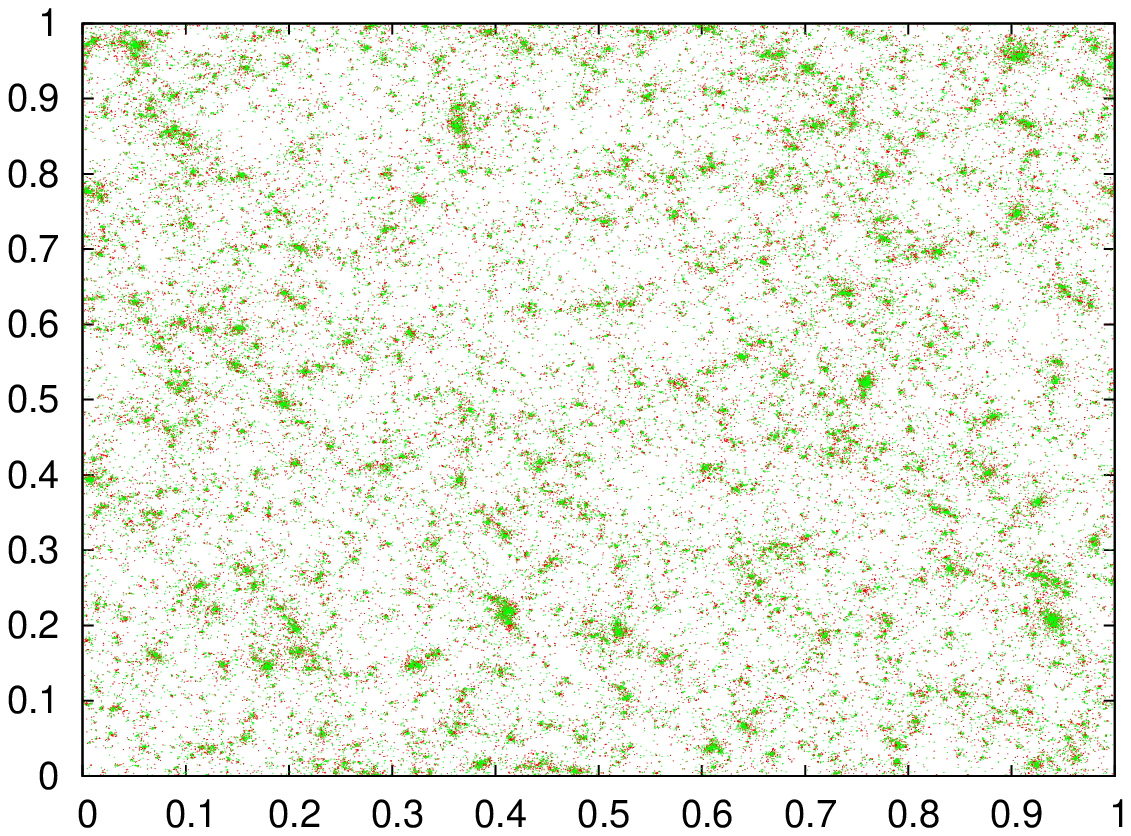}}
	\resizebox{!}{5.8cm}{\includegraphics[]{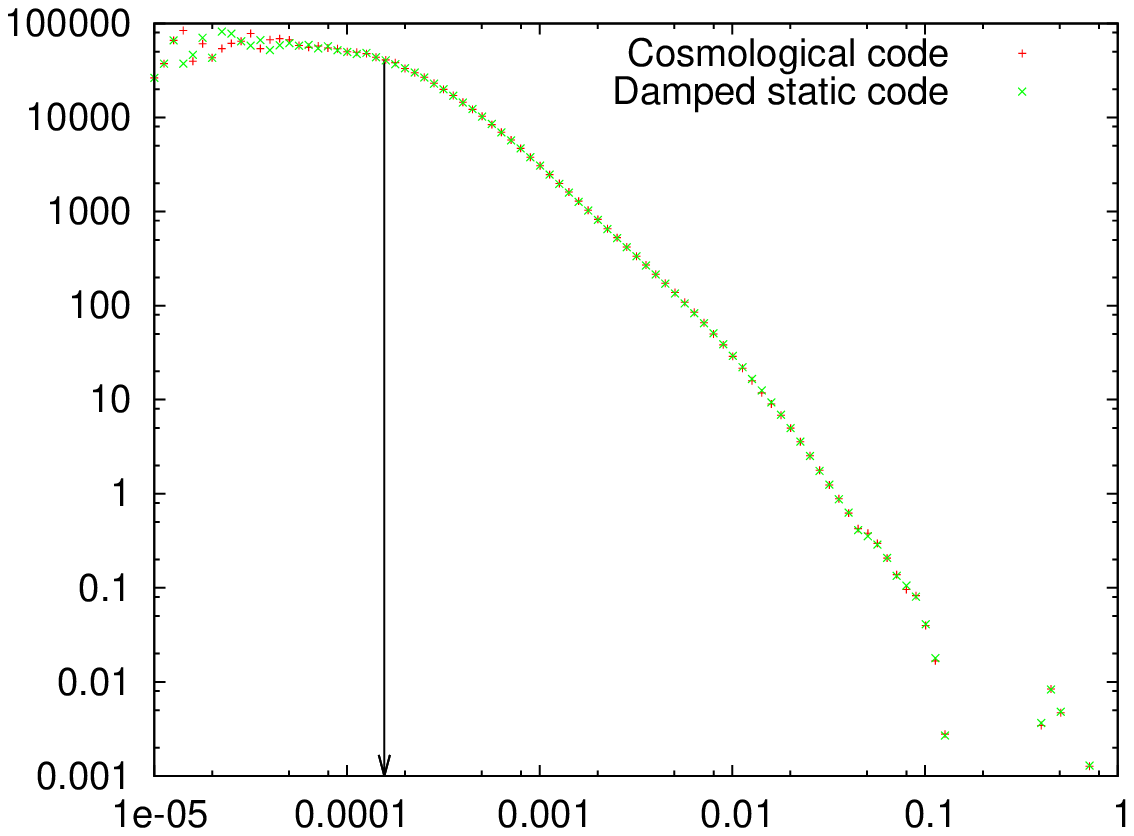}}
	\caption{Comparison of results obtained from two EdS ($\kappa=1$) simulations of identical $n=0$ 
	initial conditions: in the left panel, the particle positions in the simulation box projected onto a plane; in 
	the right panel, the measured two point correlation functions. The green points correspond to a simulation
	performed using our modification of the static version of GADGET2, and the red points a simulation
	using the existing cosmological version of the code.} 
	\label{Cosmo_vs_Damping}
\end{figure*}

Tests of this code --- notably using energy conservation --- will be discussed 
below in assessing the reliability of the results of all our simulations. One 
additional simple 
independent test of it we have done is the following: comparison between the case
$\kappa=1$, which corresponds to the usual EdS model, with the results obtained
for this case using the cosmological --- expanding universe --- version of GADGET2
for the same case (i.e. $\Omega_m=1$ and $\Omega_\Lambda=0$). We have performed
this test and found very satisfactory results. Shown, for example,
in Fig~\ref{Cosmo_vs_Damping}, is the comparison of the results obtained using the two codes,
evolved up the same scale factor,  starting from an identical initial condition
given by a realization of the case $n=0$. 
[Further details on the numerical parameters chosen will be given below,
and we note that we use here, as everywhere in the paper, {\it length units in
which the periodic box size is unity}]. The left panel shows a projection
of the particle positions, with those corresponding to the ``standard" code
in red and those of our modified static code in green; the right panel
shows the two point correlation function measured in the two cases,
with the black vertical line indicating the force smoothing scale.  
While the first reveals some visual differences between the two 
simulations --- which is to be expected as these are two different
integrations of the same chaotic system ---  the latter shows that 
the statistical properties (which is what we will measure with
such simulations) are in almost perfect agreement. 

\subsection{Simulation parameters}
Our results reported here are based on simulations from power law initial
conditions with exponents taking the values $n = -2,-1,0,2$ and values 
of $\kappa$ ranging in the $\kappa=0.23$ to close to $\kappa=3$.
Table~\ref{Table1} gives the exact values of $n$ and $\kappa$, and in each 
case also the associated predicted stable clustering exponent $\gamma_{sc}$, as
well as other parameters characterizing the initial amplitude and
duration of the simulation which we will discuss below.
As discussed in detail below this prediction turns out to be very good, and 
likewise, the estimates which we have given above for the range 
over which non-linear clustering develops. The choices of the values
of $\kappa$ simulated at each $n$ are thus appropriately guided by 
the associated value of $\gamma_{sc}$, which we have chosen to vary 
in the range $1 \leq \gamma_{sc} \leq 2.25$.  The simulations in 
Table~\ref{Table1} are for $N=128^3$ particles for the cases $n=-2$,
and for $N=64^3$ otherwise. We report separately at the end
of this section a comparison with a pair of larger simulations 
with $N=256^3$ for one case ($n=-1$, $\kappa=1$).

The lower bound on the range of $\gamma_{sc}$ has been chosen 
because, at $\gamma_{sc} \sim 1$ (corresponding to $n=-2$ for the 
case $\kappa=1$), we find that the region of self-similarity 
we can access becomes too small to allow any robust statement about 
the strongly non-linear part of the correlation function.  As we will discuss 
below, this is a simple generalisation of the same difficulty which has 
been observed in the literature for the usual $\kappa=1$ case, where 
the accessible range of self-similarity has indeed been observed to 
decrease greatly as $n$ decreases towards $-3$, with the validity 
of self-similarity below $n=-1$ a subject of discussion in the 
literature \citep{ efstathiou_88, jain+bertschinger_1997, jain+bertschinger_1998}.
As we will highlight below, one of the things we show very
clearly from our study in this larger class of models is that this difficulty
is not essentially related to the convergence properties in the 
infra-red of these spectra, but instead arises because 
 $\gamma_{sc}$ is small. Indeed at $n=-2$ we will
see that we have no difficulty observing self-similarity
when $\kappa$ is increased significantly above unity.   

The upper bound on $\gamma_{sc}$ is, on the other hand, 
related to the lower limit on the spatial resolution imposed 
by the force smoothing.  We use the version of GADGET2
with a smoothing which is fixed in comoving coordinates
(as is the practice in many large volume cosmological simulations, 
and in particular in almost all studies of the issues explored
here).  The choice of the force smoothing parameter $\varepsilon$
is an essential question as it conditions also greatly the number of 
particles which can be simulated, for given numerical resources: 
smaller smoothing requires smaller time steps. In order to determine 
whether stable clustering applies, however, we must clearly have 
the numerical  resolution necessary to detect it if it is a good
approximation, and this imposes in principle the choice of 
an $\varepsilon$ which is as small as possible. To do so,
we need to be able to  ``follow''  for as long as possible the 
(possibly stable) evolution of non-linear structures. The first 
virialized structures which can be resolved --- with of order 
one hundred particles, say --- have a size of order the initial
 interparticle distance  $\Lambda=L/N^{1/3}$. For $\kappa=1$,
 for example, the density at virialization --- following the 
 standard estimate of the spherical collapse model ---  is
of order $200$ times the mean mass density, equal 
to $1/\Lambda^3$.  Such a structure can
only be well simulated provided the force smoothing,
$\varepsilon$, is sufficiently small compared to 
the size of the structure. In the case of
stable clustering its size decreases in comoving 
coordinates, in proportion to $1/a$.  Thus,
for chosen $f=\varepsilon/\Lambda$, we can 
follow the (possible) stability of structures 
over a range of scale factor strictly bounded 
below by $1/f$. This can be seen also in 
terms of the estimate (\ref{SCrange-estimate0}) 
given above: in order to resolve fully the
regime of stable clustering we need 
$\varepsilon$ to be significantly smaller
than $x_{min}$. It is clear that, if stable
clustering applies and we wish to
resolve it well for values of $\gamma_{sc} \sim 2$,
we need to have a value of $\varepsilon$  
very considerably smaller than $\Lambda$.
We could, alternatively, evolve the system to 
times when structures containing a significant 
number of particles should, following stable clustering,
 ``shrink" below the smoothing scale. In principle 
 one should still obtain then the correct evolution 
 sufficiently far above $\varepsilon$. However
 the scale and manner in which clustering 
 above $\varepsilon$ is modified in such a
 regime is very difficult to control for and 
 would introduce another source of 
 uncertainty in our results. 

Given these considerations, and following tests of the numerical 
cost of simulations, and of energy conservation (see below), 
we have thus chosen to take the following values for
the GADGET2 parameters in the simulations reported in
Table \ref{Table1}:

\begin{itemize}
\item Force softening $\varepsilon=0.01 \Lambda$  (corresponding
to a spline softening with compact support of radius $2.8\,\varepsilon$)
\item Timestepping parameters:  {\tt ErrTolIntAccuracy}=0.001, 
{\tt MaxRMSDisplacementFac}=0.1 and {\tt MaxSizeTimestep}=0.01. 
We note that these values are smaller (by factors of
$25$ for the first, and $2.5$ for the two others) than the values suggested
in the GADGET2 userguide and treated as ``fiducial" in the literature 
(see e.g. \cite{smith_etal_2012}).
These choices were made as we found they gave significant 
improvement in energy conservation (see discussion below).
\item Force accuracy fixed by {\tt ErrTolForceAcc}=0.005 
(a typical fiducial value)
\end{itemize}

In the final section of the paper we will compare in
detail our results to the previous studies (of EdS models), in 
particular to those of \cite{smith} and \cite{bertschinger2}. We just 
note here that the most important point to remark in our parameter 
choice is that our force smoothing, in units of the initial grid
spacing $\Lambda$, is approximately the same as 
that of \cite{bertschinger2}, but about six time smaller
than that of \cite{smith}. On the other hand, the particle
numbers ($N=128^3$ for our $n=-2$ simulations
and $N=64^3$ for the others) are smaller
than those of both these other studies ($N=256^3$).
Thus, while we have considerably better resolution
of non-linear clustering at small scales than \cite{smith} --- and 
in particular, as we will see, we can follow fully the propagation 
of self-similarity to smaller scales --- our results may be
more subject to finite size effects coming from the periodic box. 
While self-similarity provides in principle a good test for both
potential biases associated with the use of a small smoothing 
parameter and with finite box size effects, we will also test
carefully below more directly for both effects using, for a few
chosen cases, additional simulations with both larger smoothing
and larger particle number. In particular we will report 
at the end of the section a comparison of our results 
with a pair of further simulations with $N=256^3$
particles for one chosen model ($n=-1$ and $\kappa=1$).
For what concerns the use of a relatively small force
smoothing, one possible issue is possible bias of the
desired mean-field evolution due to two body collisionality.
In practice the main associated difficulty 
(see e.g. \cite{knebe_etal_2000, joyce+syloslabini_2012})
is that small smoothing can leads to poor energy 
conservation if the numerical accuracy of the integration
of hard collisions is not sufficiently accurate.  With this
particular concern in mind, we have performed, as reported
in detail below, detailed tests of energy  conservation, and 
have adapted tight criteria leading to the very accurate
choice of time-stepping parameters given above.
If, on the other hand, two body collisions are integrated correctly, 
the associated effects will not be diagnosed by an analysis of energy. However, if in such a 
case such collisions actually modify the macroscopic evolution, we  
should observe a breaking of self-similarity induced by this.
Thus such effects, if they are present, should also be excluded 
from our analysis by the test of self-similarity (which we will
apply to  all our results).

Our simulations were executed using a cluster at the University of Nice using MPI 
on between 32 and 128 processors, depending on the simulation. The time necessary 
to run them varied from a few days to a few weeks. 

\begin{table}
\begin{tabular}{c||c|c|c|c|c|c|}
 n & $\kappa$ & $\gamma_{sc}$ &  $\Delta^2_L(k_{N},t_s^0)$ & $t_s^f$ & $\Delta^2_L(k_{b},t_s^f)$  \\
\hline
\hline
-2 & 1.00 & 1.00 & 0.03 & 3.00 &  1.96 $\times 10^{-1}$\\
-2 & 1.73 & 1.50 & 0.03 & 3.00 &  1.96 $\times 10^{-1}$\\
-2 & 2.32 & 1.80 & 0.03 & 2.80 & 1.31 $\times 10^{-1}$\\
-2 & 2.83 & 2.00 & 0.03 & 2.50 & 7.20 $\times 10^{-2}$\\
\hline
-1 & 1.00 & 1.50 & 0.06 & 4.00 & 1.79 $\times 10^{-1}$\\
-1 & 1.39 & 1.80 & 0.06 & 4.00 &1.79 $\times 10^{-1} $\\
-1 & 1.73 & 2.00 & 0.06 & 4.00 & 1.79 $\times 10^{-1}$ \\
-1 & 2.32 & 2.25 & 0.06 & 4.00 & 1.79 $\times 10^{-1}$\\
\hline
0 & 0.50 & 1.21 & 0.94 & 4.00 & 8.57 $\times 10^{-2}$\\
0 & 1.00 & 1.80 & 0.94 & 4.00 & 8.57 $\times 10^{-2}$\\
0 & 1.50 & 2.14 & 0.94 & 4.00 & 8.57 $\times 10^{-2}$\\
\hline
2 & 0.23 & 1.00 & 0.94 & 4.50 & 2.28 $\times 10^{-4}$\\
2 & 0.45 & 1.50 & 0.94 & 4.50 & 2.28 $\times 10^{-4}$\\
2 & 0.70 & 1.87 & 0.94 & 6.00 & 4.57 $\times 10^{-3}$\\
2 & 1.00 & 2.14 & 0.94 & 5.00 & 6.19 $\times 10^{-4}$\\
\hline
\hline
\end{tabular}
\caption{Parameters characterizing our $N$ body simulations 
with $N=128^3$ particles for the cases $n=-2$, and $N=64^3$
for the other cases.  
$t_{s}^f$ is the duration in ``static time" units of the simulation, equal to the 
logarithm of the linear growth factor; 
the quantities  $\Delta^2_L(k_N,t_s=0)$ and $\Delta^2_L(k_{b},t_s^f)$
characterize the initial and final amplitudes of the fluctuations.
Note that we use units in which the box
size is unity so that the wavenumber $k_b$ of the fundamental mode $k_b$ 
is equal to $2\pi$.}
\label{Table1}
\end{table}

\subsection{Initial conditions and duration of simulations}

We generate our initial conditions using the standard method used
in cosmological simulations(see e.g. \cite{bertschingercode, discreteness1_mjbm}): 
to particles initially on a simple cubic lattice,  
we apply a displacement field generated 
as a sum of independent gaussian variables in reciprocal 
space with variance determined by the desired  
linear power spectrum, and including all modes 
up to the Nyquist frequency $k_N=\pi/\Lambda$ 
(i.e. we sum over $\vec{k}$ such that each component
$k_i \in [-k_N, k_N]$.)      
If we denote $\vec{u}_{i,0}$ the resulting displacements
of the particles, the initial velocities 
$\vec{v}_{i,0}$ are then fixed
simply using the Zeldo'vich approximation  
\begin{equation}
	\vec u (\tau) = D (\tau) \vec u_{i, 0}
\end{equation}
where $D(\tau)$ is the linear growth factor of the
growing mode solution (\ref{linear theory-growing-general})
and the simulations starts at $\tau=0$ (and thus $t_s=0$) 
so that
\begin{equation}
 \vec{v}_{i,0} =  2 \alpha \Gamma \, \vec u_{i, 0}  ={\sqrt{\frac{\pi G \rho_0}{6}}} [-\kappa + \sqrt{\kappa^2+24}] \, \vec u_{i, 0}\,.
\end{equation}

We take an initial power spectrum $P_L(k, t_s=0)=A_0 k^n$,
and following common practice we characterize the 
initial amplitude of fluctuations by specifying the
value of $\Delta_L^2 (k_N)= \frac{A_0 k_N^{3+n}}{2 \pi^2}$,
which is (approximately) the normalized mass variance 
in a gaussian sphere of radius $\Lambda$. 
In fixing the initial amplitude of our simulations as given 
in Table~\ref{Table1}, we have used as 
guidance the previous work notably of \cite{jain+bertschinger_1998} and 
\cite{knollmann_etal_2008} which report tests showing that self-similarity 
is recovered better for the cases of smaller $n$ if low amplitudes
are used. Thus the amplitude for our simulations with  
$n=0$ and $n=2$ corresponds at the starting time 
to $\Delta^2(k_{N}) \approx 1$, while for the two other
cases they are significantly smaller. Also given in
Table~\ref{Table1} are the final times $t_s^f$ 
considered for our analysis in each of the simulations, and 
the corresponding values of the
linear theory extrapolated amplitude $\Delta_L^2 (k_b, t_s^f)$ 
at the fundamental mode of the periodic box $k_b=2\pi/L$. 
The latter corresponds approximately to the normalized  
mass variance in a gaussian sphere of order the size
of the box. In the cases $n=-2$ and $n=-1$ our 
simulations thus extend to times when this quantity
is no longer much smaller than unity, and one would
expect this to lead to significant finite size effects.
We will indeed detect such effects clearly and consider
carefully the limitations they place on our results. 
The final times $t_s^f$ in the other simulations are, on
the other hand, significantly smaller than those at which
such effects might be expected to become significant, 
and they are determined in most cases rather by the 
numerical cost of integration or considerations of energy 
conservation which we discuss below.

\subsection{Monitoring of Energy}
\label{Monitoring of Energy}

In $N$ body simulations in a non-expanding space energy conservation 
is the most fundamental control on numerical precision, and poor energy
conservation (less than a few percent) is known to be indicative typically
of a poor representation of macroscopic properties (see e.g. 
\cite{hockney}).
In simulations in an expanding background total energy is not conserved, and 
one thus no longer disposes of this robust and simple instrument of control 
on simulation accuracy.  
Nevertheless one can exploit and test a constraint on the evolution of 
energy, given by the so-called Layzer-Irvine equation, which is 
usually written (see e.g; \cite{peebles}) as 
\begin{equation}
\frac{dE_p}{da} = - \frac{2K_p+U_p}{a} \,.
\label{LI-standard}
\end{equation}
where  $K_p=\frac{1}{2}\sum_i m a^2 (\frac{d {\bf x}_i}{dt})^2$ is
the peculiar or ``physical" kinetic energy, $U_p$
is gravitational potential energy in physical coordinates and
$E_p=K_p+U_p$. 

With the equations of motion written in the time variable $\tau$, as
in (\ref{3d-equations-3}), it is very trivial by integration to 
derive this equation in the form 
\begin{equation}
		\frac {dE}{d \tau}  =  - 2 \Gamma K
\label{energy-evolution-gamma}
\end{equation}
where $E=K+U$, and 
$K = \frac{1}{2} \sum_i m \left(\frac{d {\bf x}_i}{d \tau} \right)^2$
and  
$U = \frac{1}{2} \sum_{i,j} \phi (|{\bf x}_i - {\bf x}_j|)$ 
where $\phi (|{\bf x}|)$ is the two body potential from which the force 
$\bf F_i$ is derived. Note that to derive this relation we need only 
assume that the two body potential is time independent 
(in comoving  coordinates), so it remains valid including the force 
smoothing (which is fixed in these coordinates in our simulations). 
Now, using (\ref{tau-definition}), and thus $K=aK_p$ and
$U=aU_p$, we see that (\ref{energy-evolution-gamma}) and
(\ref{LI-standard}) are equivalent. 

Given the equation in the form (\ref{energy-evolution-gamma}), 
a natural definition for a parameter to characterize the precision
of the numerical evolution of the energy evolution is
\begin{equation}
A_0( \tau) = \frac {E ( \tau) + 2 \Gamma \int_{0}^{\tau} K d \tau}{E_0}=\frac {E (a) +  \int_{1}^{a} \frac{K}{a} da}{E_0}
\end{equation}
where $E_0=E(\tau=0)$ is the initial energy, and the last equality holds for any 
$\Gamma \neq 0$. While this parameter clearly reduces to the usual monitoring of 
energy conservation in the static limit, the choice is clearly not unique, 
nor necessarily optimal, when we consider an expanding background. 
Indeed, starting from (\ref{LI-standard}), one might instead take
\begin{equation}
	A_1 (a) =
		\frac {E_p +  \int_{1}^{a} \frac{2K_p+U_p}{a} da}{E_0}\,.
\end{equation}
Even more generally, for any $\Gamma \neq 0$, and 
defining  $E_\beta=E/a^\beta$,  $K_\beta=K/a^\beta$, 
$U_\beta=U/a^\beta$ the equation  (\ref{energy-evolution-gamma}) can be 
written as 
\begin{equation}
		\frac {dE_\beta}{d a}  =   - \frac{\beta}{a}E_\beta -  \frac{1}{a}K_\beta \,.
\label{energy-evolution-gamma-generalized}
\end{equation}
with an associated family of possible parameters
\begin{equation}
	A_\beta (a) =
	 \frac {\frac{E}{a^\beta}  +  \int_{1}^{a} \frac{\beta E +K}{a^{\beta+1}} da}{E_0} \,.
\end{equation}
While all of these parameters are equal to unity when (\ref{energy-evolution-gamma})
is valid, their deviations from unity in a numerical integration are not trivially related
to one another and it is not a priori clear which, if any of them, provides the most suitable 
measure of the accuracy of a simulation. The problem with this kind of measure
is that we do not dispose (at least currently) of any absolute calibration which tells us 
how  much deviation from unity of such indicators can be tolerated. 
In short, while in a non-expanding 
simulation we know we should tolerate only percent level deviations of the energy,
we do not know what deviation from unity of the parameters $A_\beta$ should be
considered acceptable. Most studies \footnote{An exception is a recent
study \cite{winther_2013} applying the Layzer Irvine  equation to monitor 
the accuracy of N body simulations of scalar-tensor theories of modified gravity.}
in the cosmological 
literature which report results for monitoring of the energy evolution 
(see e.g. \cite{REFS_layzerIrvine}, \cite{REFS_layzerIrvine2}), \cite{smith}) 
consider the parameter 
\begin{equation}
	A^\prime =  \frac {E_p(a) - E_0  + \int_{1}^{a} \frac{2K_p+U_p}{a} da}{U_p} \equiv \frac{E_0}{U_1} (A_1-1)\,,
\end{equation}
i.e., the integrated fluctuation is normalized with the 
physical potential energy $U_p$ rather than the initial total energy. While
in absence of any absolute calibration for any of these parameters, one
cannot know which parameter is the most appropriate to use, it appears
to us, compared to the parameter $A_1$, that this canonical normalisation
is probably not an astute one. Firstly, extrapolated to the non-expanding limit by taking
$\Gamma \rightarrow 0$, it corresponds to normalizing 
the total energy fluctuation to an energy which evolves, and typically 
increases in magnitude in time,  due to the development of clustering. 
Thus one can obtain arbitrary variation (and typically decrease) in 
the measured ``energy error"  measured with $A^\prime$ which
would appear to have a priori little to do with (integrated) numerical 
error.  Further we have found, tracing its behaviour, that $A^\prime$  
can even diverge because the potential energy can change sign 
(as it may be positive in the almost uniform initial configuration). 

The crucial point is that, in any case, with the current absence of any
absolute calibration, we can use these parameters only as a tool
to {\it compare} the accuracy of different simulations, but not to make any
useful inference about their absolute accuracy. In the course of this study 
we have, in our choice both of numerical simulation parameters and
the range of $n$ and $\kappa$ simulated, made use of 
both $A_0$ and $A_1$, in this way. In particular, as mentioned above 
we found in test simulations that their difference from unity could be reduced
significantly taking the time step parameters we have chosen, 
compared to fiducial values. Further, and more interestingly, we
have found that large deviations from unity of these parameters 
are often clearly correlated with a breakdown of self-similarity.
This opens up the possibility of using this class of
models as an absolute calibrator for accuracy of numerical
simulations as probed by parameters like $A_\beta$. 
These issues will be discussed at length elsewhere.
We report here, for 
brevity, only measurements with the indicator $A_1$, because
it is the one closest to the often used $A^\prime$. Further it has the
nice feature that, for the case of a single isolated virialized 
structure (for which $2K_p+U_p=0$, if the effects of force smoothing are 
negligible), it reduces to the fractional energy error in 
the physical energy $E_p$, which is the error measurement
one would usually use for this case. 

\begin{figure*}
	\resizebox{!}{5.8cm}{\includegraphics[]{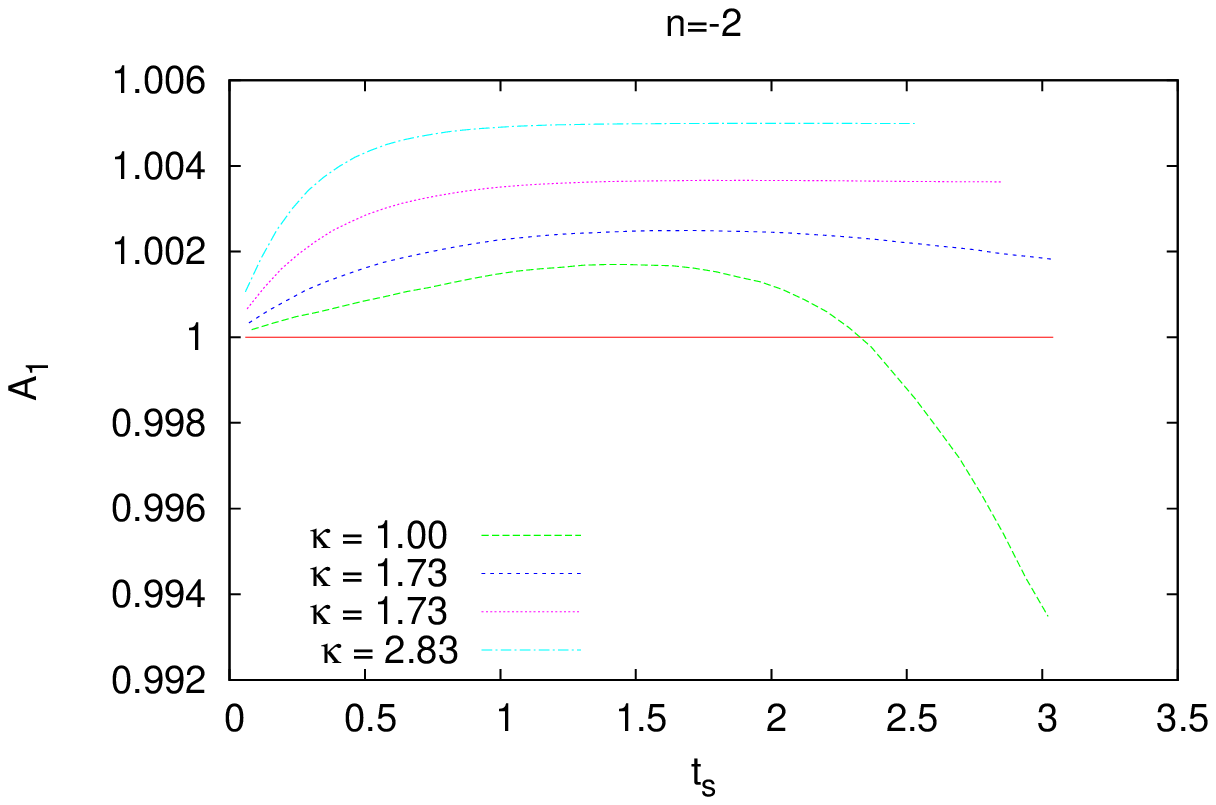}}
	\resizebox{!}{5.8cm}{\includegraphics[]{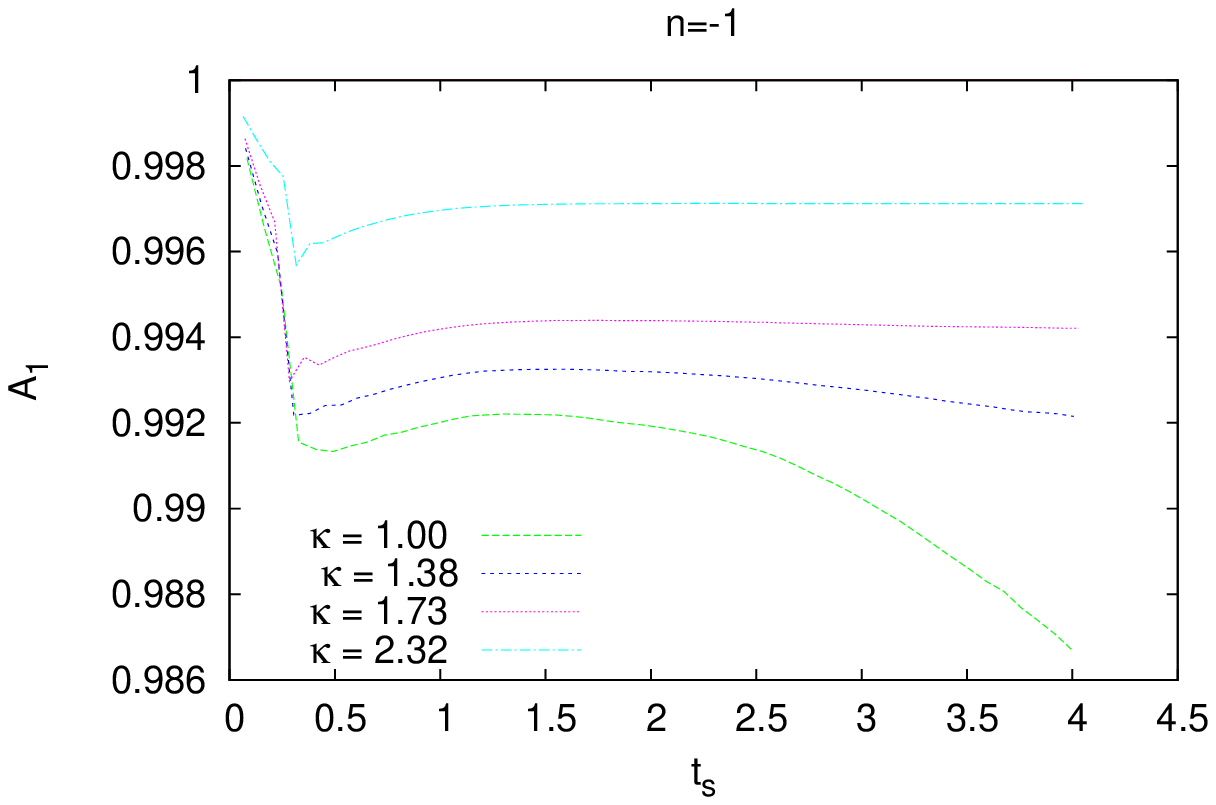}}\\
	\resizebox{!}{5.8cm}{\includegraphics[]{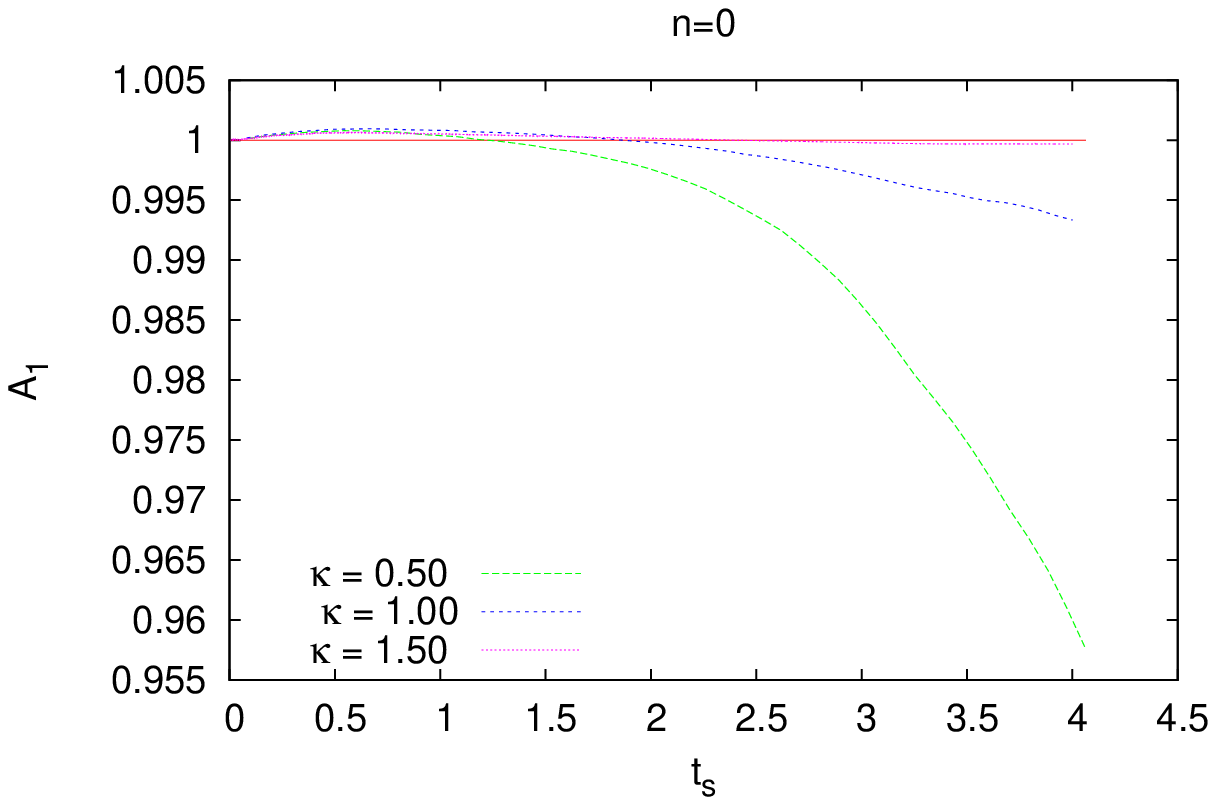}}
	\resizebox{!}{5.8cm}{\includegraphics[]{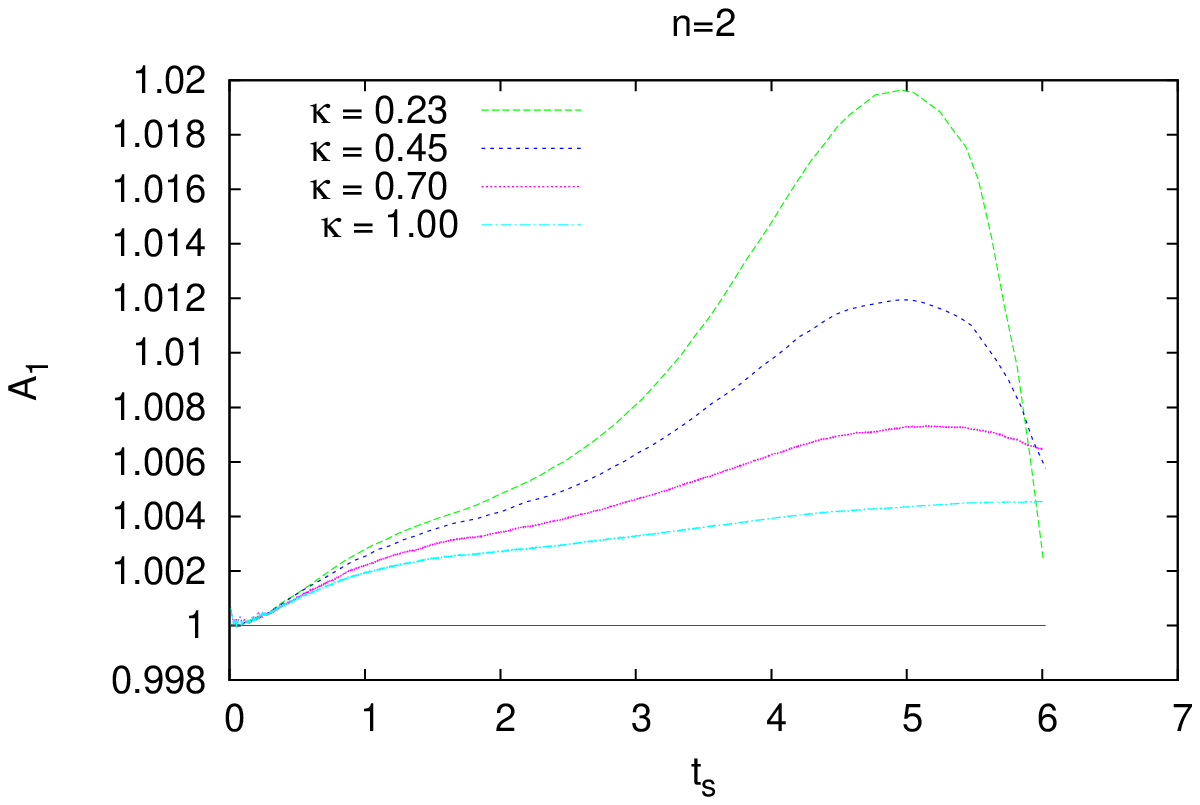}}
	\caption{Evolution of the energy evolution parameter $A_1$ defined in the text as
	a function of $t_s$.}
	\label{fig-energy-evolution}
\end{figure*}

Shown in Fig. \ref{fig-energy-evolution} is the evolution 
of the parameter $A_1$ in our different simulations, each panel 
showing the simulations of a given $n$ for the different
$\kappa$. For the case $n=2$ we plot data in all models 
up to $t_s=6$, which for three models is beyond the final 
$t_s^f$  given in Table~\ref{Table1}, i.e., we include
(for the purposes of illustration) some simulation data 
which we have excluded from our analysis. From 
Fig. \ref{fig-energy-evolution} we see that 
according to this measure the accuracy of the simulations 
varies, but is of comparable order, with maximal deviations 
from unity of order at most a few percent. The slightly smaller 
amplitude of deviations in the case $n=-2$ are a reflection of
the larger particle number compared to that in
the other cases. For each  $n$, we see also that
the poorest precision in $A_1$ is obtained 
for the model with the smallest $\kappa$.
The model $(n=0, \kappa=0.5)$ shows
a significantly larger amplitude deviation from
unity than any other, while the models with $n=2$
and smaller $\kappa$ show the onset of a more
rapid evolution of $A_1$ after reaching a peak.
We have excluded from our analysis the later 
time data in these models precisely because
we have concluded that this behaviour is
correlated with an unphysical evolution of
macroscopic quantities, and specifically
a breakdown of self-similarity (which, as
we will see, holds at earlier times).  To 
illustrate this a little more, we retain here in our 
analysis the later time data in the 
model $(n=0, \kappa=0.5)$ which, we will
see, also manifests such a deviation from 
self-similarity which we believe is a result
of the poorer numerical precision indicated
by the behaviour of $A_1$. 

\subsection{Effects of force smoothing}
\label{Effects of force smoothing}

\begin{figure*}
	\centering\includegraphics[scale=0.7]{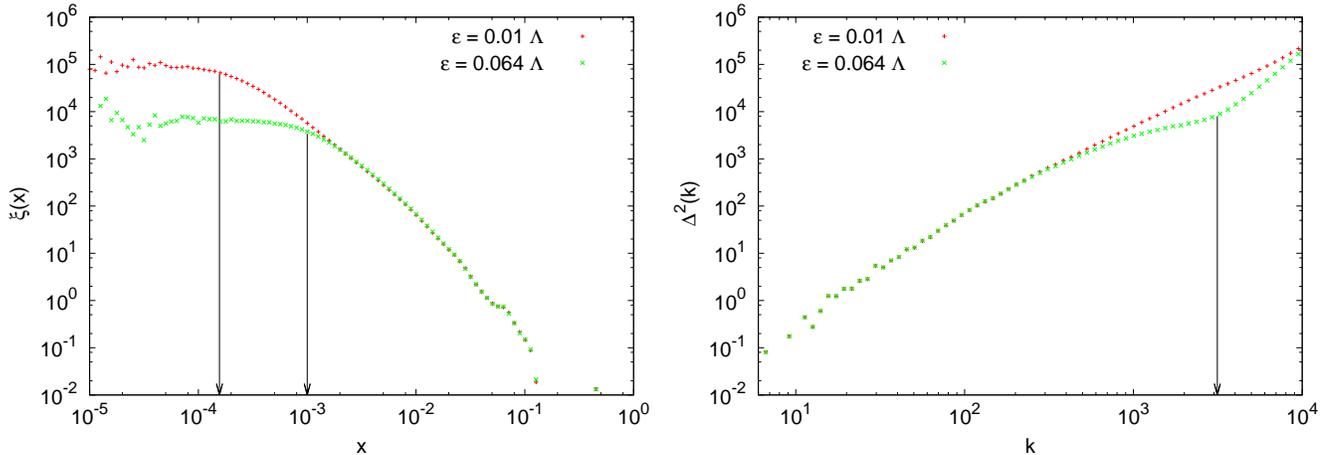} 
	\caption{Comparison of results for correlation functions (top) and power spectra (bottom) in
	two $N=64^3$ simulations for the case $n=0$ and $\kappa = 1$ with different force 
	smoothing: one with  $\varepsilon_1 =0.01 \Lambda$, and the other
	$\varepsilon_2 =0.064 \Lambda$; both of these $\varepsilon$ are indicated by	
	the vertical lines on the plots of $\xi$, while only the scale $\pi/\varepsilon_2$
	is marked on the power spectra plots ( $\pi/\varepsilon_1$ lies out of the plotted
	range).}
	\label{fig_compare_Gadget_smoothing}
\end{figure*}

As the limitation on the spatial resolution associated
with force smoothing is an important issue in interpreting
the results of simulations, we have performed some
studies of the effect of varying $\varepsilon$. 
Shown in Fig. \ref{fig_compare_Gadget_smoothing}
are results for the
correlation functions and power spectra measured in
two $N=64^3$ simulations, for the case $n=0$ and 
$\kappa=1$,  which are identical other than for the 
value of $\varepsilon$ used: one simulation uses our chosen 
value,  $\epsilon_1= 0.01 \Lambda$, and the other 
one $\epsilon_2 = 0.064 \Lambda$, as in  \cite{smith}. 
For the correlation function we see 
that the result for the lower resolution simulation agrees 
very well with that of the higher resolution down to a 
approximately $2\varepsilon$, while below this scale 
the clustering is (as one would expect) very 
suppressed compared to that in the higher resolution 
case. For the power spectrum we observe a similar 
behaviour. We note, however, that the scale $k$
in reciprocal space at which we observe clear
deviation of the low resolution simulation is 
{\it almost an order of magnitude} smaller than 
$\pi/\varepsilon_2$, which is naively where one 
might expect this deviation to be observed. 
The reason for this is evidently that the power 
spectrum at any $k$, which is the Fourier transform 
of $\xi(x)$, clearly ``mixes" a certain range of scales 
around $\pi/k$ and thus the suppression of 
the correlations below $x \sim \varepsilon$ lead to 
a significant suppression in power well below 
$\pi/\varepsilon$.  Our conclusion from this 
analysis is that it is more straightforward to 
identify in real space the range in which results 
are unaffected by smoothing. Specifically
we will assume below that such effects are sufficiently 
small beyond  $2\varepsilon$. In Fourier space 
great care should be taken in identifying
the scale at which force smoothing modified
results, and we will take as indicative 
the result of 
Fig. \ref{fig_compare_Gadget_smoothing}
showing that significant suppression of power 
is observed above $k \approx 0.1 (\pi/\varepsilon)$ 
in the model with $n=0$ and $\kappa=1$\footnote{ 
We will see below in comparing with larger 
simulations ($N=256^3$), that in the model 
with $n=-1$  and $\kappa=1$ a visible 
suppression of the power due to smoothing
indeed sets in at about the same scale.}.

\subsection{Results: visual inspection}

\begin{figure*}
	\centering\resizebox{!}{23.5cm}{\includegraphics[]{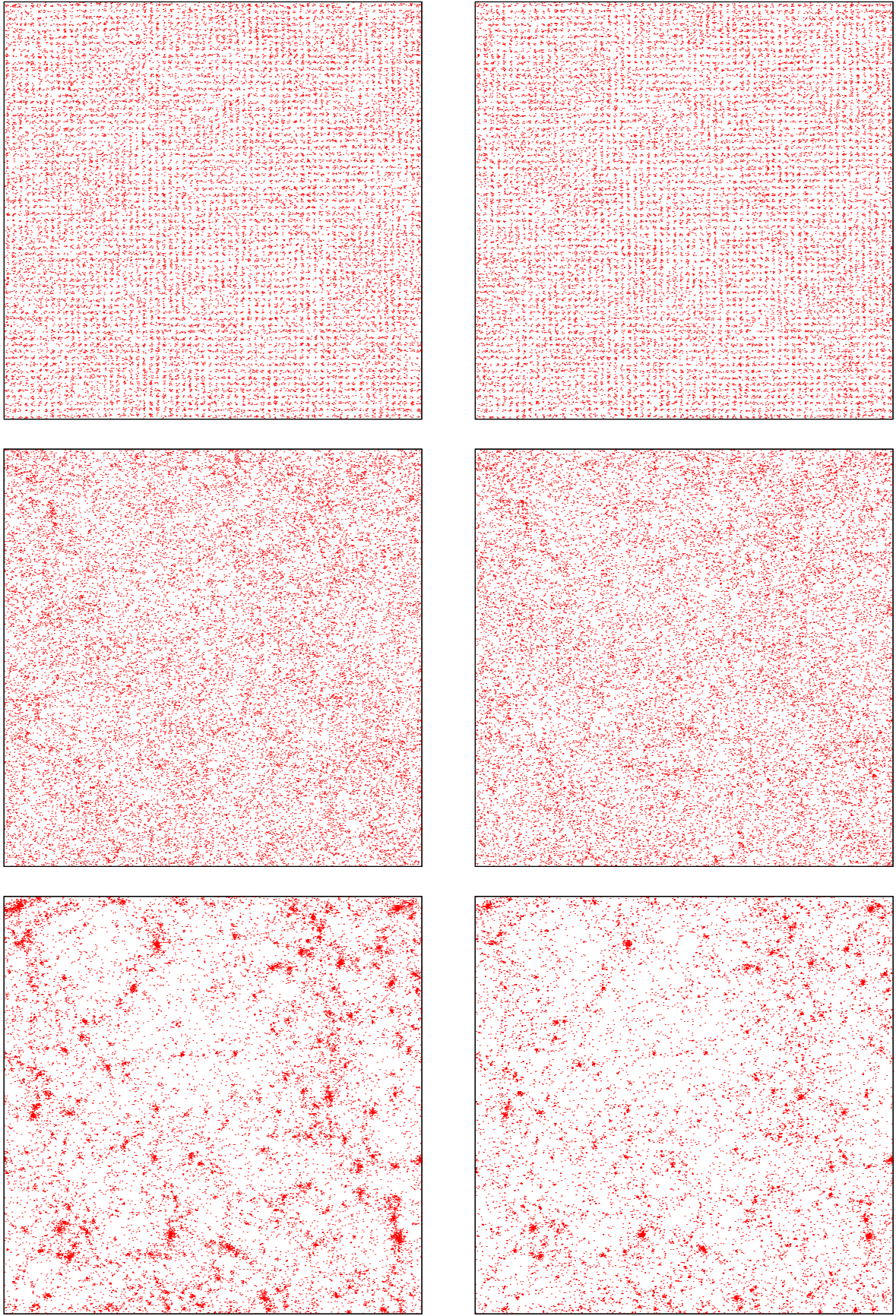}} 
	\caption{Snapshots (projections on plane) for 
	initial conditions 
	$P(k) \propto k^{-1}$, at the times $ t_s = 1.0, 2.5 , 4.0 $ for
	the models with $\kappa = 1, 2.3$ (from left to right)}
	\label{fig_VisualInspection_n=-1}
\end{figure*}

\begin{figure*}
	\centering\resizebox{!}{23.5cm}{\includegraphics[]{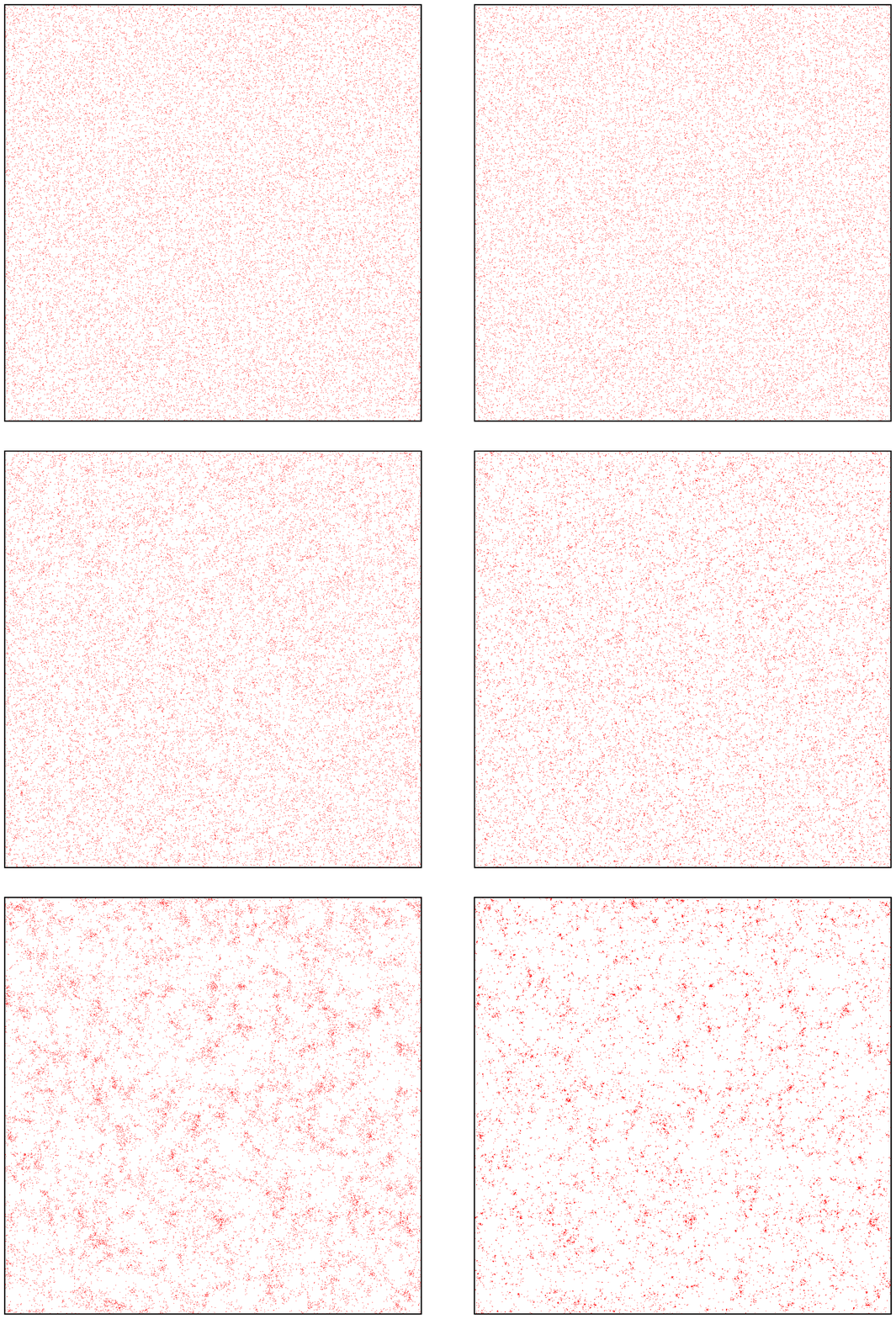}} 
	\caption{Snapshots (projections on plane) for initial conditions 
	$P(k) \propto k^{2}$, at the times $ t_s = 1.0, 3.0 , 5.0 $  for
	the models with $\kappa = 0.23, 1$ (from left to right)}
	\label{fig_VisualInspection_n=2}
\end{figure*}

Shown in Fig.s \ref{fig_VisualInspection_n=-1} and \ref{fig_VisualInspection_n=2}
are some snapshots of the particle configurations in a few chosen simulations. 
In each case we show, for a given initial power spectrum, the configurations at 
several times $t_s$ for simulations with the largest and smallest simulated value
of $\kappa$. As discussed above, the time variable $t_s$ is defined so that it 
corresponds to the same linear amplification of the growing mode in any model.
Thus, for the same initial condition, the evolution in linear perturbation
theory in the different $\kappa$ models should be identical. Further
following our discussion above, we expect the non-linear structures to
become more compact as $\kappa$ increases. Both expectations are
evident qualitatively in the snapshots: in all cases the structures at
larger scales --- where perturbations are small --- are indeed very 
similar, and in all cases we see that the effect of increasing 
$\kappa$ is to make the non-linear structures more compact.

\subsection{Results: dependence of two point correlation properties on $\kappa$}

\begin{figure*}
	\resizebox{!}{23.5cm}{\includegraphics[]{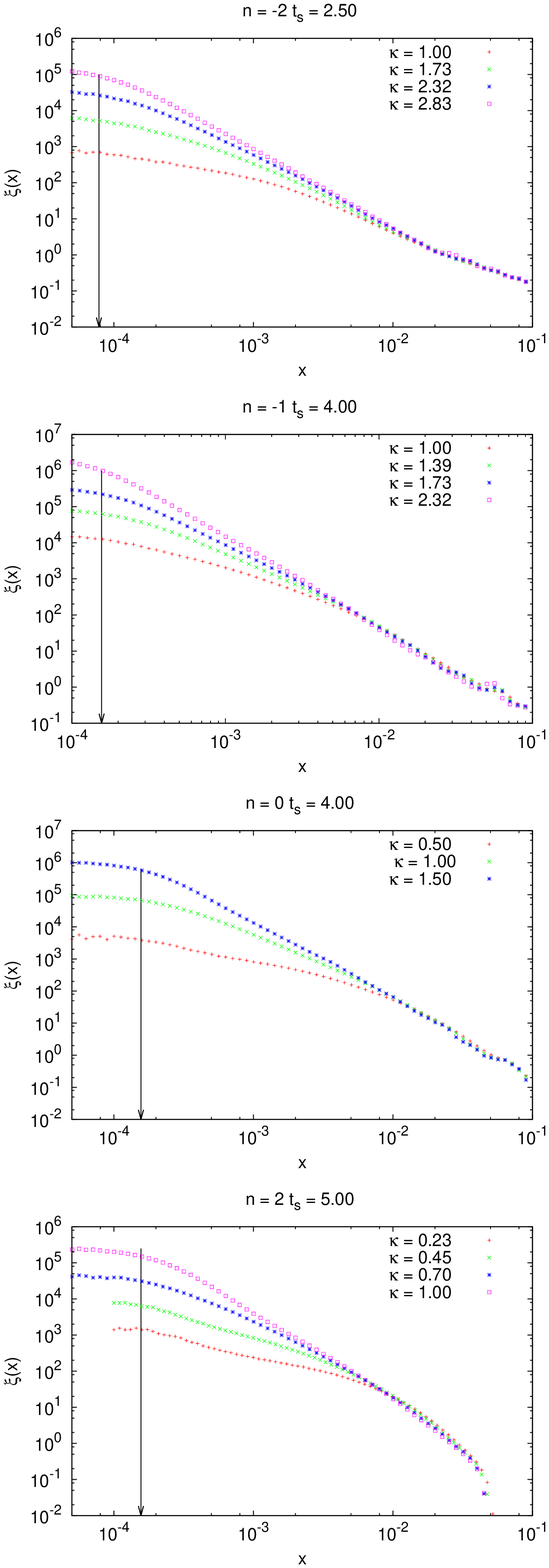}}
	\resizebox{!}{23.5cm}{\includegraphics[]{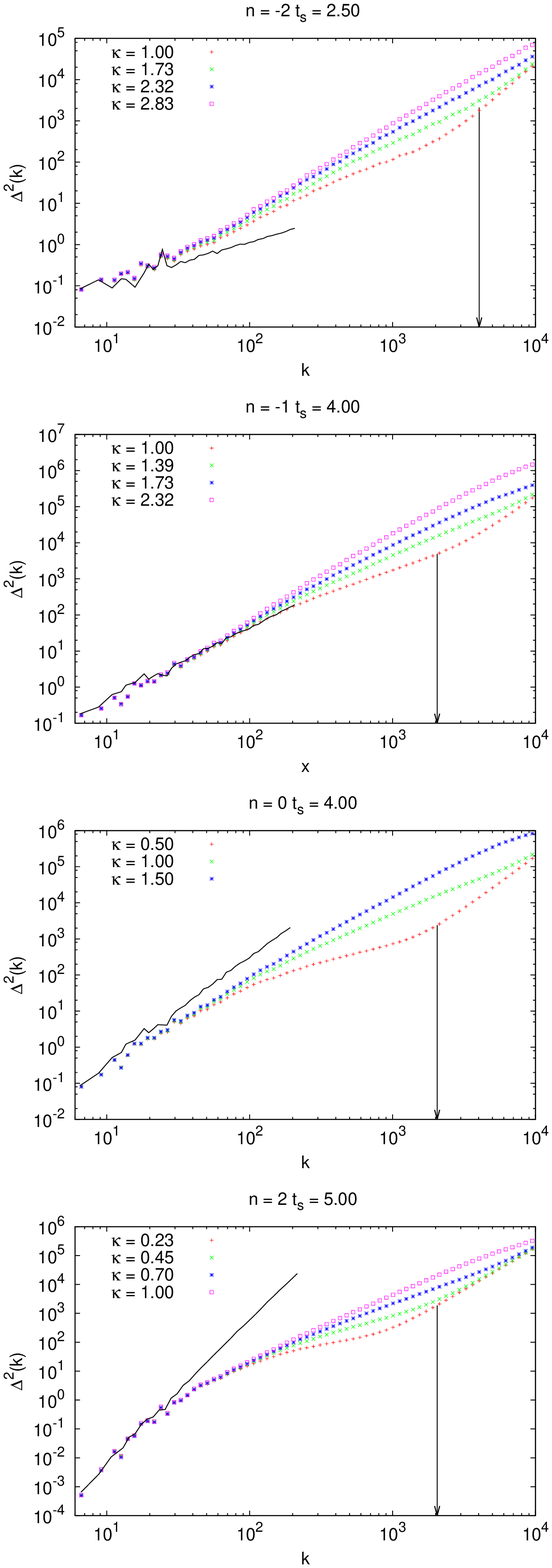}}
	\caption{Plots of $\xi$ and $\Delta^2 (k)$ at indicated times $t_s$ in different models. 
	The black vertical line in each plot of $\xi(x)$ indicates the smoothing parameter $\varepsilon$.
	 The black curve in each plot of $\Delta^2 (k)$ shows the prediction of linear theory. }
	\label{fig_Gamma_effect_xi+Delta2_all}
\end{figure*}

In Fig. \ref{fig_Gamma_effect_xi+Delta2_all}
are shown our results for $\xi(x)$ and $\Delta^2(k)$ for all simulations when
they are highly evolved. Each plot shows, for the subset of models with a 
fixed $n$ (but different $\kappa$), one of the two quantities at the 
indicated times $t_s$. Also shown in each of the $\Delta^2(k)$ plots
is the prediction of linear theory, obtained by multiplying the measured
$\Delta^2(k)$ for the initial conditions (identical for all models at given $n$)
by the predicted linear theory amplification $e^{t_s}$. The black vertical  
line indicates the smoothing parameter 
$\varepsilon$ in the plots of $\xi(x)$, and $0.1 (\pi/\varepsilon)$ in the plots of 
$\Delta^2(k)$.

These plots confirm quantitatively what was anticipated 
above in the visual snapshots: the evolved configurations have 
indeed almost identical two point correlation properties at the 
larger scales at which linear theory is valid\footnote{  The
small but noticeable ``bump" feature in $\xi(x)$ at  
$x \approx 0.5$ in the $n=-1$ models is, it will be 
seen below, a finite box size effect.}, while in the 
non-linear regime the effect of increasing $\kappa$ is to lead 
to greater relative power at smaller scales, with both $\xi(x)$ (and 
$\Delta^2(k)$) clearly increasing much more rapidly 
with decreasing $x$ (and increasing $k$) 

These behaviours --- notably a non-linear correlation function 
which steepens as $\kappa$ increases ---  are thus {\it qualitatively}
in line with those predicted in the stable clustering 
hypothesis. We note that these results are also qualitatively 
in line with what one would anticipate from
results of previous numerical studies of the effect of
modifications of cosmology, notably
by curvature and/or a cosmological
constant. As discussed in Section
\ref{A family of 3D scale-free models},   
the case of an open universe or
cosmological constant correspond to
a $\kappa$ which becomes larger
than unity at later times, and indeed
it has been observed in previous
studies (see e.g. \cite{peacock, padmanabhan_etal_1996})
that in these cases non-linear power is 
increased when one compares the
models at times at which the linear
fluctuations are identical. 

Results like the ones just mentioned,
and more generally the analysis of
two point correlation properties, are 
often presented in the cosmological 
literature in terms of representations 
of $\xi$, or more often $\Delta^2$, as 
functions of variables $\xi_L$ or $\Delta_L^2$, 
representing the linearly evolved $\xi$  or $\Delta^2(k)$
at a length scale $x_L$ (or $k_L$) related to $x$ (or $k$) 
through a mapping described, e.g., 
in \cite{hamilton, peacock}. 
We have performed this analysis for all our
models to obtain $\Delta^2$ [$\Delta_L^2$],
using $k_L = k (1+ \Delta^2(k))^{-1/3}$.
For brevity, we do not report the results here,
as they do not reveal any particular simplicity 
additional to what we have already obtained by mapping 
the linear evolution working in time units defined by $t_s$. 
In particular we note that the stable clustering hypothesis,
which leads to the ``universal" behaviour
$\Delta^2(k) \propto [\Delta_L^2(k)]^{3/2}$
for the usual EdS model, generalizes, given 
the generalized linear evolution Eq.~(\ref{linear theory-growing-expanding}), 
to $\Delta^2(k) \propto [\Delta_L^2(k)]^{3/2\alpha}$.
In line with stable clustering,
we indeed observe such a steepening of $\Delta^2(k)$
plotted as a function of $\Delta_L^2(k)$ in the strongly 
non-linear regime ($\Delta^2(k) \gtapprox 10^2$)
\footnote{ Thus, in the stable clustering 
approximation, the functional form of the dependence 
of $\Delta^2(k)$ on $\Delta_L^2(k)$ is ``universal" 
(i.e. model-independent) only in its dependence
on $n$,  but explicitly depends on $\kappa$
i.e. on the cosmology. For this reason this 
particular representation of the non-linear
correlation properties does not appear to
be a particularly useful or relevant one
for our class of models.}.

\subsection{Results: self-similarity}
\label{Results: self-similarity}

\begin{figure*}
	\resizebox{!}{5.8cm}{\includegraphics[]{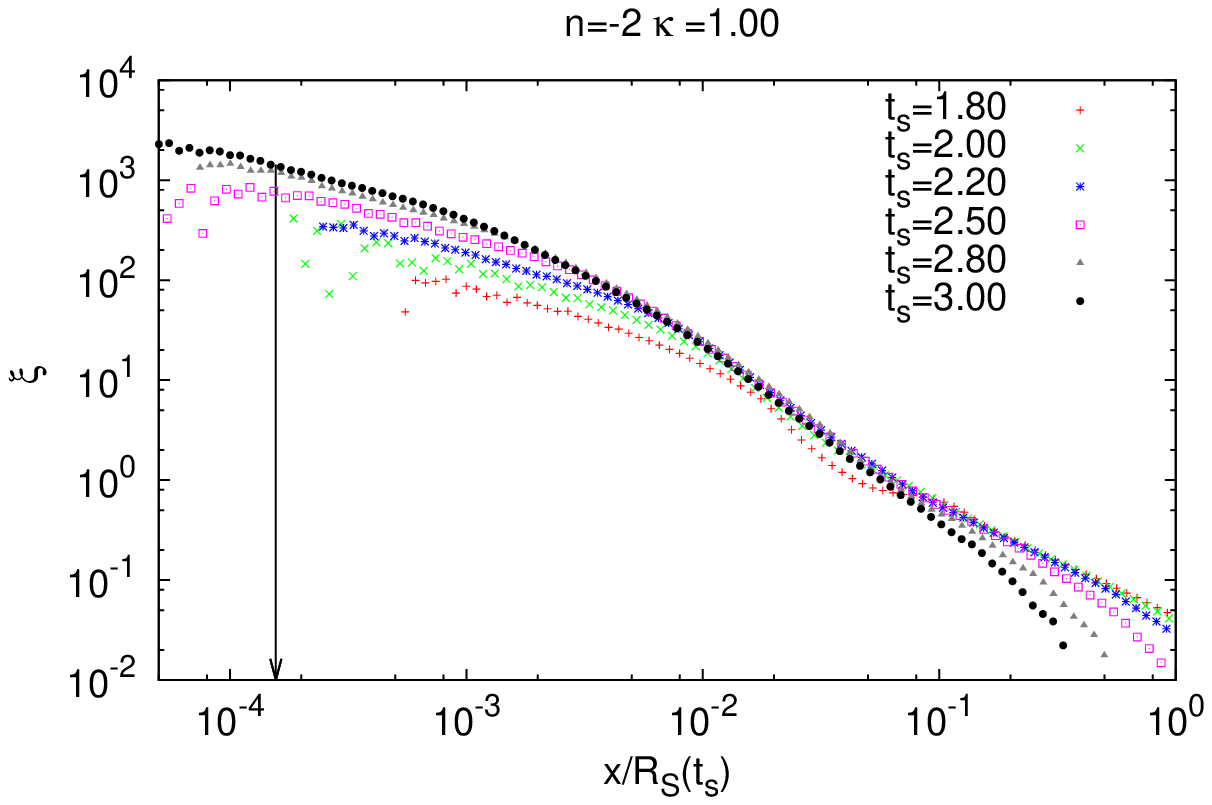}}
	\resizebox{!}{5.8cm}{\includegraphics[]{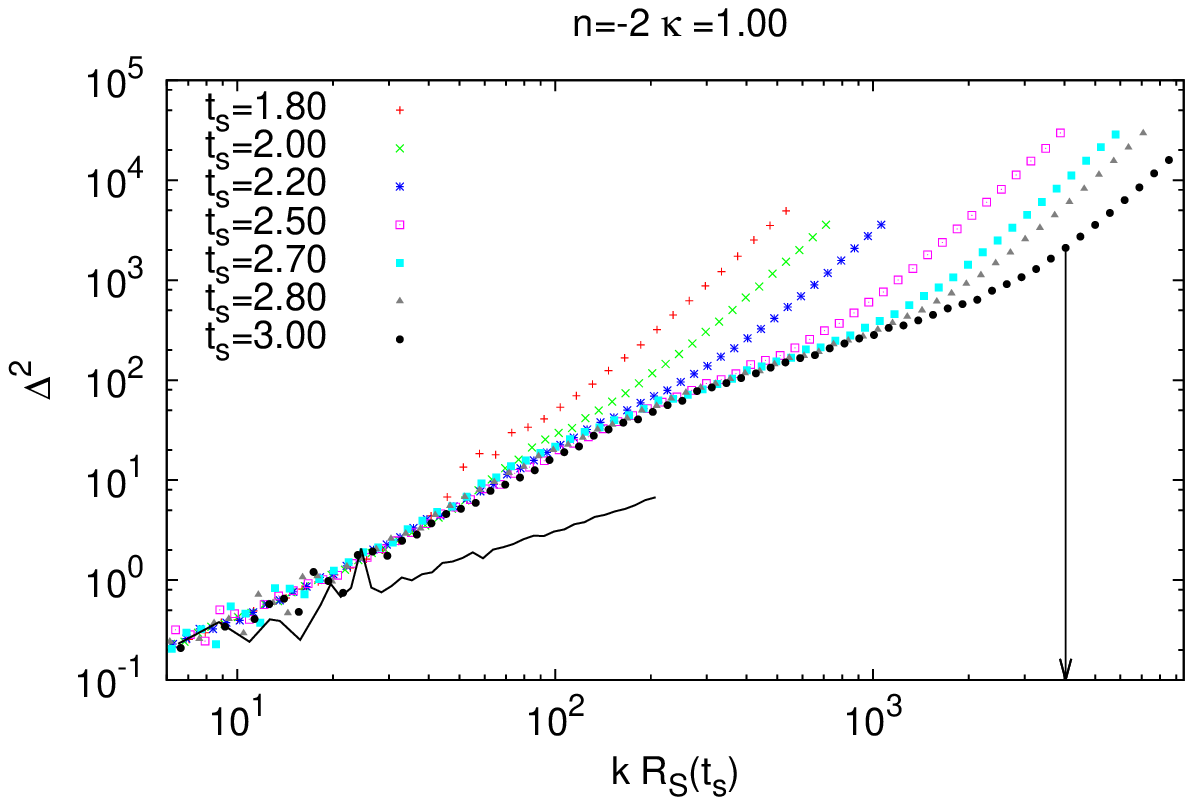}}\\
	\resizebox{!}{5.8cm}{\includegraphics[]{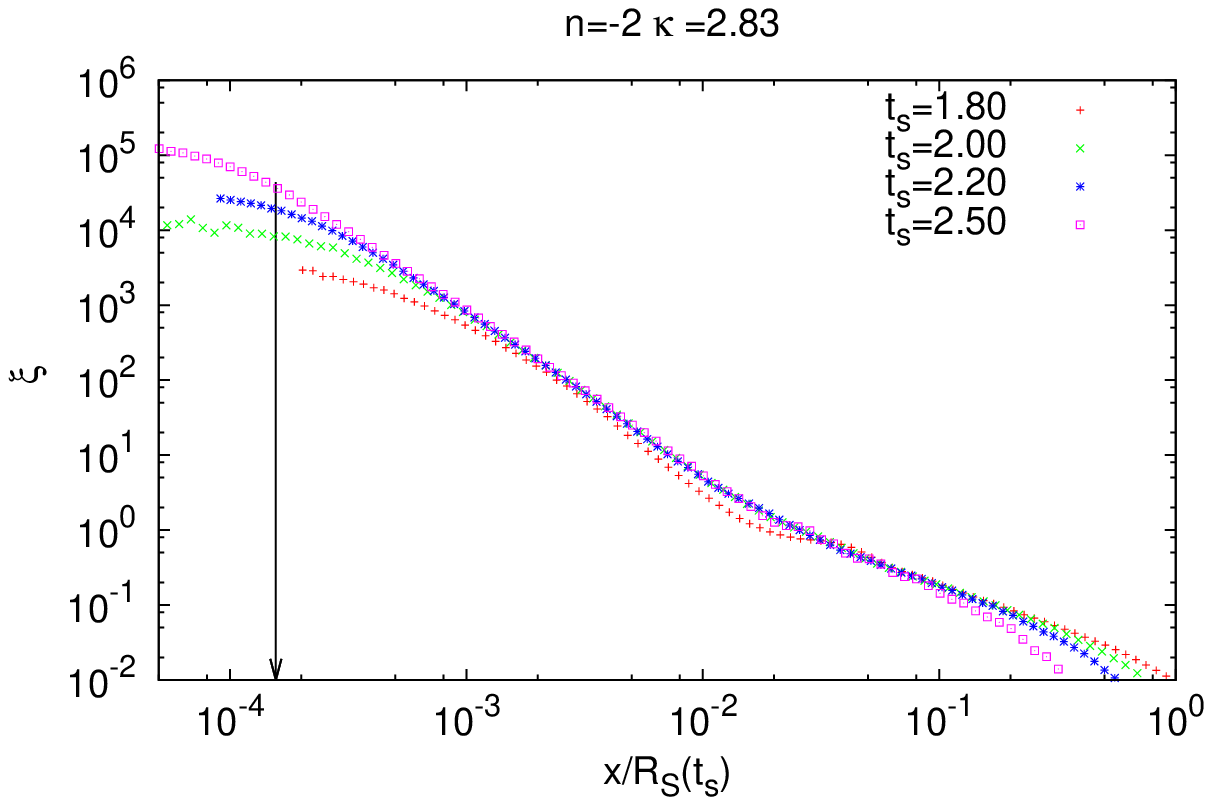}}
	\resizebox{!}{5.8cm}{\includegraphics[]{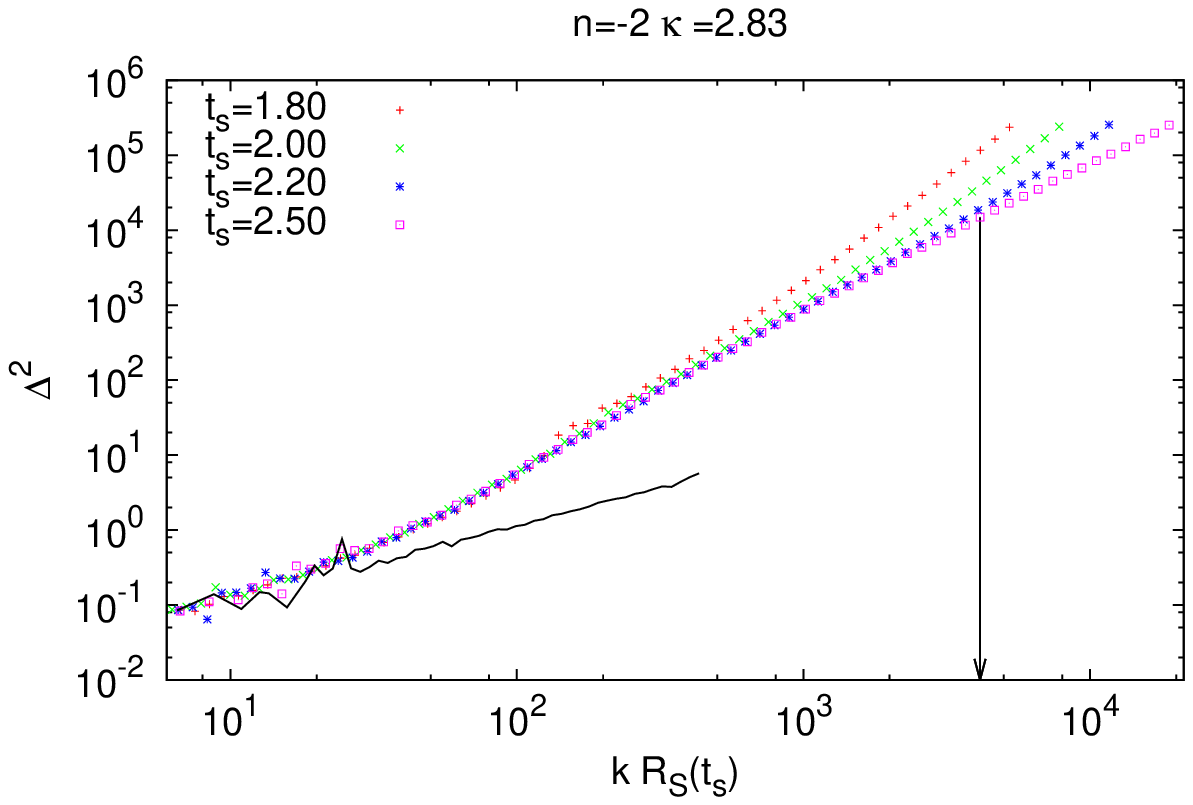}}\\
	\resizebox{!}{5.8cm}{\includegraphics[]{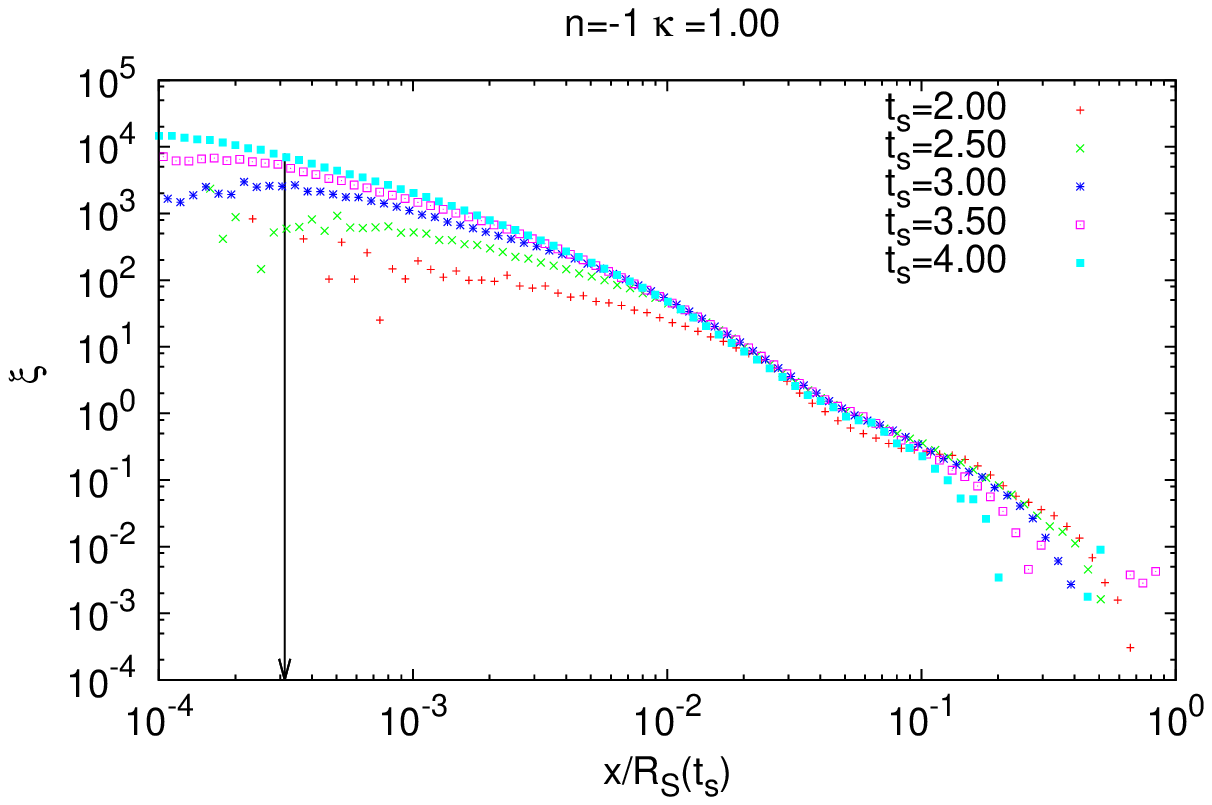}}
	\resizebox{!}{5.8cm}{\includegraphics[]{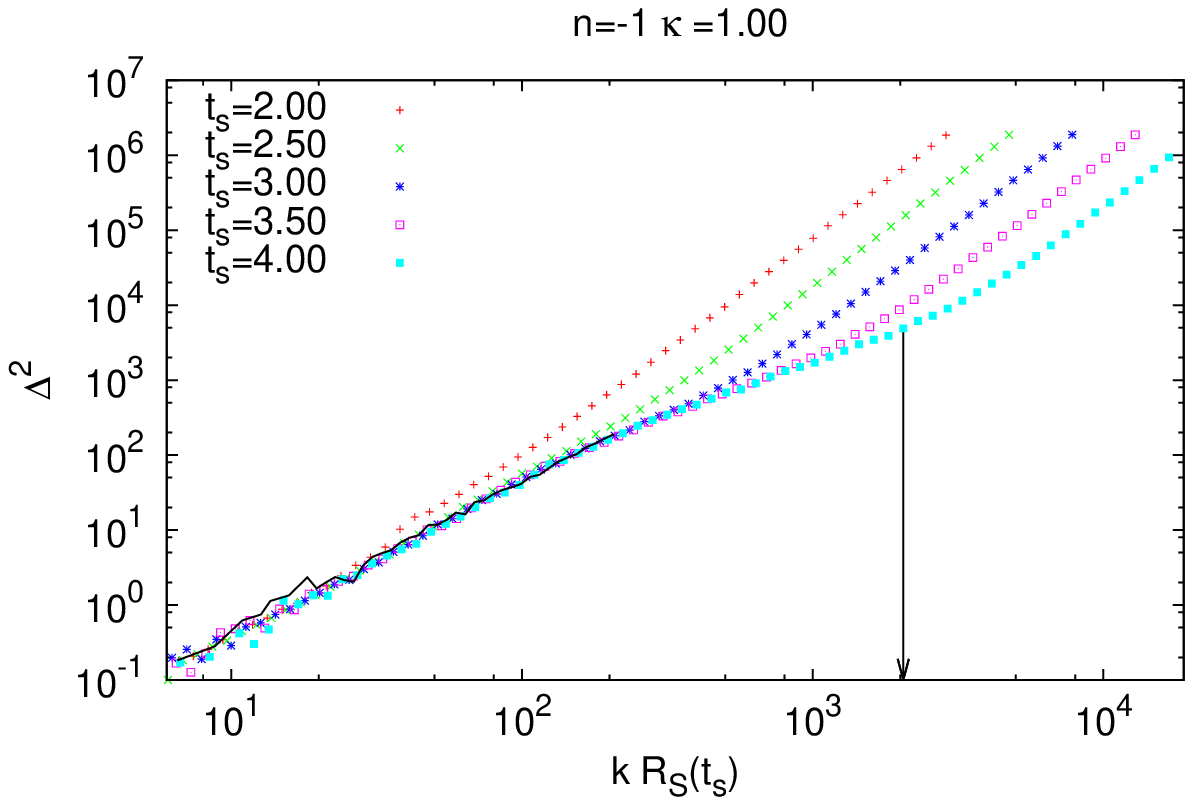}} \\
	\resizebox{!}{5.8cm}{\includegraphics[]{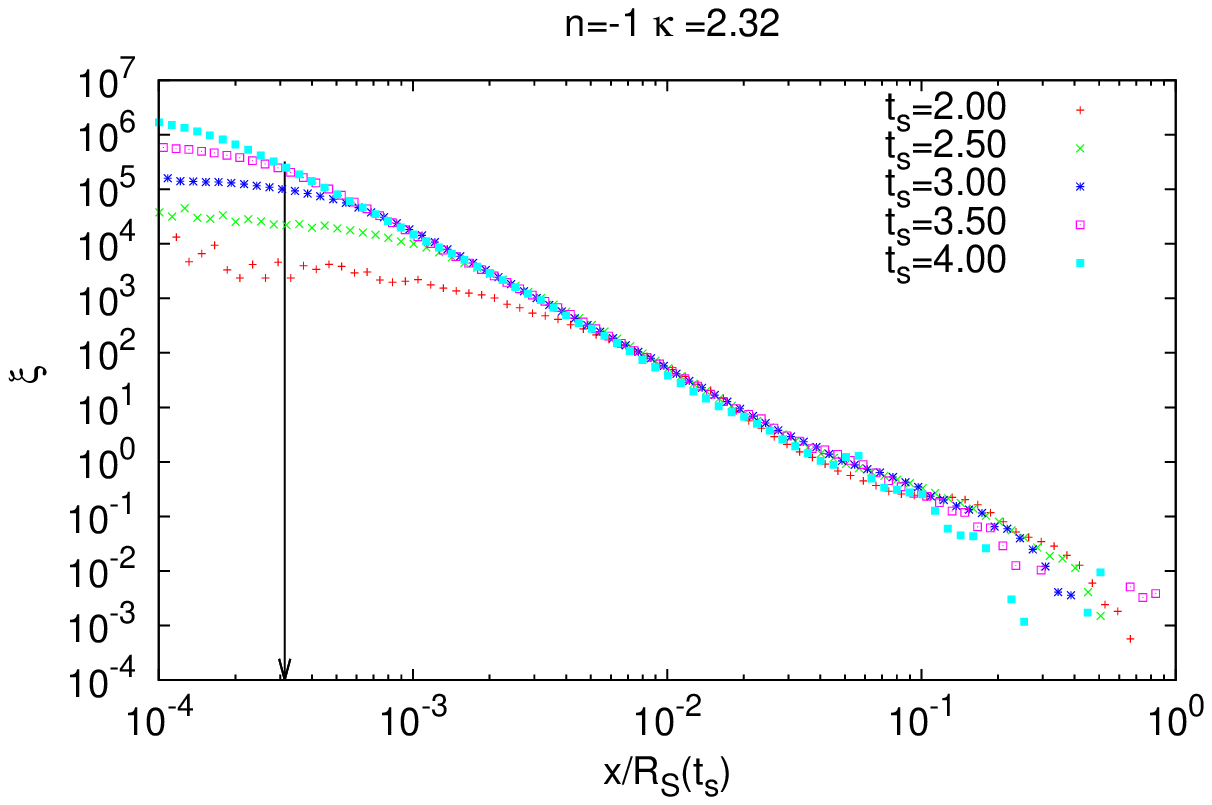}}
	\resizebox{!}{5.8cm}{\includegraphics[]{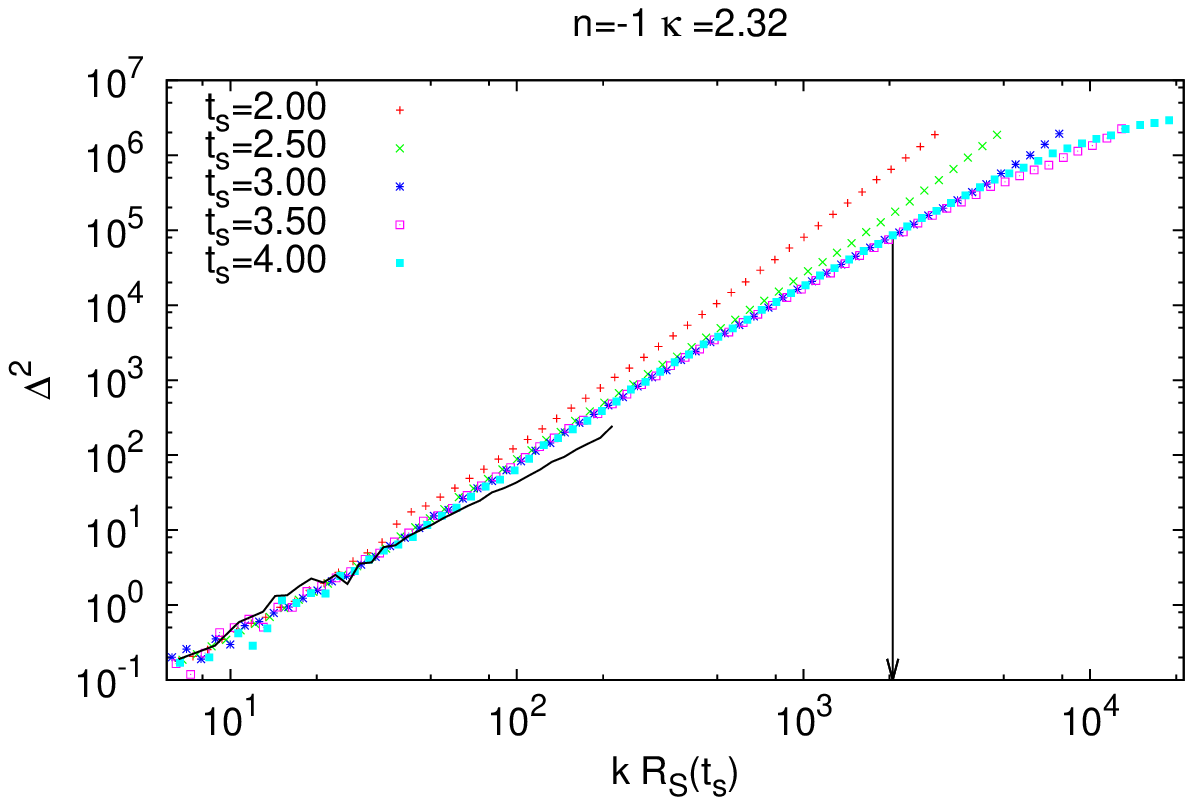}}
	\caption{Each plot shows $\xi$ or $\Delta^2$ for a model with $n=-2$ or $n=-1$, for different times and 
	in rescaled space variables appropriate to test for self-similar evolution. The black vertical lines indicate 
	the scale $2\epsilon$ in the plots of $\xi$, and the scale $0.1(\pi/\varepsilon)$ in the plots of $\Delta^2(k)$. The black curve in each plot of $\Delta^2 (k)$ shows the prediction of linear theory. }
	\label{fig-selfsimilarity-n-2_n-1}
\end{figure*}

\begin{figure*}
	\resizebox{!}{5.8cm}{\includegraphics[]{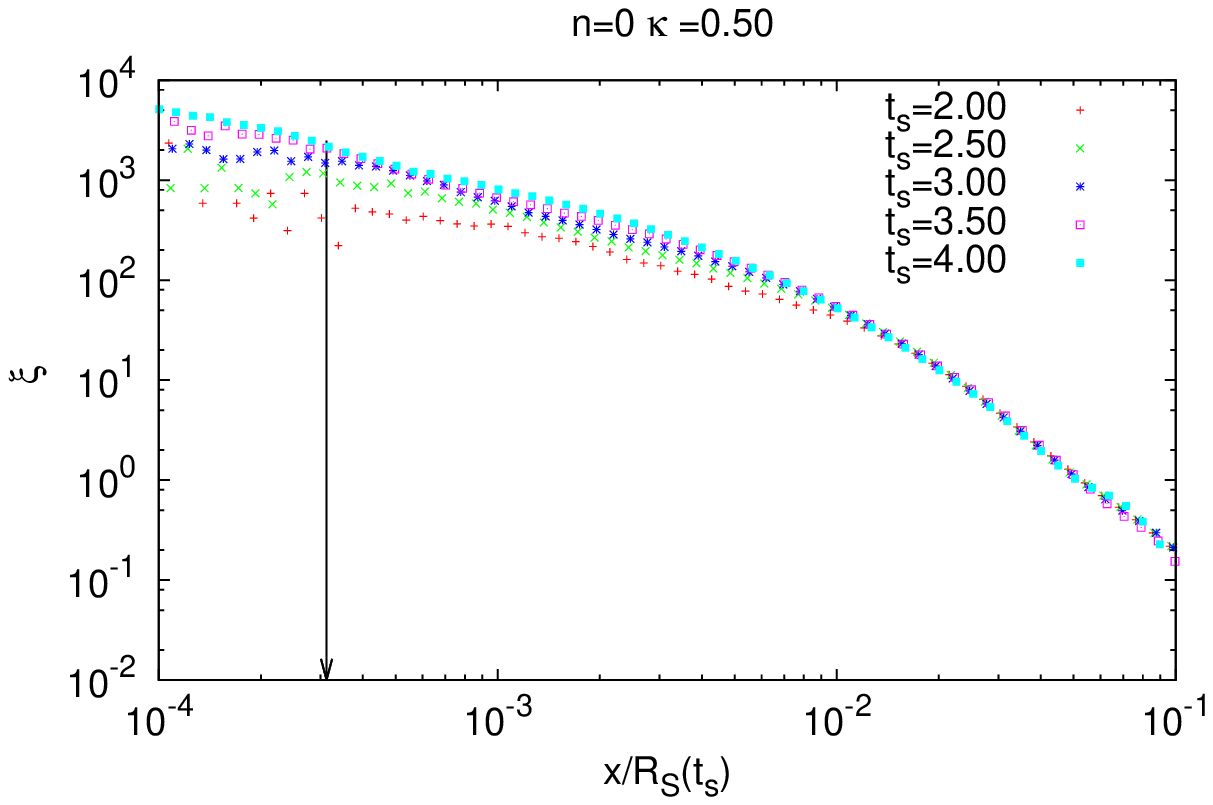}}
	\resizebox{!}{5.8cm}{\includegraphics[]{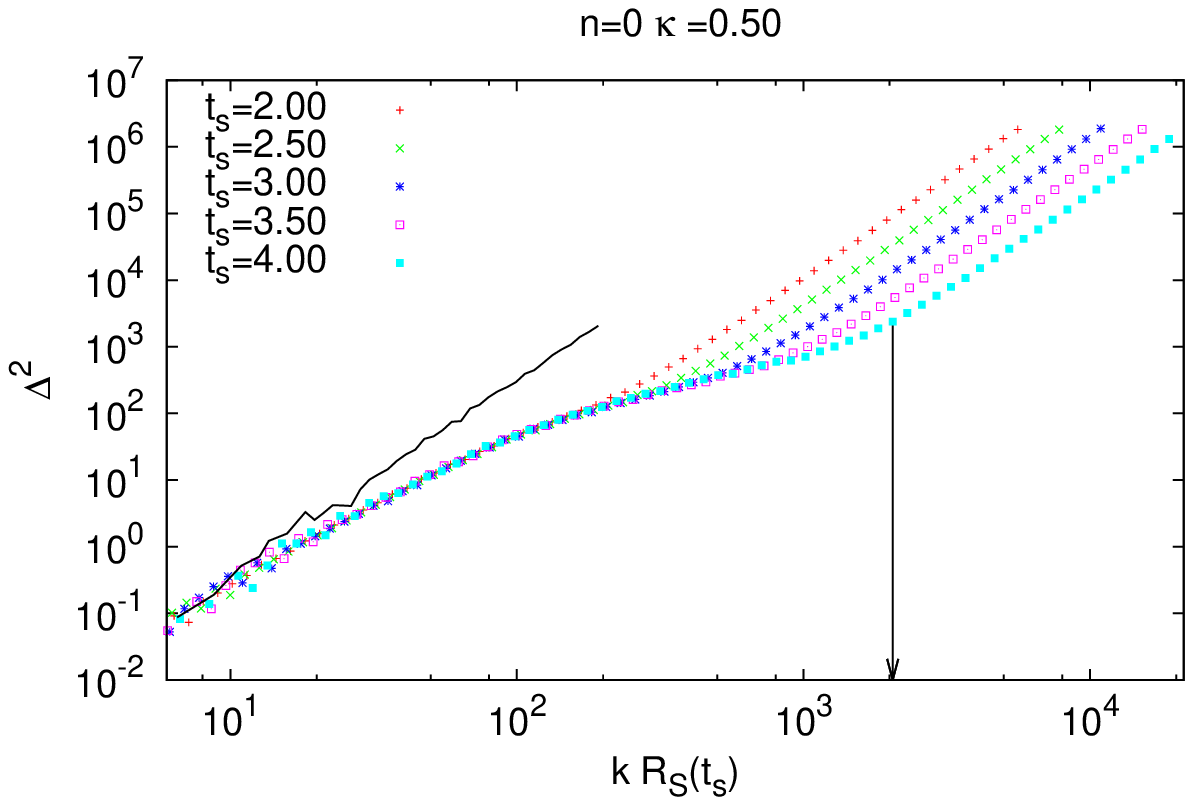}}\\
	\resizebox{!}{5.8cm}{\includegraphics[]{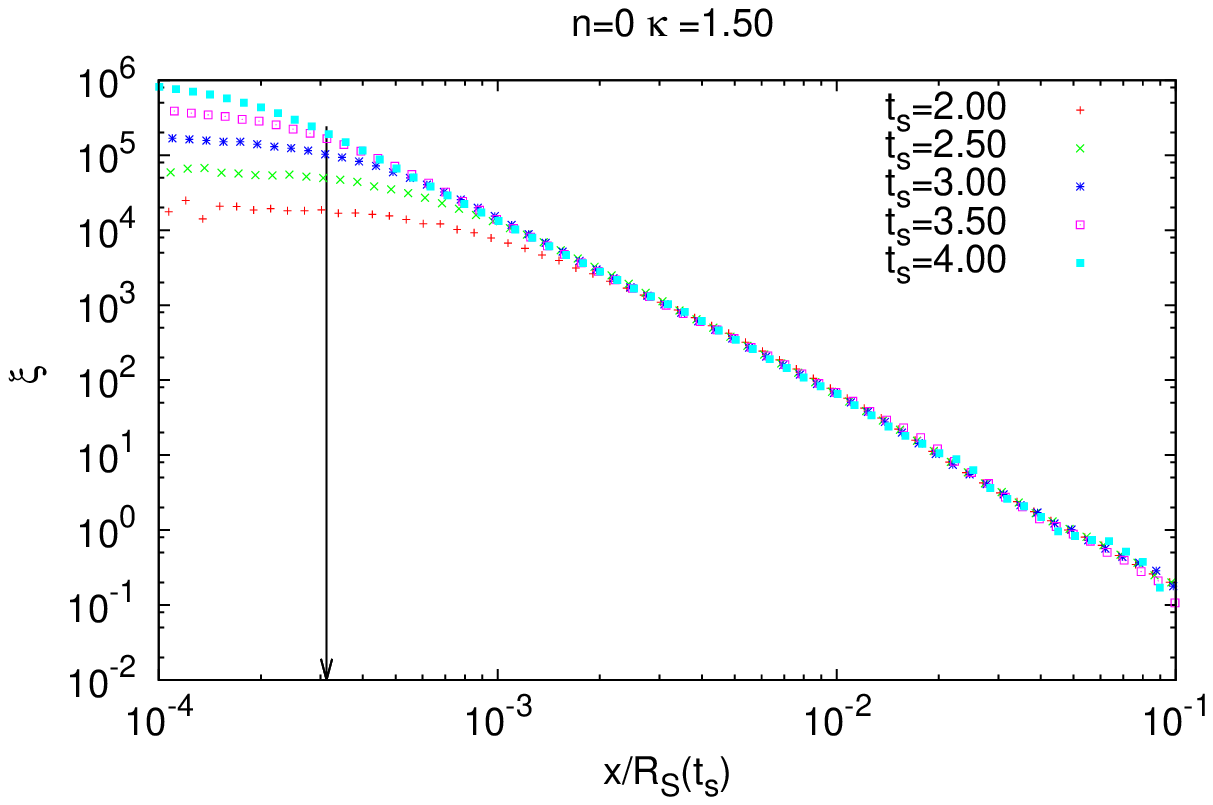}}
	\resizebox{!}{5.8cm}{\includegraphics[]{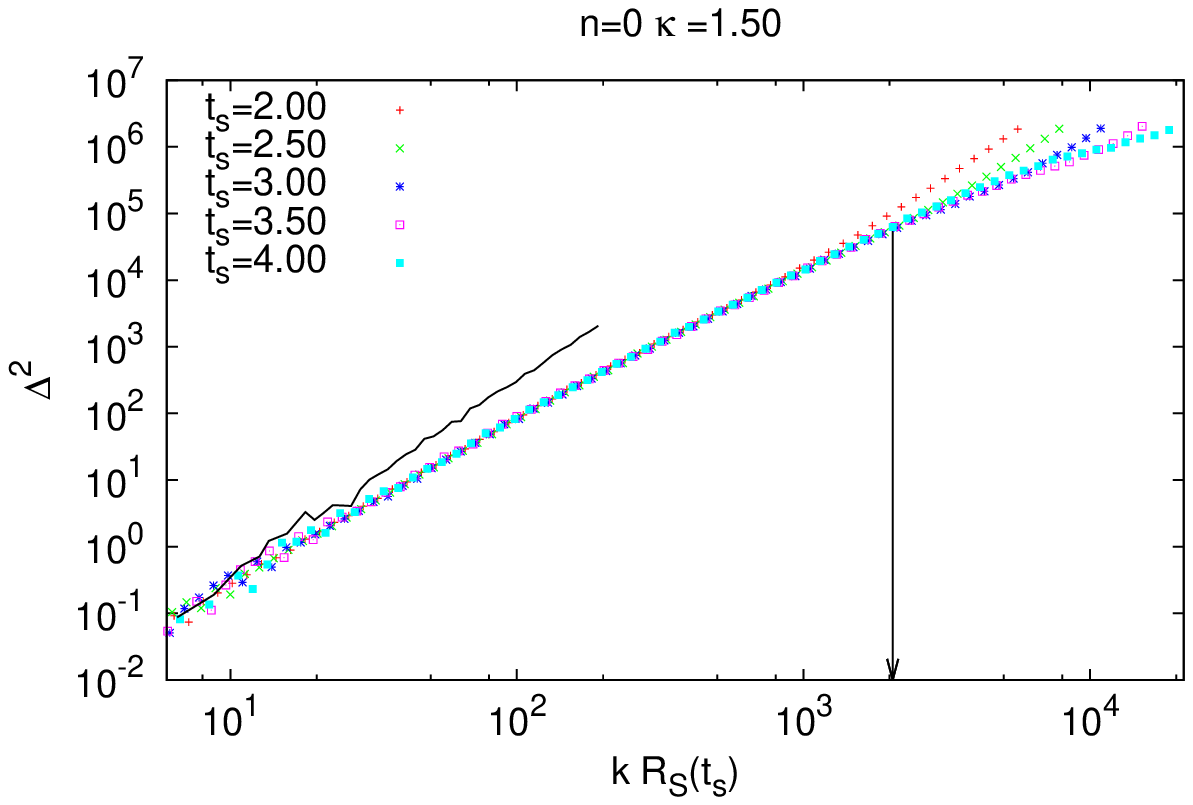}}\\
	\resizebox{!}{5.8cm}{\includegraphics[]{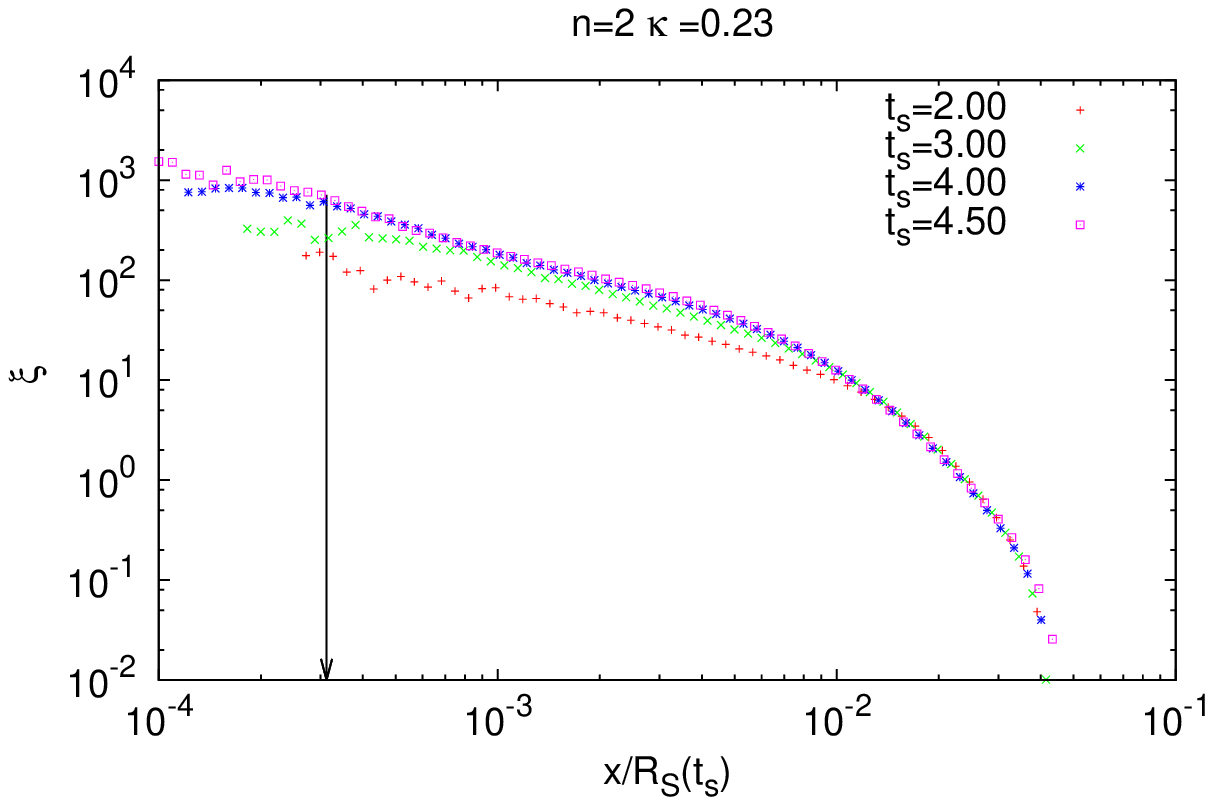}}
	\resizebox{!}{5.8cm}{\includegraphics[]{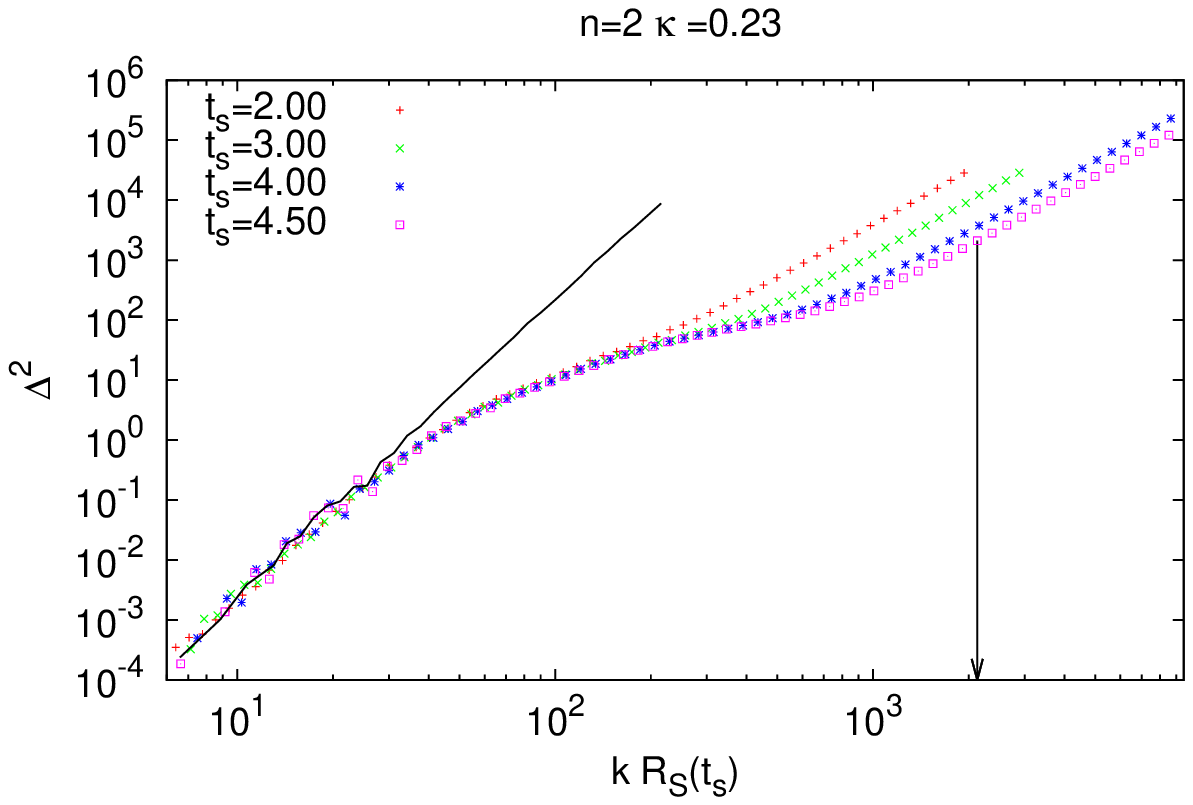}}\\
	\resizebox{!}{5.8cm}{\includegraphics[]{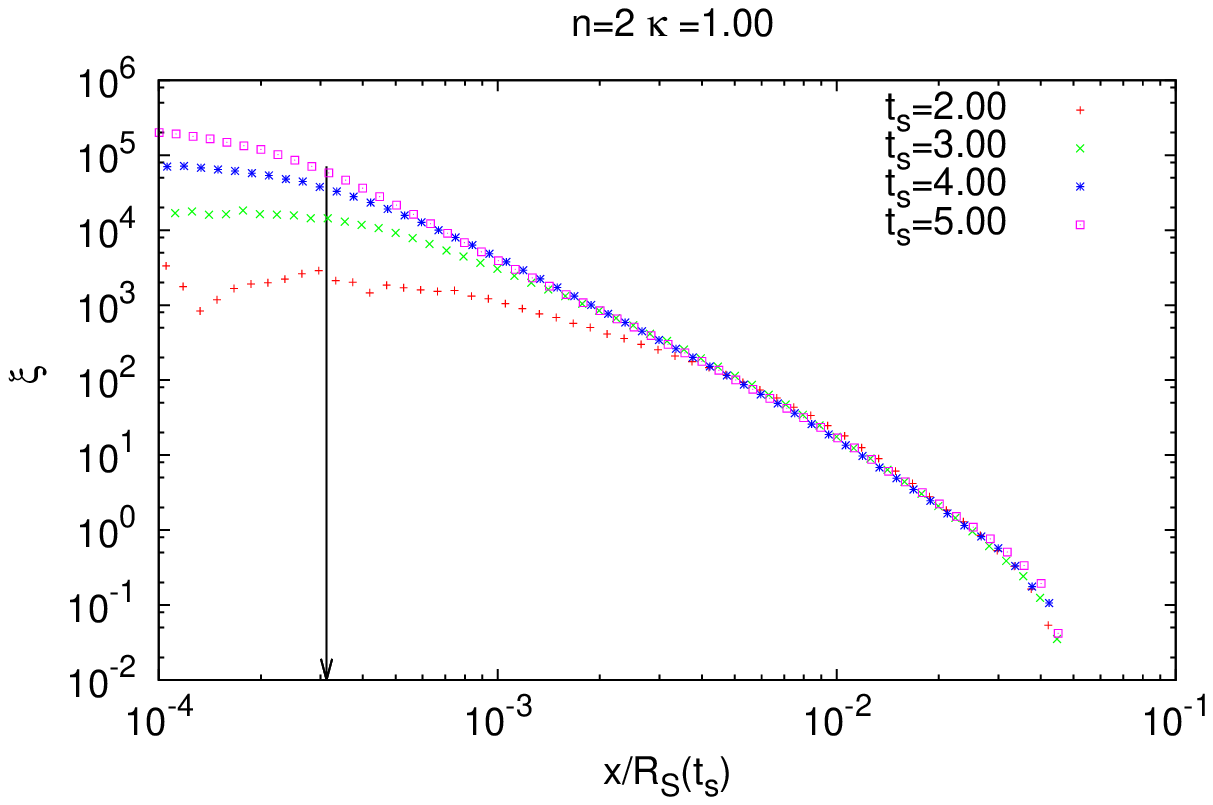}}
	\resizebox{!}{5.8cm}{\includegraphics[]{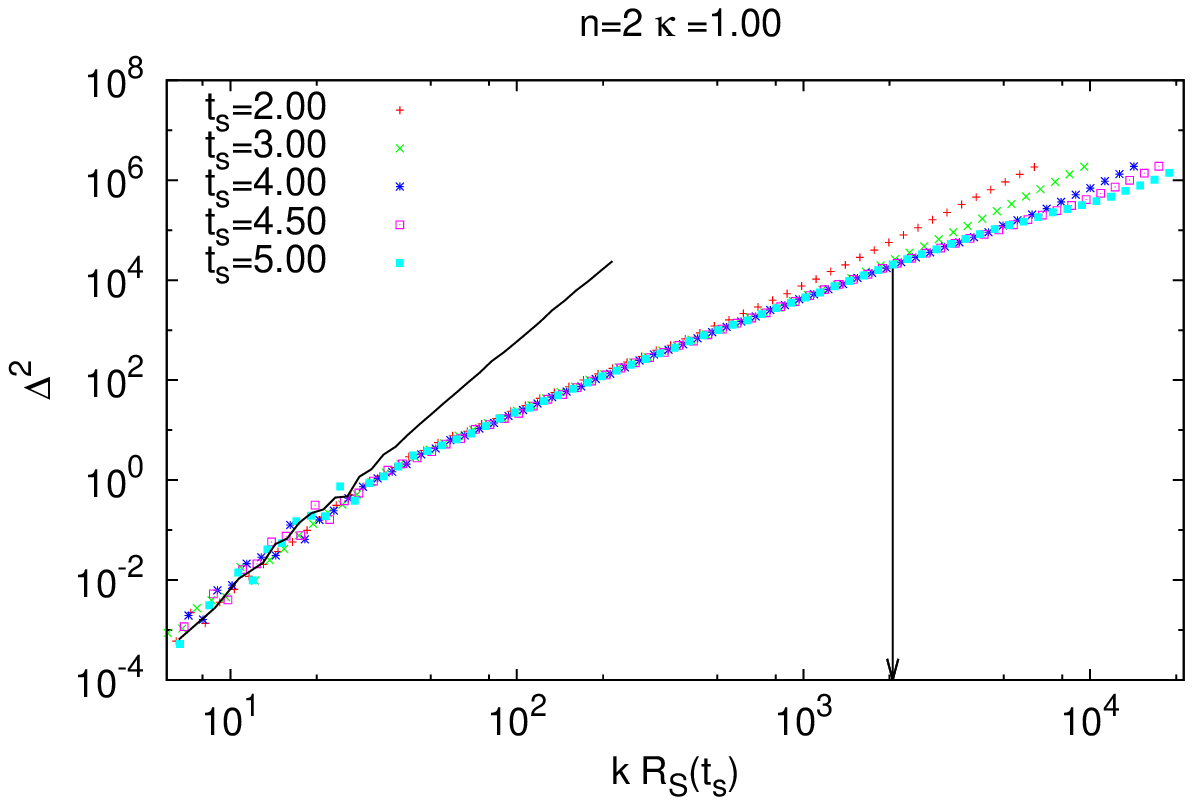}}
	\caption{Each plot shows $\xi$ or $\Delta^2$ for a model with $n=0$ or $n=2$, for different times and 
	in rescaled space variables appropriate to test for self-similar evolution. The black vertical lines indicate 
	the scale $2\epsilon$ in the plots of $\xi$, and the scale $0.1(\pi/\varepsilon)$ in the plots of $\Delta^2(k)$. The black curve in each plot of $\Delta^2 (k)$ shows the prediction of linear theory.  }
	\label{fig-selfsimilarity-n0-n2}
\end{figure*}

To test for self-similarity in the two point correlation properties, as 
expressed by (\ref{self-similarity-xi}) and (\ref{self-similarity-Pk}), it is 
convenient to plot for each simulation $\xi(x)$ and $\Delta^2(k)$ in
the rescaled length units ($x/R_s$, $kR_s$) in which they should 
be identical if self-similarity holds. Shown in Fig. \ref{fig-selfsimilarity-n-2_n-1} 
and Fig. \ref{fig-selfsimilarity-n0-n2} ,
are these plots for a number of our simulations. Specifically for 
each $n$ we show results for the simulation with the smallest 
and largest $\kappa$ (and thus the smallest and largest $\gamma_{sc}$). 
The rescaling has been done taking the final time of our simulation as 
the  reference time (i.e. for the latest time shown the length scales are 
the untransformed  ones). Also shown in each of the 
$\Delta^2(k)$ plots is the prediction of linear theory, obtained
by rescaling according to linear theory the measured value in 
the initial conditions. The black vertical lines indicate the
scale $2\varepsilon$ in the plots of $\xi$, and the scale $0.1(\pi/\varepsilon)$
in the plots of $\Delta^2(k)$. Following the results and discussion of 
Section \ref{Effects of force smoothing}, these are the scales at 
which we anticipate that the results begin to be significantly 
affected by force smoothing. 

At any given time we can infer the measured $\xi(x)$ or 
$\Delta^2(k)$ to be self-similar over the range of scale 
over which they are well superimposed with their 
values at other times. In all the  plots we indeed observe 
that, starting from the initial time, there is a region of 
superposition of each of $\xi(x)$ and $\Delta^2(k)$ 
with its value at the subsequent time step, and, in 
some cases, at all subsequent time steps. In the plots 
of $\xi$ the corresponding range at each time has a
very clearly identifiable {\it lower} cut-off for 
$\xi(x)$,  at a value of $\xi$ which increases 
monotonically in time.  and,  correspondingly, 
in the $\Delta^2(k)$ plots an upper cut-off at a 
value of $\Delta^2(k)$ which increases in time.
These behaviours reflect the progressive 
establishment of self-similarity via the mechanism  
of hierachical structure formation well documented
in the usual cold dark matter models: the transfer to 
smaller scales of the initial power at larger scales by non-linear 
evolution, which leads to  self-similarity when the initial 
fluctuation spectrum contains only a single characteristic scale. 
Conversely the dependence of clustering on the details of the 
initial fluctuations at scales around and below $\Lambda$ 
is progressively wiped out (but at a rate which,
as we will discuss below, clearly depends strongly on the model)
\footnote{ The asymtotic behaviour at large $k$ in
the plots of $\Delta^2(k)$ is simply the shot noise intrinsic
to the stochastic point process: from the definition of the
power spectrum, we have $\lim_{k\rightarrow \infty} P(k)=1/n_0$,
and therefore $\Delta^2(k) \propto k^3$ at large $k$. This
behaviour evidently corresponds to the inevitable breaking of
self-similarity at small scales (in real space) by the particle 
discreteness.}
Thus there is clear evidence for the asymptotic establishment of 
self-similar evolution in {\it all} these models. We note in particular 
that our results show self-similarity to apply  also for the models 
with $n=2$ (and also for the two others not shown here). Thus, as 
anticipated in our discussion above in Sect. \ref{section-SS}, it is clear that 
there is no breakdown of self-similarity above $n=1$ as has 
been suggested (on theoretical grounds) in some 
works. 

The degree to which self-similarity is established varies, however,
quite markedly from model to model: 

\begin{itemize}
\item We observe, for each $n$, that
comparing the two values of $\kappa$ at 
the same $t_s$,  the {\it lower cut-off} to 
self-similarity is very significantly smaller
for the case with larger $\kappa$, and
the amplitude in $\xi$ (or $\Delta^2$)
to which it extends is larger. Indeed
for the cases with the smallest $\kappa$, 
and most notably for ($n=2, \kappa=0.23$) 
and ($n=-2, \kappa=1$), the region in which 
self-similarity can be observed is very limited,
barely extending beyond $\xi \sim 10^2$.
These results are clearly  qualitatively in line with the 
estimates we made in Section \ref{range of SC} based 
on the hypothesis  of (self-similar) stable clustering,
with the range of non-linear self-similar correlations
clearly strongly increasing as $\gamma_{sc}$ does.
Analysis of the same plots for the (seven) other models, 
corresponding  at each $n$ to the models with values of $\kappa$ intermediate 
between those shown here, confirm very clearly 
these  trends, and even show rough quantitative agreement
with the estimates given in Section \ref{range of SC}. 
Notably, at given $n$, we indeed 
observe self-similarity develop in a logarithmic range 
of scale which is very consistent with a 
proportionality to $3/(3 - \gamma_{sc})$, as predicted
by Eq.~(\ref{SCrange-estimate}).
Comparison of these estimates for simulations at 
the same $\gamma_{sc}$, but different $n$, 
shows also good agreement, although it
is complicated by the fact that the scale denoted
$x_f$ in Section \ref{range of SC}, the largest scale 
which has gone fully non-linear, in fact varies 
quite significantly as a function of $n$ in our 
simulations: examining, for example, the scale 
at which $\xi=1$ in Fig. \ref{fig-selfsimilarity-n-2_n-1}
and Fig. \ref{fig-selfsimilarity-n0-n2}
we see it is substantially larger for the cases 
$n=-2$ and $n=-1$ than for the two other 
values of $n$.

\item We see also other very clear differences,
in the plots of $\xi$, in the behaviour at larger scales:  for 
the cases  $n=-2$ and $n=-1$ one can clearly see
that self-similarity has at each time
an {\it upper cut-off}
at scales well inside the box, at a 
value of $\xi$ which monotonically increases 
in time; for $n=0$ and $n=2$, on the other hand, no 
such upper cut-off can be detected 
(and this is true also in plots not shown 
extending to the scale of the box). 
In the ($n=-2, \kappa=2.83$) model
self-similarity is thus visibly broken at the
latest time for $\xi \ltapprox 1$, while in the 
other $n=-2$ model (which extends to a 
larger $t_s$)  and the two $n=-1$ models,  
this break extends almost to
$\xi \sim 10^2$, with a curve at the final time 
in these cases slightly lower than that defined 
in this region by the superposition of the curves at
the earlier times. 
This breakdown of self-similarity at larger 
scales is  precisely in line with
what one anticipate due to finite box size effects:
indeed the deviations from self-similarity are
much more significant for the models with 
$n=-2$ and $n=-1$, for which the amplitude 
of fluctuations at the scale of the box 
(cf. Table \ref{Table1}) are largest 
at the latest times, and which are 
expected to be most sensitive to the 
``missing power" at larger scales
\footnote{ For a recent study quantifying such
effects carefully see \cite{orban_2013}.}.
Indeed for these cases, the higher amplitude
makes it not only possible for us to detect
clearly, in $\xi$, the breakdown of self-similarity
at quite early times in the linear regime, but
also its propagation to the point where it 
affects the correlations in the non-linear 
regime. The fact that this behaviour can be traced 
clearly in $\xi$, but not in $\Delta^2(k)$, is due
to the fact that the latter is more sensitive to 
the contributions from the ``missing modes"  
(i.e. below the  fundamental  of the periodic box), 
increasingly so as $n$  decreases  (reflecting the 
infra-red divergence in the integral defining 
$\xi$ as $n \rightarrow -3$). Thus for $n=0$
and $n=2$ we would need to evolve the 
simulation much further even to be able
to detect such effects in the linear regime.

\item We note that 
the ($n=0, \kappa=0.5$) model, and
to a much lesser extent the  
($n=2, \kappa=0.23$) model, show a qualitatively 
different behaviour to the other (thirteen) models,
in the strongly non-linear regime at the last time shown:
there is, comparing the last two times, apparently good 
self-similar superposition down to a scale of order $\varepsilon$ 
but broken by a slight ``bump" in $\xi$ at intermediate
scales. We believe that these results, at least at 
amplitudes significantly above $\xi \sim 10^2$, are 
probably unphysical because they correlate
precisely with the poorer numerical precision
indicated by the data in 
Section \ref{Monitoring of Energy}, and we will
take account of this in the discussion of our 
final results below. 

\end{itemize}

It is interesting to comment on these results in relation
to discussion in the previous literature (for the case $\kappa=1$)
of establishing self-similarity (and evaluating the
validity of stable clustering) below $n=-1$  (e.g. \cite{efstathiou_88, jain+bertschinger_1997, jain+bertschinger_1998, smith}). Through
the study of our family of models, we see very
clearly that the difficulty in establishing self-similarity (and
thus testing for stable clustering) is unrelated to any
intrinsic problem posed by the infra-red properties
of the spectrum. Indeed we see that we have no difficulty
observing self-similarity (and, as we will see, establishing
that the stable clustering approximation is good) for
$n=-2$ when $\kappa$ is increased, and conversely
we encounter difficulty, for example, when $n=2$ 
when $\kappa$
is much less than unity. At the same time, we do see 
very clearly the effect of greater sensitivity to finite 
size effects for decreasing $n$, and can control 
for them: if self-similarity is broken only in the 
regime $\xi <1$, we can be very confident that
these finite size effect do not modify the 
non-linear regime at all; and even when
they lead to breaking of self-similarity to 
significantly larger $\xi$ --- as is the case, as 
we have seen, in some of the simulations 
with $n=-2$ and $n=-1$ at the latest times ---
we can still always identify the part of the
correlation function or power spectrum
affected by it.  Further direct
tests for finite box size effects reported
below confirm the reliability of this procedure.

\subsection{Non-linear self-similar clustering: two point correlation properties}
\label{Non-linear self-similar clustering: two point correlation properties}

We now analyse in detail the strongly non-linear regime.
 More specifically {\it we isolate the self-similar
region} of the measured two point correlations, and 
assess to what extent the two point correlation properties 
are in line with, or deviate from, the predictions of 
stable clustering.
The latter, as we have 
discussed, predicts a region of power-law clustering $\xi(x) \sim x^{-\gamma}$
with the exponent $\gamma= \gamma_{sc} (n, \kappa)$ as  given 
in (\ref{sc-prediction}). In the power spectrum such a region, if
sufficiently extended in real space, will be expected to lead to
a region in which $P(k) \sim k^{-3 + \gamma}$, or  
$\Delta^2(k) \sim k^{\gamma}$. 
 We privilege the real space 
analysis, using the two point correlation function $\xi(r)$,
for the reasons illustrated by our analysis in 
Section \ref{Effects of force smoothing}: we
can be confident the effect of the 
force smoothing is clearly localized in direct space,
while this may not be the case in reciprocal space.
However we perform also the  reciprocal space analysis, to 
see the consistency of our results, and also 
for the purposes of comparison with other studies.

\begin{figure}
	\centering\includegraphics[scale=0.6]{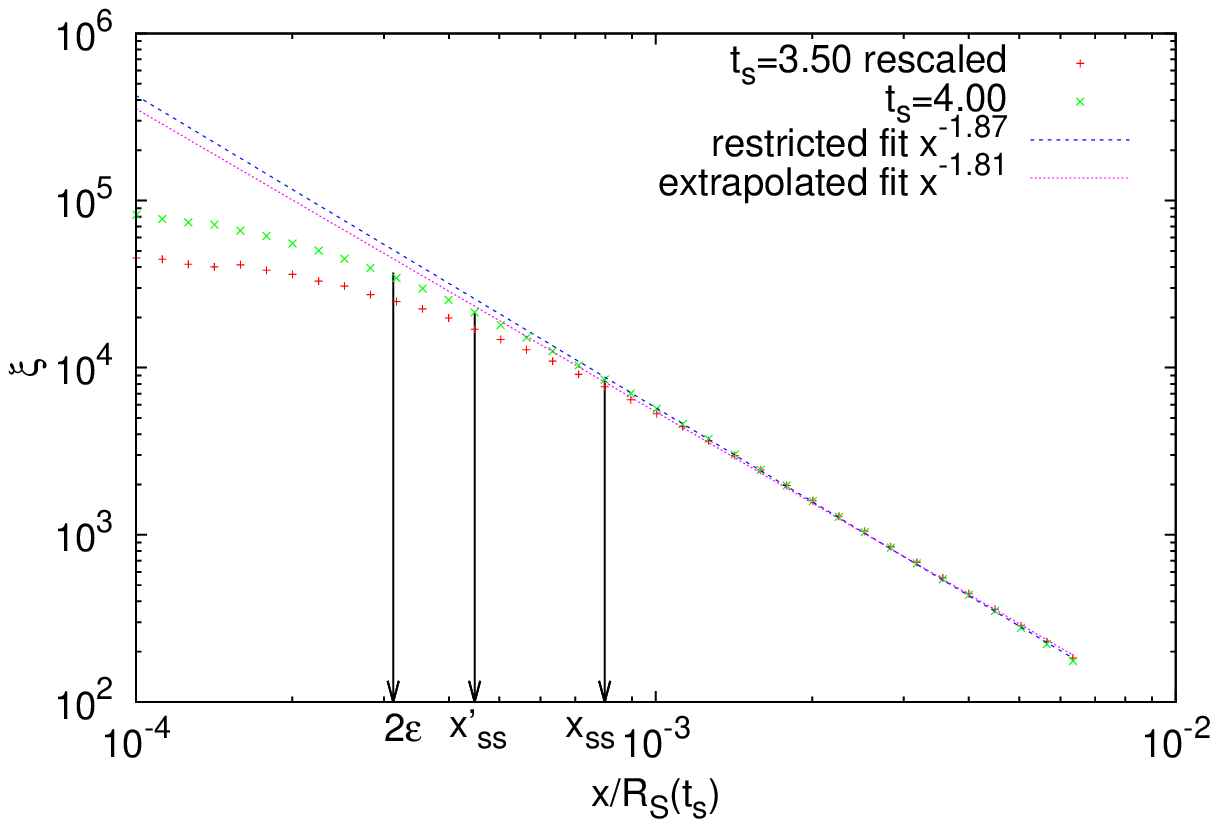} \\
	\centering\includegraphics[scale=0.6]{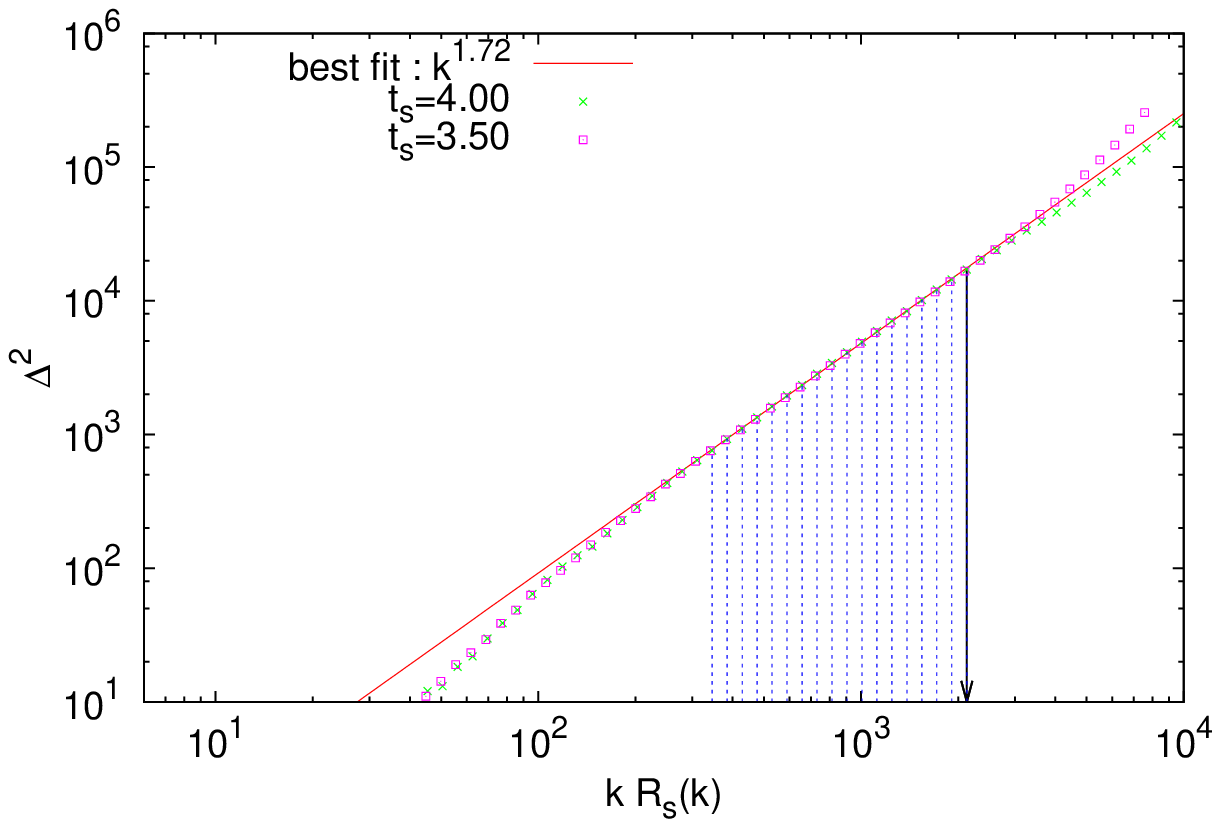}	
	\caption{Plots showing our fitting procedure for the non-linear self-similar correlations for one model ($n=0, \kappa=1$):  for $\xi$ we perform a ``restricted fit" and an ``extended fit'' using two lower cut-offs (indicated by arrows) determined as described in the text; for $\Delta^2$ we perform a fit over the range of self-similarity as determined by visual inspection.} 
	\label{fig-fittingSSregionsXi}
\end{figure}

In each simulation we perform our analysis in the following steps:
\begin{itemize}
\item We extract the correlation function $\xi_f (x)$ at the final
time $t_s^f$ (as in Table~\ref{Table1}), and the correlation
function  $\xi_{f-1} (x)$ at a slightly earlier time, 
$t_s^f - \Delta t_s $. For the $n=-2$ simulations we have
taken $\Delta t_s = 0.3$, and  $\Delta t_s = 0.5$ for
the others. 

\item We perform the self-similar scaling on $\xi_{f-1} (x)$ 
to the time $t_s^f$, and comparing it with $\xi_{f} (x)$,
we determine a scale $x_{ss}$ below which the 
rescaled $\xi_{f-1} (x)$ deviates by more than 
ten percent from $\xi_{f} (x)$. This defines what we 
take to be the lower cut-off to self-similarity. 

\item
We examine the correlation function $\xi_f (x)$ above the 
scale $x_{ss}$ and perform a power law fit in the region
in which such a fit appears by eye to be reasonable.
In all simulations this is the case if we take an 
upper cut-off for the fit at $\xi \sim 10^2$, which is very 
consistent with the assumption that this upper cut-off
marks the transition to a quasi-linear regime\footnote{The
weak dependences on $\kappa$ expected from 
the spherical collapse model will be described in future work.}.
In the cases $n=-2$ and $n=-1$ this upper cut-off scale 
to the power law fit is always smaller than the upper 
cut-off to self-similarity due to finite size effects.

\item
We determine a second scale  $x_{ss}^\prime < x_{ss}$ 
by extrapolating $x_{ss}$ assuming stable clustering 
applies between $t_s^f - \Delta t_s $ and $t_s^f$ 
at the physical scale associated with it at 
$t_s^f - \Delta t_s $. We perform a new power 
law fit, using max [$x_{ss}^\prime$, $2\varepsilon$]
as the lower cut-off  and the same upper cut-off 
as for the first fit. 
\end{itemize}
This procedure is illustrated in Fig. \ref{fig-fittingSSregionsXi}
for one case ($n=0, \kappa=1$).  We note that in this procedure 
the interval $\Delta t_s$ needs evidently to be chosen large 
enough to allow the identification of a lower cut-off scale to 
self-similarity above $2 \varepsilon$.  To optimize the extraction 
of information about self-similarity in our data, on 
the other hand, it should be chosen as small as possible. 
The values we chosen for $\Delta t_s$ given are
in practice sufficiently close to such an optimum:
in the (lower $\gamma_{sc}$) cases where 
$x_{ss}^\prime$ is larger than $2\varepsilon$, 
we have checked that our measured values 
(and the error bars we estimate) below do not 
change significantly if we use smaller 
$\Delta t_s$ (we have data at times intervals
in $t_s$ of $0.1$).
 
For the $k$ space analysis, we have identified by eye 
in each self-similarity plot a value of $k$ up to which 
the $\Delta^2(k)$ at the final time overlaps well with 
that at $t_s^f - \Delta {t_s}$ (using the same values
of $\Delta {t_s}$). A power law 
is then fitted between this point over the region where 
such a fit appears reasonable, which again corresponds 
in all cases to a cut-off at an amplitude consistent with 
the transition to a strongly non-linear virialized regime.
We have also limited ourselves to the region 
$k < 0.1 (\pi/\varepsilon)$.

\begin{table*}
\begin{tabular}{c||c|c|c|c|c|c|c|c|c}
n & $\kappa$ & $\gamma_{sc}$  & $\gamma_{\rm restricted}$ & range  & $\gamma_{\rm extrapolated}$ & range & $\gamma_{\Delta^2}$ & $\overline{\gamma} $ & $\Delta \gamma $ \\
\hline
\hline
-2 & 1.00 & {\bf 1.00} & 1.13 & 0.25 & 1.04 & 0.47 & 1.05 & {\bf 1.07} & {\bf 0.05} \\
-2 & 1.73 & {\bf  1.50} & 1.55 & 0.75 & 1.51 & 1.10 & 1.52 &  {\bf 1.53} & {\bf 0.02} \\
-2 & 2.32 & {\bf  1.80} & 1.79 & 1.00 & 1.78 & 1.35 & 1.78 &  {\bf 1.78} & {\bf 0.01}  \\
-2 & 2.83 & {\bf  2.00} & 2.11 & 0.65 & 2.07 & 1.10 & 2.08 &  {\bf 2.09} & {\bf 0.02} \\
\hline
-1 & 1.00 & {\bf 1.50} & 1.61 & 0.35 & 1.51 & 0.65 & 1.40 &  {\bf 1.51} & {\bf 0.1}  \\
-1 & 1.39 & {\bf 1.80} & 1.89 & 0.80 & 1.83 & 1.20 & 1.78 &  {\bf 1.83} & {\bf 0.06} \\
-1 & 1.73 & {\bf 2.00} & 2.13 & 0.85 & 2.10 & 1.10 & 2.00 & {\bf 2.08} & {\bf 0.07} \\
-1 & 2.32 & {\bf 2.25} & 2.42 & 0.95 & 2.42 & 1.00 & 2.27 & {\bf 2.37} & {\bf 0.08} \\
\hline
0 & 0.50 & {\bf 1.21} & 1.37 & 0.25 & 1.29 & 0.40 & 1.06 & {\bf 1.24} & {\bf 0.16} \\
0 & 1.00 & {\bf 1.80} & 1.87 & 0.90 & 1.81 & 1.20 & 1.72 & {\bf 1.80} & {\bf 0.08} \\
0 & 1.50 & {\bf 2.14} & 2.27 & 1.25 & 2.27 & 1.30 & 2.12 & {\bf 2.22} & {\bf 0.08} \\
\hline
2 & 0.23 & {\bf 1.00} & 0.94 & 0.76 & 0.95 & 0.80 & 1.10 & {\bf 1.00} & {\bf 0.08}  \\
2 & 0.45 & {\bf 1.50} & 1.58 & 0.16 & 1.46 & 0.32 & 1.49 & {\bf 1.51} & {\bf 0.06}  \\
2 & 0.70 & {\bf 1.87} & 1.89 & 0.80 & 1.86 & 1.15 & 1.80 &{\bf 1.85} & {\bf 0.04}  \\
2 & 1.00 & {\bf 2.14} & 2.26 & 0.80 & 2.31 & 1.10 & 2.18 & {\bf 2.25} & {\bf 0.07} \\
\hline
\hline
\end{tabular}
\caption[ ]{Theoretical prediction $\gamma_{sc}$ for exponent
charactering non-linear self-similar two point correlations in
each model, and the three values measured from simulations
as described in the text. For the two fits to the correlation
function the spatial range of the fit is given in the form
 $log \frac{x_{max}}{x_{min}} $. The last two columns
 give the average of these three measures, and an  
 error bar given by half the maximum difference
 between the three measures.}
\label{Table2}
\end{table*}

The results of this analysis are given in Table \ref{Table2}.
For each simulation it gives the exponent obtained from
the two fits to the correlation function (``restricted" and ``extrapolated"), 
as well as the range in which these fits are performed
following the procedure described. Also given is the 
best fit exponent obtained in the $k$ space analysis.
For all of these exponents a standard fitting procedure
gives a statistical error bar ranging between $\pm 0.02$
to $\pm 0.005$. Examining the results we see that 
overall the  measured exponents show in all cases, within
roughly $\pm 0.1$, agreement with the prediction
of stable clustering. Some slightly
larger deviations are seen in a few cases (up
to $\pm 0.2$), but they are associated with 
a dispersion of the same order in the 
exponents estimated in reciprocal space.
In other words, taking the dispersion of
the exponents obtained with the different fitting
methods as indicative of a (dominating) systematic 
error bar, all our results are quite consistent with
the stable clustering hypothesis, and exclude
deviations of at most about $\pm 0.15$.
To quantify the comparison between the predicted
and measured exponents succinctly, the last
two columns in Table \ref{Table2}. give the results 
of the measurements in the form of an estimated 
exponent and  error bar, with the former given by 
the average of the three measured exponents and 
the latter by half the maximum dispersion of the three 
mesures of the exponent. In several cases,  agreement with the 
stable clustering prediction within $\pm 0.05$,
or even considerably less. The largest error
bar --- reflecting a large dispersion between the
real and Fourier space estimates --- is
for the case $(n=0, \kappa=0.5)$, for which
we have noted the late time behaviour is
probably affected by poor numerical precision.
The only case showing possibly significant 
discrepancy with the theoretical exponent
are the models $\gamma_{sc} \geq 2$, 
which all have estimated values marginally
above the theoretically predicted one. On the basis 
of a more extended analysis of the energy evolution 
parameters and other simulations, which we will report in detail elsewhere, 
we believe that there is indeed such a systematic
effect but that it is numerical in origin, linked
to the difficulty of simulating accurately into
the regime where the smallest structures 
``shrink" below the force smoothing,
which is the case in these models.

\begin{figure*}
	\resizebox{!}{5.5cm}{\includegraphics[]{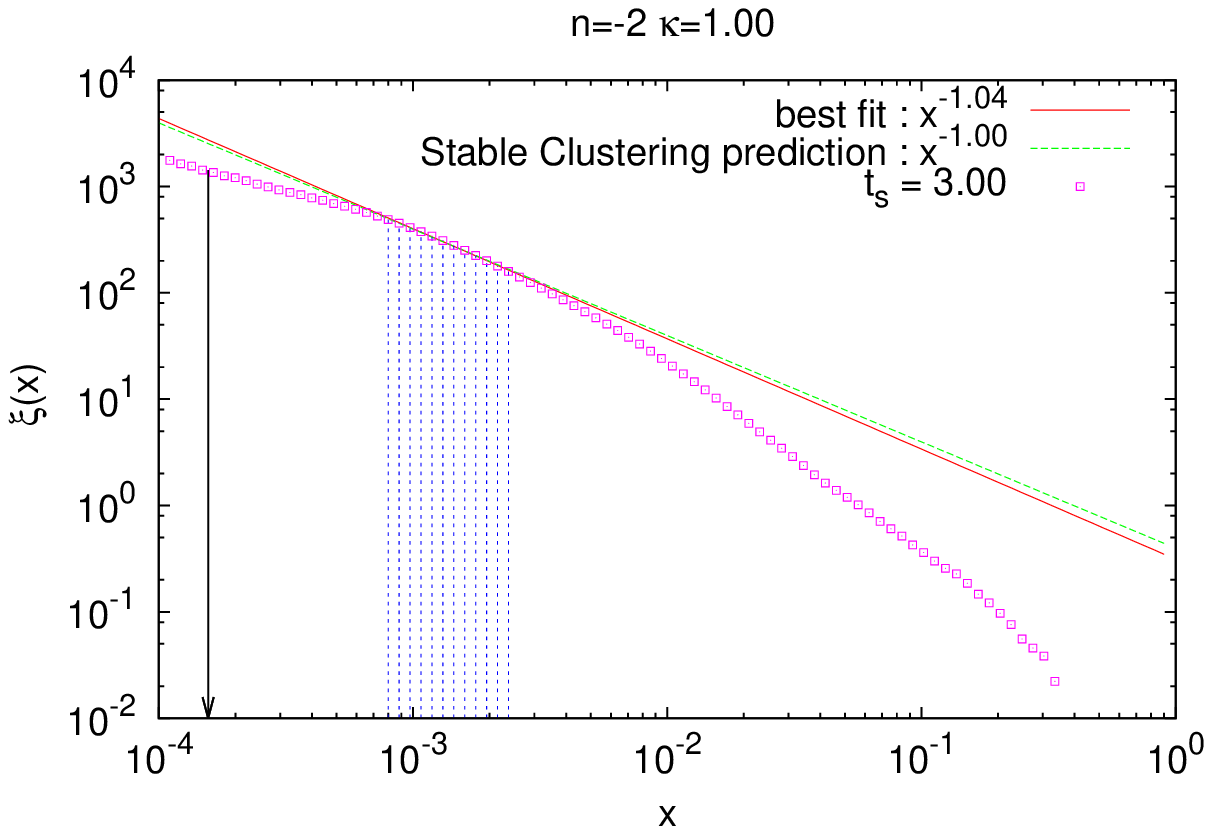}}
	\resizebox{!}{5.5cm}{\includegraphics[]{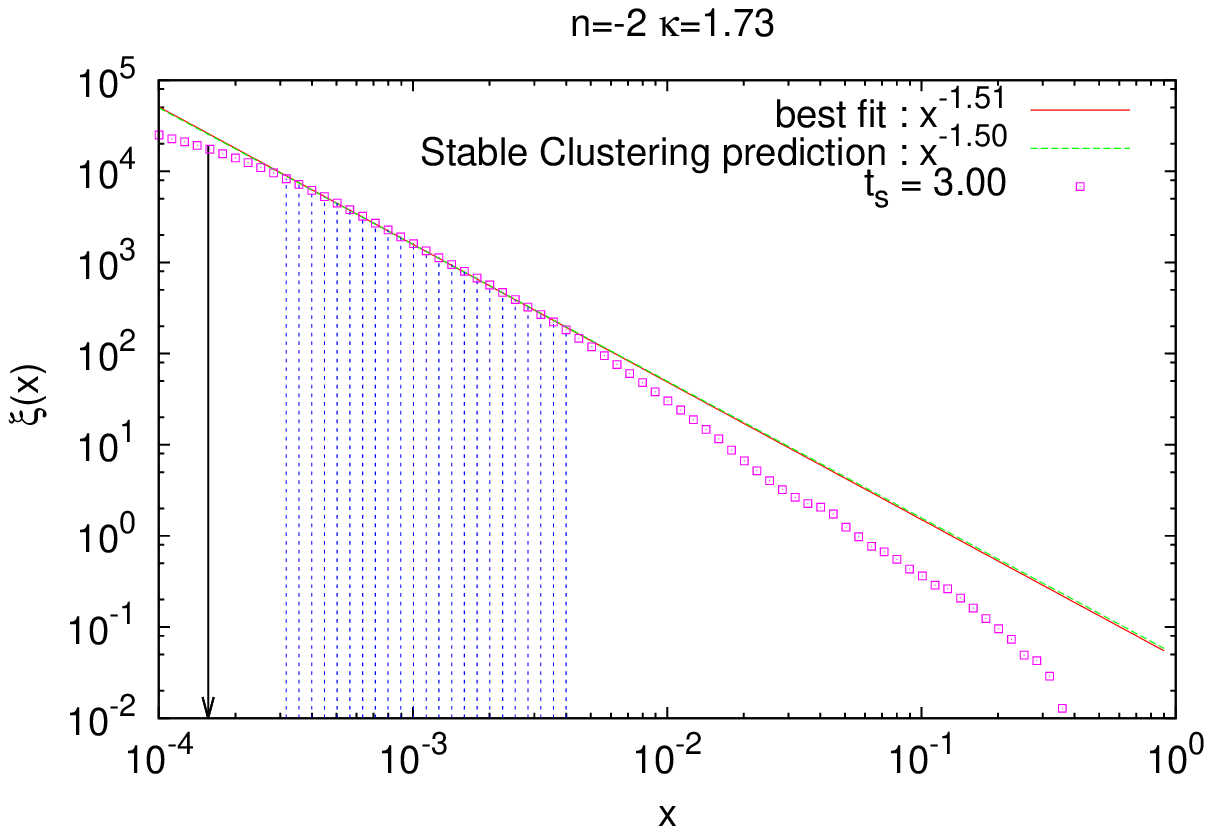}}\\
	\resizebox{!}{5.5cm}{\includegraphics[]{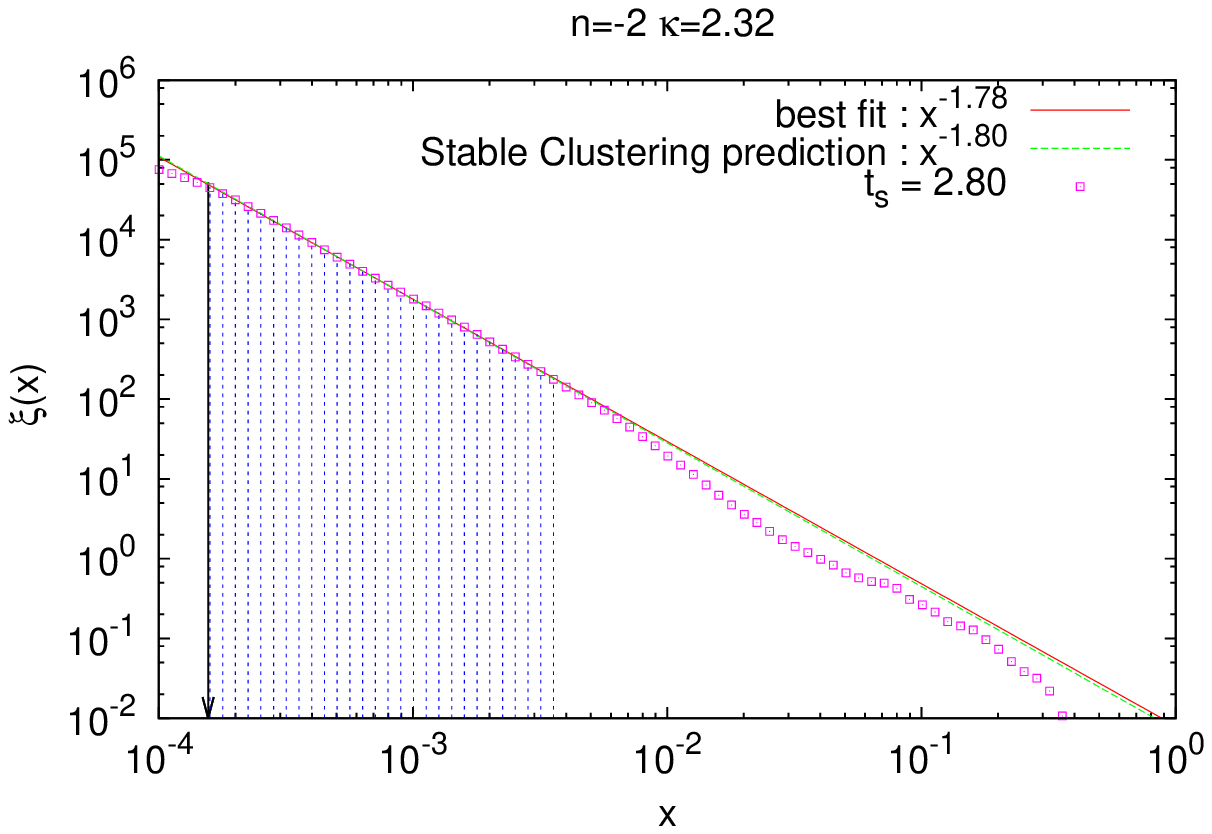}}
	\resizebox{!}{5.5cm}{\includegraphics[]{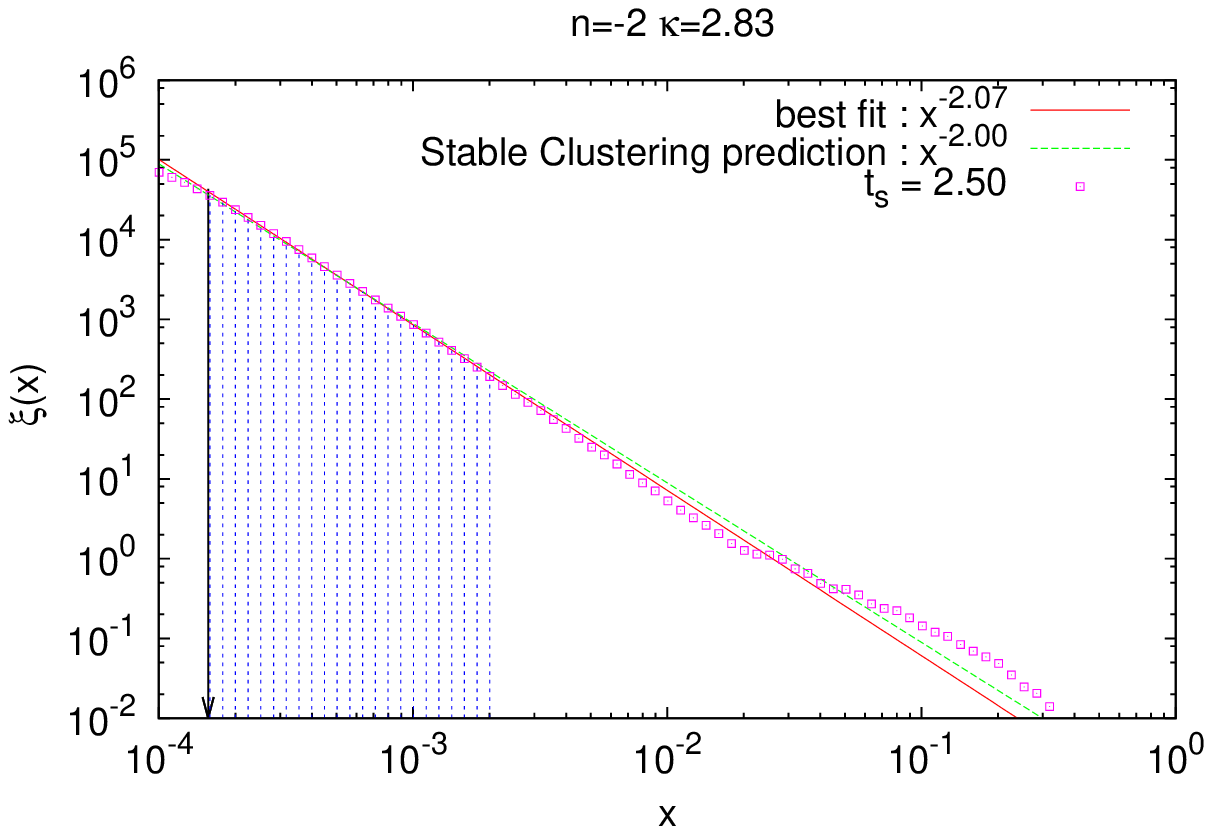}}\\
	\resizebox{!}{5.5cm}{\includegraphics[]{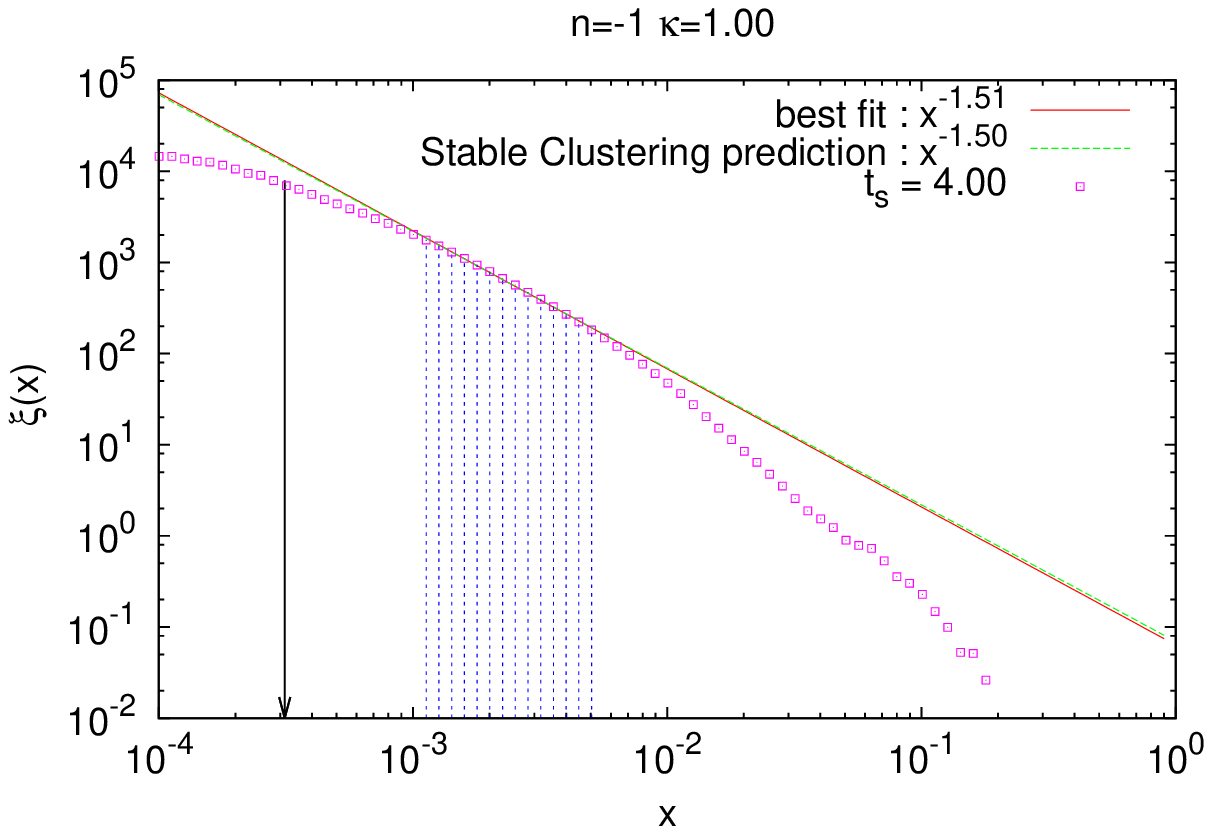}} 
	\resizebox{!}{5.5cm}{\includegraphics[]{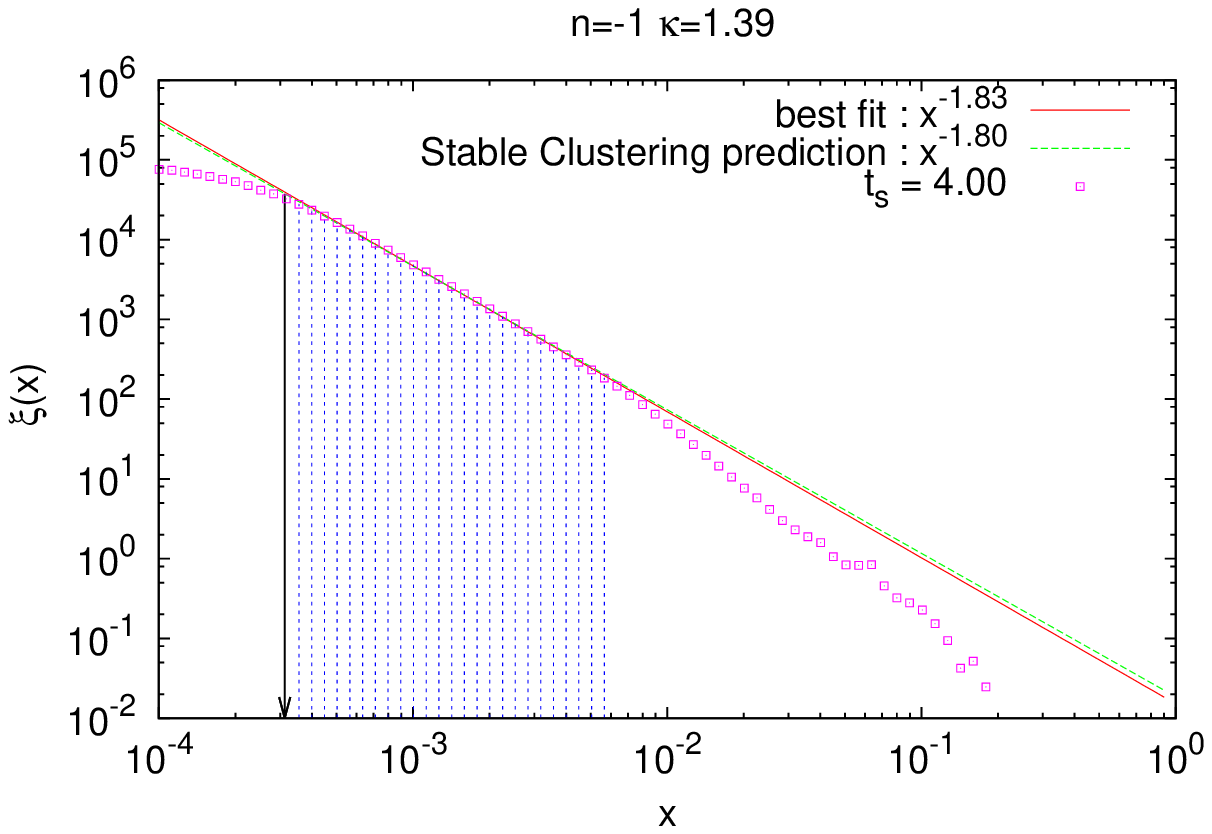}}\\
	\resizebox{!}{5.5cm}{\includegraphics[]{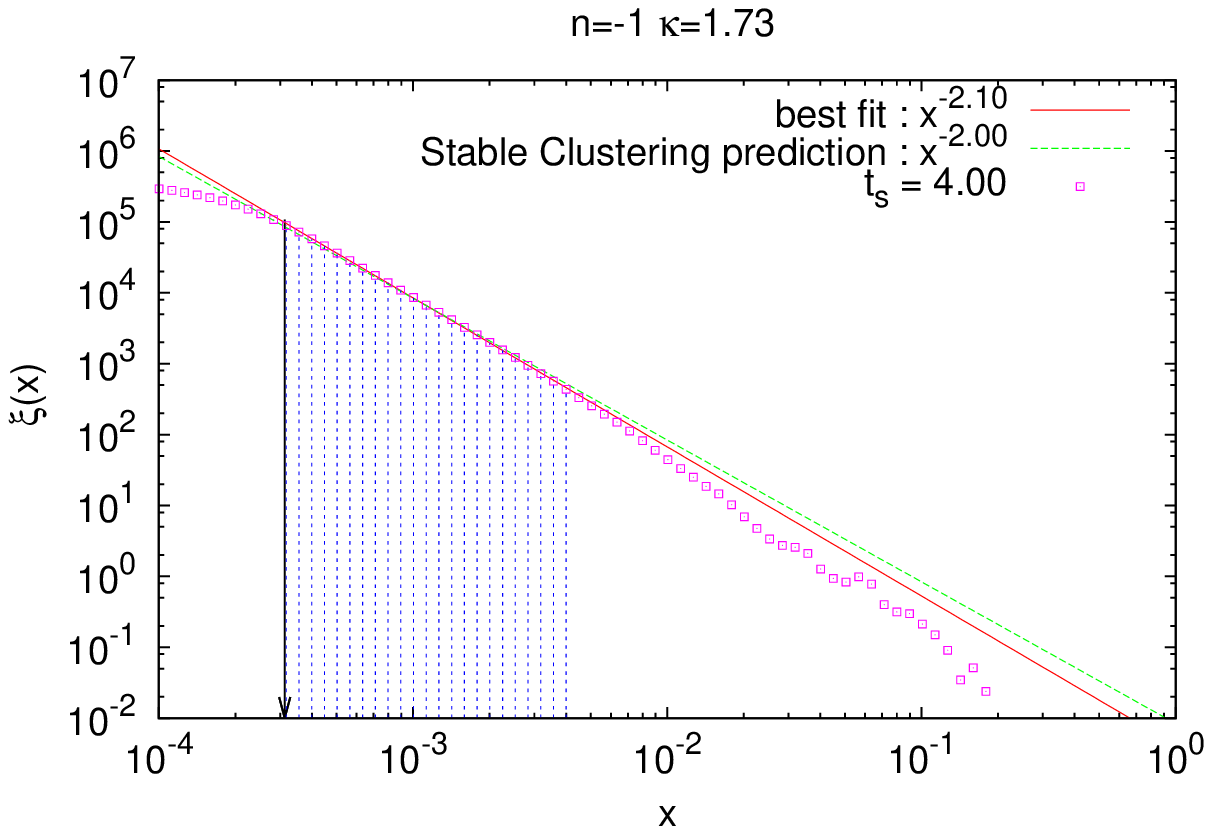}}
	\resizebox{!}{5.5cm}{\includegraphics[]{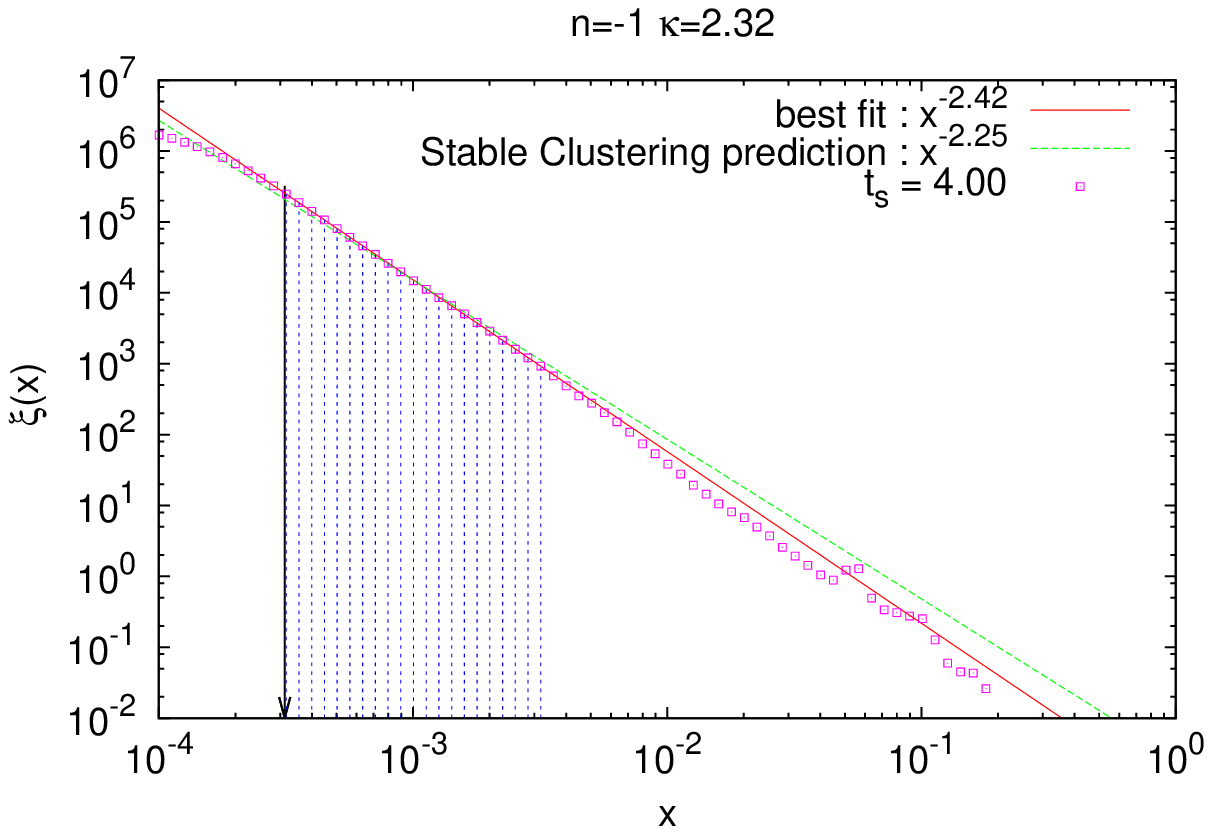}}
	\caption{Plots for $\xi$ at the latest simulation time in each given model, along with lines
	corresponding to our best ``extrapolated fit" and the prediction of stable clustering, in the range 
	indicated by the vertical blue lines. The black vertical line gives the scale $2\varepsilon$. }
	\label{fig-fits-two-xis_n-2_n-1}
\end{figure*}

\begin{figure*}
	\resizebox{!}{5.5cm}{\includegraphics[]{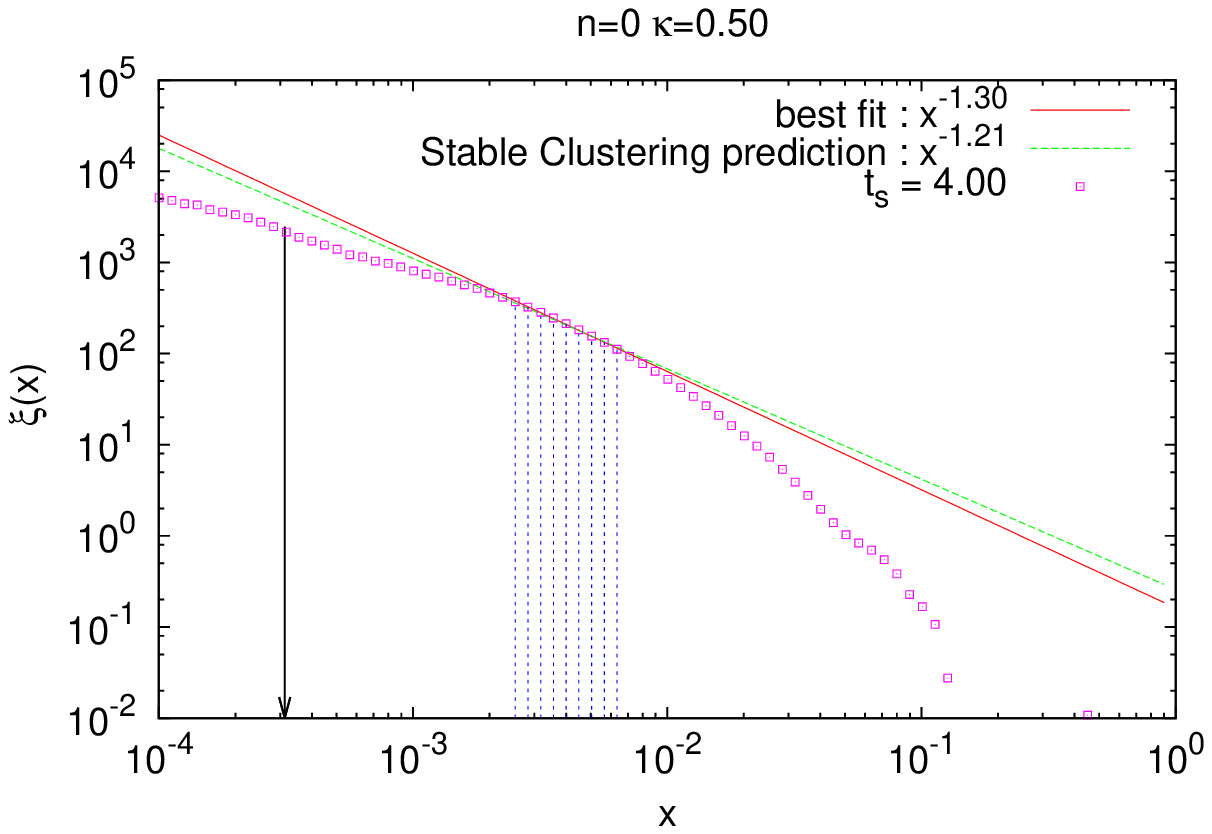}}
	\resizebox{!}{5.5cm}{\includegraphics[]{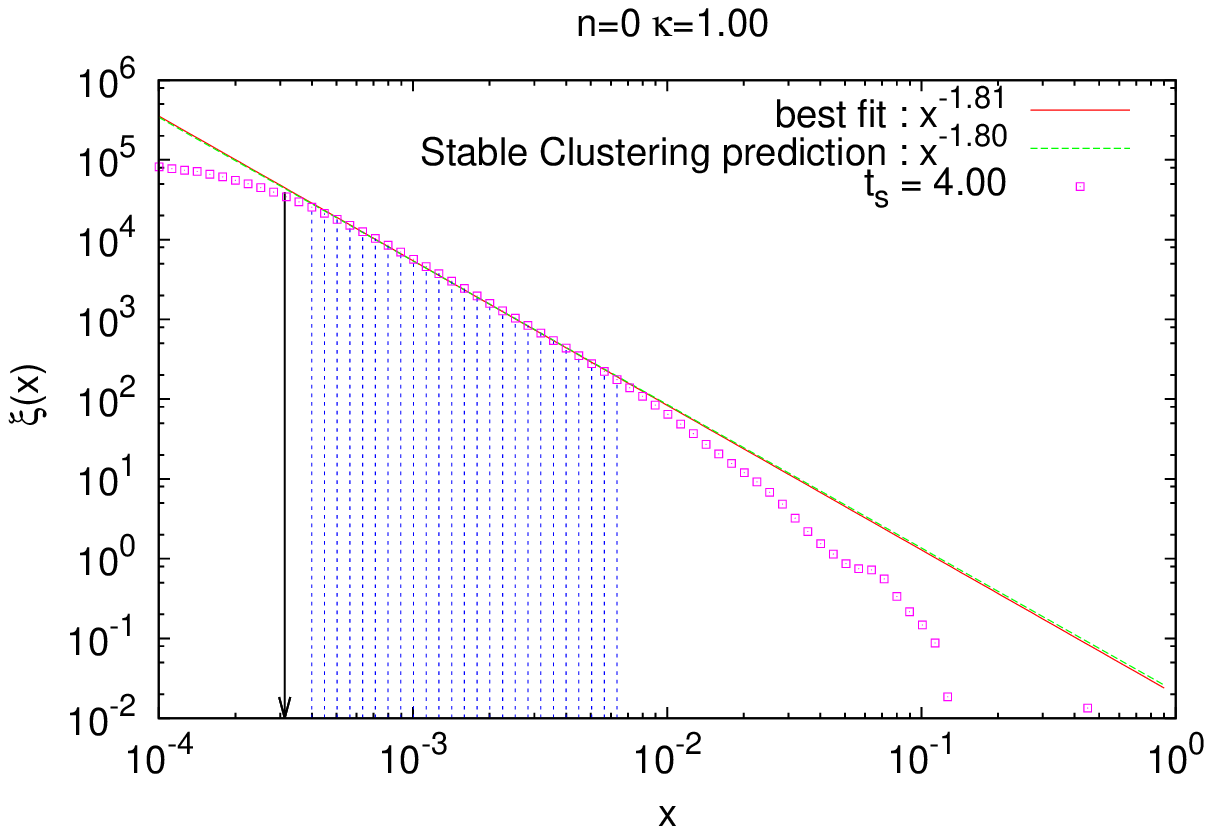}} \\
	\resizebox{!}{5.5cm}{\includegraphics[]{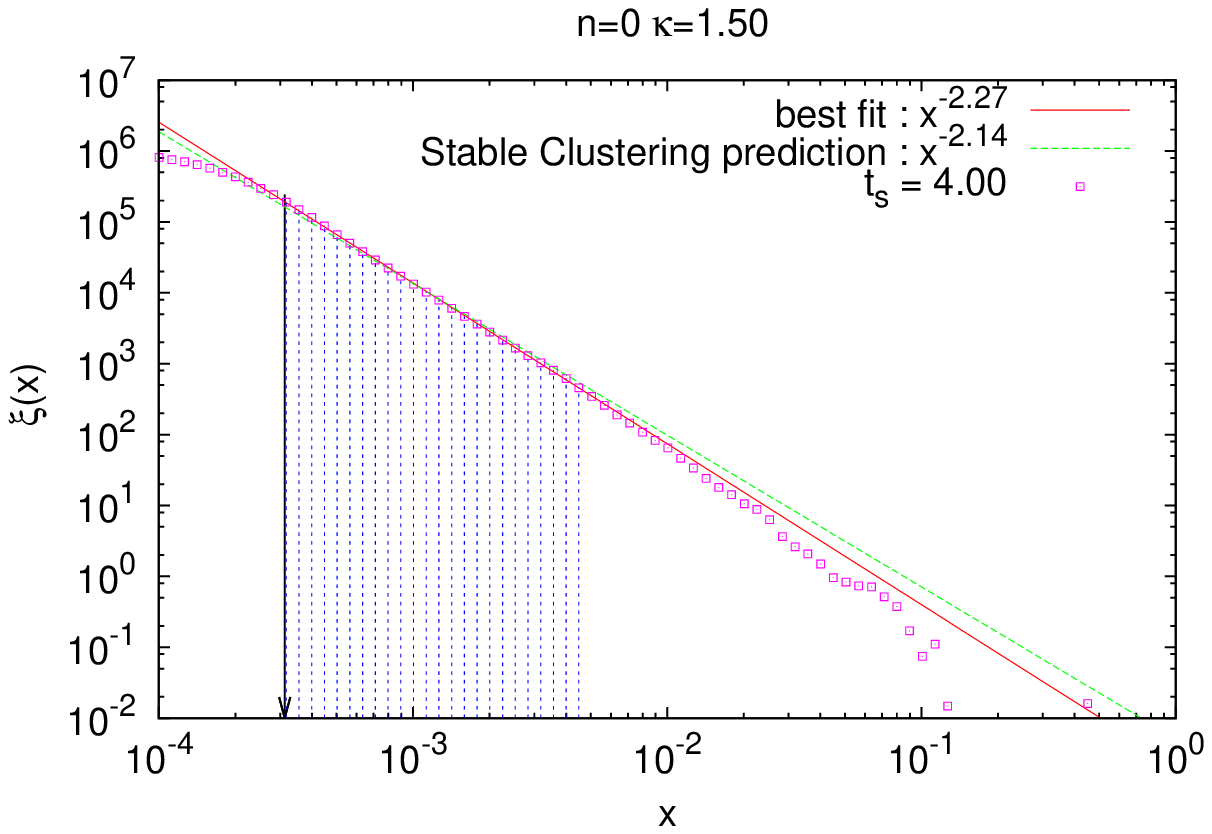}}\\
	\resizebox{!}{5.5cm}{\includegraphics[]{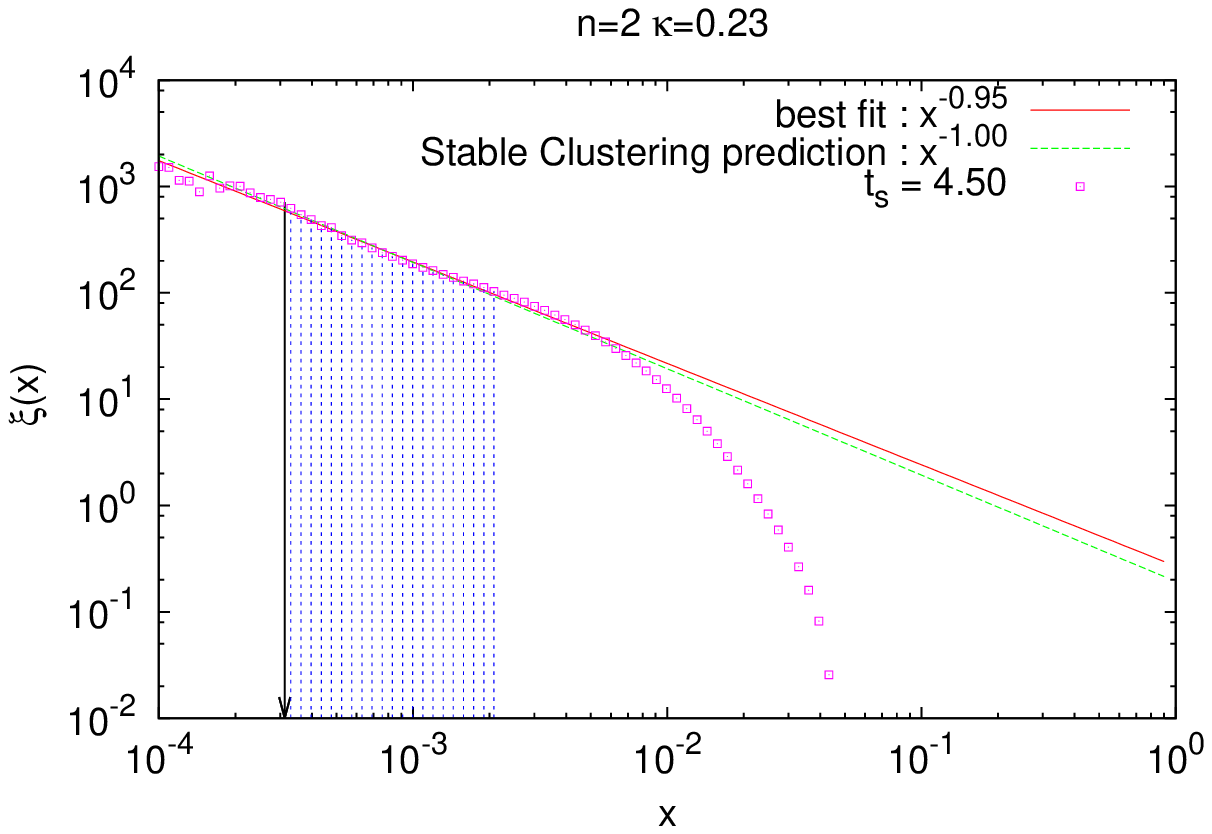}}
	\resizebox{!}{5.5cm}{\includegraphics[]{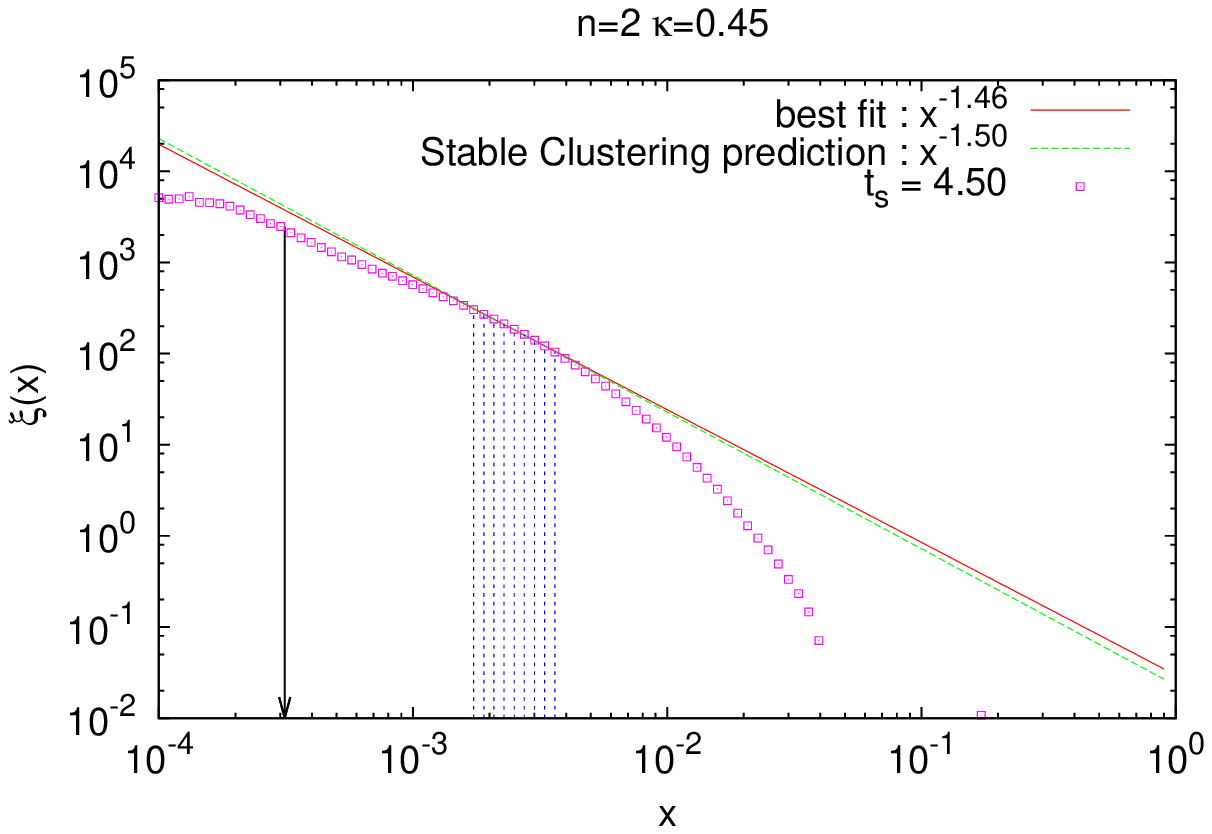}}\\
	\resizebox{!}{5.5cm}{\includegraphics[]{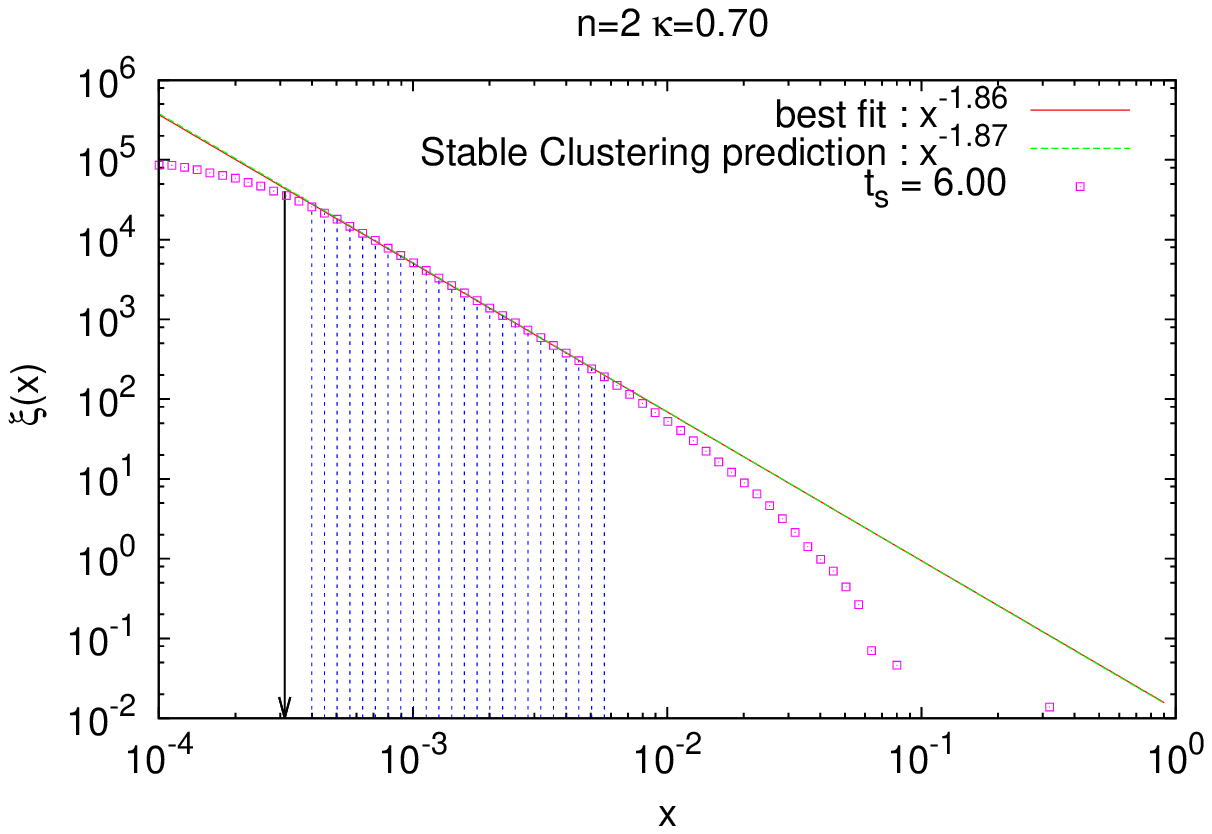}}
	\resizebox{!}{5.5cm}{\includegraphics[]{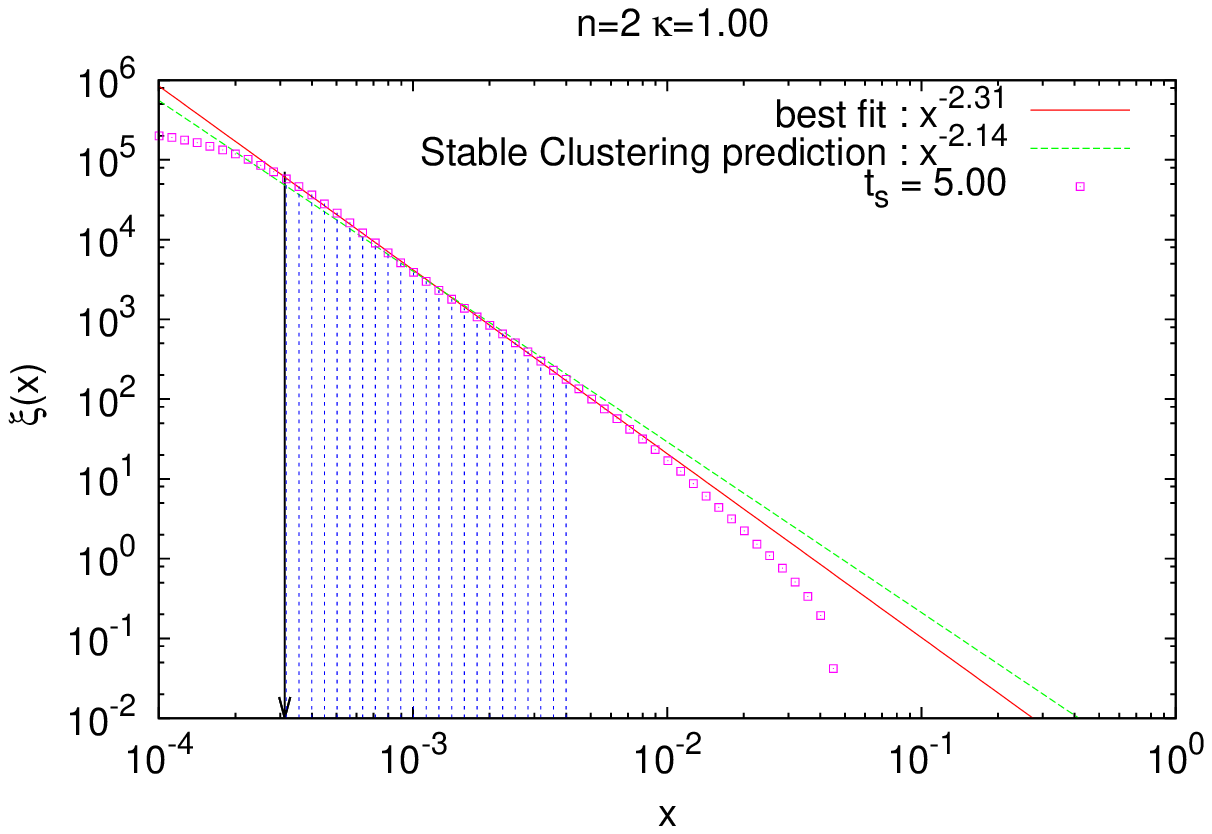}}
	\caption{Plots for $\xi$ at the latest simulation time in each indicated model, along with lines
	indicating our best ``extrapolated fit" and the prediction of stable clustering.}
	\label{fig-fits-two-xis_n0_n2}
\end{figure*}

Shown also in Fig. \ref{fig-fits-two-xis_n-2_n-1}
and  \ref{fig-fits-two-xis_n0_n2}
are the $\xi(x)$ at the latest 
time, for all simulations.
In each case are plotted also
a line showing 
our best ``extended fit" as well
as a best fit (by eye) to the exact 
stable clustering prediction, and
the range of scales in which the
fit is performed. The results of the quantitative analysis
using the estimated error bars are clearly
very coherent with what be observed by
visual inspection of these curves:
excellent agreement with stable
clustering within error bars which
are typically larger for the smaller
$\gamma_{sc}$ simulations, 
for which the physical (i.e. self-similar)
non-linear correlations may be measured
only in a very narrow range of scale. 
We note
that the ($n=0, \kappa=0.5$)
simulation is fitted in a region
excluding most of the range 
in which we observed 
unusual behaviour in our
discussion in 
Section \ref{Results: self-similarity},
while the model ($n=2, \kappa=0.23$)
is fitted in a surprisingly large range,
as the small ``bump" feature noted
also in this model is not excluded from
the self-similar fit by our criterion
of a ten percent discrepancy.
We believe that the real systematic
error in this particular case --- due probably
to poor numerical precision indicated
by $A_1$ --- is therefore larger
than that estimated in Table \ref{Table2}.

 We underline that,
as can be seen in Figs. \ref{fig-fits-two-xis_n-2_n-1}
and  \ref{fig-fits-two-xis_n0_n2}, 
the criterion of self-similarity is in
practice crucially important in 
allowing us to identify the appropriate 
lower cut-off to the range of scale 
in which the simulated correlation
function can be taken to represent
the physical (self-similar) correlation
function. Indeed, while for the 
larger $\gamma_{sc}$
models the determined 
lower cut-off extends down 
to $2\varepsilon$, i.e., down
to the scale which we would
assume automatically be 
a lower cut-off on resolution
of clustering due to force
smoothing, this in not
the case for the models
at smaller $\gamma_{sc}$.
In most of these models 
there is a very significant 
range of scale above 
$2\varepsilon$
--- up to as much as a 
decade in a few cases --- 
in which the clustering 
signal measured in the 
simulation is not self-similar,
and therefore (we assume)
not physical. This means
that in this range of scale
the measured correlations
are unphysical transients
from the initial conditions,
in which notably the
characteristic scale $\Lambda$
in the initial conditions is imprinted.
Conversely, for these simulations, 
we can conclude that no
extra physical information has
been gained by using the small
force smoothing we have 
employed, while for the
larger $\gamma_{sc}$ the
strong constraints on the 
non-linear correlations we
have obtained are a result
of this choice.

In summary, our conclusion is thus that, in the regime where we can
measure with confidence the non-linear {\it and self-similar} 
correlation function of these models using our simulations, it can be fit well 
in the models we have simulated by the stable clustering hypothesis.

\subsection{Direct tests for finite box size effects}
\label{Direct tests for finite box size effects}
As we have discussed above, the restriction of our
measurements of the correlation function and power
spectrum to the scales in which these quantities
are self-similar should ensure that our results
are unaffected by effects arising from the finite 
size of our simulation box. Indeed, for our models
with $n=-1$ and $n=-2$ in which we expect such
effects to be most important, our tests for 
self-similarity have identified clearly the 
presence of such effects at larger scales.  
It is interesting to test also more directly 
for finite size effects by using simulations 
which are identical other than for the
simulation box size, and in particular to
verify that the correlation function and
power spectrum are indeed unaffected 
by the box size in the range of scale
indicated by the test of self-similarity.

\begin{figure*}
	\centering\includegraphics[scale=0.7]{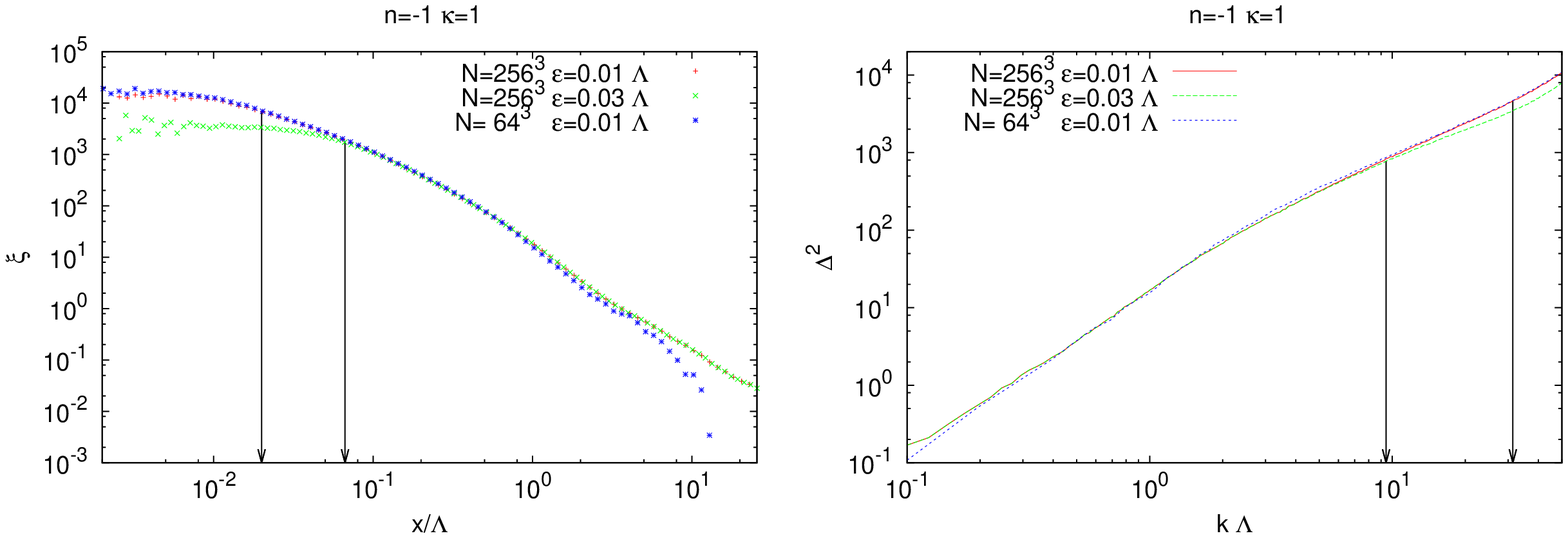} 
	\caption{ Comparison of $\xi(r)$  (left) and $\Delta^2(k)$ (right) 
	 in three simulations of the model  $n=-1$ and $\kappa = 1$,
	 at the time $t_s=4$, and starting from an amplitude of 
	 initial fluctuations given in Table \ref{Table1}. One pair 
	 of simulations differ only in particle number (i.e. box size),
	 while the pair at $N=256^2$ differ only in force smoothing.
	 The $x$-axis in both plots has been rescaled to the initial
	 grid spacing $\Lambda$, and the black vertical lines 
	 indicate the values of $\varepsilon$ on the plots of $\xi$, 
	 and those of  $(0.1)\pi/\varepsilon$ on the plots of $\Delta^2(k)$.}
	\label{fig_compare_N=256}
\end{figure*}

To perform such a test we have resimulated
the model with $n=-1$ and $\kappa=1$ (i.e. EdS)
with $N=256^3$ particles (starting from the same
amplitude of initial fluctuations as given in 
Table \ref{Table1}). For one simulation we have
used the same smoothing ($\varepsilon=0.01$)
as in our fiducial simulations, and for the
other a larger smoothing ($\varepsilon=0.03$).  
Shown in Fig. \ref{fig_compare_N=256} are the
measured $\xi(r)$ and $\Delta^2(k)$ in these
two simulations at the time $t_s=4$ alongside
those for the simulation with $N=64^3$ used
in our analysis above, at the same time.The 
comparison between the two simulations 
with $\varepsilon=0.01$ shows that there are
indeed significant finite size effects at 
scales which are large but well inside 
the box size at this time. Indeed, in
excellent agreement with what we have 
inferred from our analysis using self-similarity 
of the smaller simulation, these effects
extend at this time (the latest time used 
for our analysis above) down to scales 
in which $\xi \sim 10^1$ -- $10^2$. 
It is interesting to note also that the 
peculiar ``bump" feature at $\xi \sim 1$ in the
$64^3$ simulation clearly disappears
in the larger box simulation, and is 
indeed unphysical as one would anticipate
(in a scale-free model).

The comparison of the two simulations with
$N=256^3$ with different smoothing is shown as 
it confirms further the conclusions of the analysis in 
Sect. \ref{Effects of force smoothing}: 
we see that the correlation functions for the 
two simulations agree very well down to close
to the larger $\varepsilon$, while in $\Delta^2(k)$
the effect of the smoothing in decreasing the
power is clearly visible already at  
$k \approx 0.1 (\pi/\varepsilon)$. 
The exponent $\gamma$ obtained fitting 
either simulation in the self-similar region 
gives a value very consistent with that 
obtained from the $64^3$ simulations.

\section{Discussion and conclusion}

In this final section we first give a brief summary of our
most important results and conclusions. We then 
compare our results to some previous studies 
in the literature (of the $\kappa=1$ case). We
conclude with a brief discussion of the possible
implications of our study, in particular in relation
to the issue of ``universality" of gravitational 
clustering in cosmological models, and then
trace finally some directions for future research. 

\subsection{Summary of main results}

\begin{itemize}
\item We have pointed out the interest in studying 
structure formation in a family of EdS models, 
noting that, like the canonical case, they are expected 
to display, for power law initial conditions, the property 
of self-similarity. The latter provides a 
powerful tool to distinguish the physically relevant 
results of a numerical simulation: where the
observed clustering is not self-similar, it is necessarily
dependent on the length and time scales 
which are introduced by the finite simulation. This
family of models allows us to investigate in such a 
context not just, as in the usual EdS model, the 
dependence on initial conditions of clustering, but also 
the dependence on cosmology. 

\item We have derived in these models both the 
theoretical predictions for the self-similar behaviour, 
and those for the exponents $\gamma_{sc} (n, \kappa)$
characterising the non-linear regime in the additional
hypothesis of stable clustering. 

\item We have explained that the (theoretical)
exponent $\gamma_{sc} (n, \kappa)$ has a very 
simple physical meaning: it controls the relative 
size of virialized structures compared to their 
initial relative comoving size. Larger
$\gamma_{sc} (n, \kappa)$  corresponds thus to a
greater ``shrinking" of substructures contained
in a given structure. It follows that we expect
$\gamma_{sc} (n, \kappa)$ to be a good control
parameter for the validity of stable clustering,
and also for the range in which it can be observed
in a finite simulation (if it is a good approximation).

\item We have reported the results of an extensive
suite of $N$ body simulations, performed using an
appropriate modification of the GADGET2 code which
we have implemented and tested. To control the
code's relative accuracy in different simulations 
we have defined a suitable  generalized variant of 
the usual test based  on the Layzer-Irvine equation.

\item We have analysed in detail the two point 
correlation properties in real and reciprocal space 
of the evolved distributions for models covering
a broad range in the $(n, \kappa)$ parameter
space. Our chosen range in $n$ extends to the 
region $n>1$ which has not been studied 
previously even in the standard ($\kappa=1$) 
case. Our results show that all our models
display a clear tendency to establish
self-similar evolution over a range of 
comoving scale which grows 
monotonically in time. 
Further, we find that in all cases the 
self-similar region of the fully 
non-linear part of the correlation 
function (and power spectrum) are
well fit by a simple power law from
with an exponent in good agreement
with the prediction of stable clustering.
As anticipated theoretically, the robustness
and range of validity of these numerical 
results turns out itself to be a function 
of $\gamma_{sc}$:  as  it decreases 
below $\gamma_{sc} \approx 1.5$ the 
range of strongly non-linear self-similar 
clustering diminishes considerably, so much 
so that at $\gamma_{sc} \approx 1$
the prediction of stable clustering can 
only be marginally tested. Given these
limitations we are not able to detect 
any deviation from the stable clustering 
prediction in this class of  models, and
conclude that significantly larger simulations
would be needed to do so.

\end{itemize}

\subsection{Comparison with other studies}
\subsubsection{Three dimensions}
Previous studies have considered the usual EdS model
for different values of $n$ in the range $-3<n \leq 1$,
focussing both on the validity of self-similarity and
stable clustering. 

Concerning self-similarity our results are in agreement 
with all previous studies for this range, but help to clarify
two questions. Firstly they show that for the usual EdS model
self-similarity indeed extends to $n>1$. We have studied
only the case $n=2$, but we can be extremely confident,
given the observed trend with $n$, numerical results
in the one dimensional case (\cite{joyce+sicard_2011, benhaiem+joyce+sicard_2013}), and the theoretical considerations concerning $\gamma_{sc}$,
that this result extends up  to $n=4$.  Secondly, our results 
throw light on the difficulty, documented and discussed in the previous
literature, of observing self-similarity in numerical
simulations in the range $-3 < n < -1$, which even
led to the suggestion that it might not apply for this
case because of infra-red convergence of possibly
relevant physical quantities 
(\cite{jain+bertschinger_1997, jain+bertschinger_1998}).
Our study shows very clearly
that the difficulty in observing self-similarity is not
due to any intrinsic problem related to infrared
properties of these spectra. Indeed we have no
difficulty observing self-similarity over an extensive
range even for a model with $n=-2$ when the
cosmology is modified. The difficulty of observing
self-similarity is due essentially just to the
very limited range in which strongly non-linear 
clustering can develop in simulations of limited
size as $n$ decreases: quite simply the range 
of scale factor which can be simulated is so 
small that the increase in the density 
contrast in stable clustering is very small.
The relative degree of this difficulty for
different models is, as we have explained, 
well characterized by the theoretical stable 
clustering exponent of the model. 

Concerning the validity of the stable clustering hypothesis as
tested by comparison with its predictions for the two point
correlation properties, our conclusions are in 
disagreement with the most recent major study 
of the issue, that of \cite{smith}. This article considers the
cases $n=0, -1, -1.5,  -2$  in the usual EdS model, and 
reports results of an analysis of two point properties
showing, in all cases, significant deviations from the
predictions of stable clustering, characterized by 
the measurement of exponents significantly {\it smaller}
in all cases than the ones predicted by stable clustering:
for $n=0$, $\gamma=1.49 \pm 0.05$ (compared to $\gamma_{sc}=1.8$);
for $n=-1$, $\gamma=1.26 \pm 0.1$ (compared to $\gamma_{sc}=1.5$);
for $n=-2$, $\gamma=0.77\pm 0.15$ (compared to $\gamma_{sc}=1$).
Comparing these results\footnote{The error bars 
quoted are inferred approximately from those shown in Fig. 6 in \cite{smith}. 
Fig.s are given in the text of the paper for quoted errors on the exponents $\alpha$
obtained by fitting $\Delta_{NL}^2$ as a function of $\Delta_{L}^2$, which lead 
to errors in $\gamma$ a factor of two to three times smaller. The derivation of
error bars are not explained in the article.} and our results in Table \ref{Table2}, 
it is not evident, given the roughness of the error bar estimates, whether
there is really a strong discrepancy with our results as described 
above for the case $n=-2$ and $n=-1$. The results for 
the case $n=0$ is, however, are most definitely strongly at odds.
Indeed we found in this case that all three fitting procedures give 
an exponent within $0.1$ of the predicted value, and 
evidently \cite{smith} very strongly exclude a fit to an exponent 
in this range. On the other hand, the slightly earlier study of the case 
$n=-2$ of \cite{bertschinger2}  concludes, like us, that the 
two point correlation properties are in this case in line with 
the prediction of stable clustering, while \cite{smith} clearly
state that their result is discrepant also with this result
(and propose an explanation for this discrepancy which
we will discuss further below).

Let us compare more closely the different studies.
 The analysis and conclusions of 
both \cite{smith} and \cite{bertschinger2} are obtained 
from simulations with $N=256^3$ particles,
while we have used $N=256^3$ for only
one model and $N=128^3$ or $N=64^3$ otherwise.
On the other hand, we have used a softening  $\varepsilon$ 
which, in units of the initial  interparticle spacing $\Lambda$, 
is approximately the same as that of \cite{bertschinger2}
but smaller by a factor of six than that of \cite{smith}.
The overall spatial range {\it potentially} resolved, strictly 
bounded below by $\varepsilon$ and above by the box size, 
is thus largest for the $n=-2$ ($\kappa=1$) simulation 
of \cite{bertschinger2} and our ($n=-1$, $\kappa=1$) 
simulation, and smallest for \cite{smith}, with our simulations
simulations intermediate between the two. In practice, however, 
we must compare the different results taking account of the 
range of scale in which non-linear structure is actually 
simulated in each case, and assess whether systematic effects 
can explain the differences between them.  Laying aside the 
possibility of any significant discrepancies in the numerical integration, 
such systematic effects are either (i) finite size effects associated 
with the periodic box, or (ii) force resolution effects associated 
with the force smoothing.  

In their discussion of \cite{bertschinger2},  \cite{smith} argue
that the discrepancies of their results (for the case $n=-2$) 
arise from finite size effects: on the basis of a simple theoretical 
estimate of the effect of ``missing power" (associated with
modes below the fundamental) in the mass variance, they argue 
that to avoid significant finite size effects a simulation should be 
evolved at most until $\Delta_L^2 (k_b) \approx 0.04 (n+3)$.
They note that \cite{bertschinger2} evolves, and measures the
compatibility of the non-linear power spectrum with stable
clustering, up to a time when $\Delta_L^2 (k_b) \approx 0.4$, 
an order of magnitude larger than this bound admits,
while their own simulations for $n=-2$ are stopped just when this
bound is absorbed.  As can be seen from Table~\ref{Table1}, 
our final value for $\Delta_L^2 (k_b)$ is about half that 
of \cite{bertschinger2} for $n=-2$ and five times larger than 
the bound proposed by \cite{smith}, for $n=-1$ our final 
value is about twice that of the bound, while for $n=0$
($n=2$) our final value is significantly (far) below the 
bound.  

In our analysis we have seen that finite size effects may
be observed directly in simulations through the breaking
of self-similarity at large scales in the correlation function 
$\xi$,  and we have confirmed the reliability
of this method for identifying these effects through
direct comparison with a large simulation in 
Section \ref{Direct tests for finite box size effects}.
We have noted that the behaviours we observe
of these effects are qualitatively in line
with what would be anticipated from the criterion of  
\cite{smith} --- finite size effects are clearly detected 
for $n=-2$, and also for $n=-1$, and
they not detectable for the other cases. However, even
though at the final time in the $n=-2$ and $n=-1$ 
simulation, the  breaking of self-similarity has spread 
beyond $\xi \approx 1$ well into the non-linear region, 
these effects have not propagated into the region of 
very strong clustering, which remains self-similar.
It is in this range which we measure the correlation
function exponent.
Further we have checked that, when we perform
our analysis using as final time one at which
the bound of \cite{smith} is satisfied, we still
obtain an exponent --- albeit over an even more
limited range --- consistent with our quoted result
(and with stable clustering). In short our analysis 
leads us to conclude that, although finite size 
effects are clearly present in the simulations 
with $n=-2$ and $n=-1$ at the final times, 
they cannot explain the discrepancy between our
measured results and those of \cite{smith}.
 
As we have noted the clearest discrepancy between
our results and those of \cite{smith} is for the case
$n=0$, in which clearly finite size effects have, in any
case, no role. If the numerical results of both studies
are correct,  the only possible systematic effect
which can explain the difference in the measured
exponents is one arising from the force smoothing.
In principle we have controlled  carefully for
such effects by always identifying  
a lower cut-off scale at which self-similarity
applies, and down to which we measured 
exponents. Further we have always checked
that, in both real and Fourier space, that
this scale is above that at which we
expect smoothing to cause deviations.
As discussed in Section \ref{Effects of force smoothing},
and illustrated in Fig. \ref{fig_compare_Gadget_smoothing},
this scale in real space can be taken to be about
$2\varepsilon$, while in reciprocal space it may
be very much smaller than one might naively 
expect. Indeed  we have seen through
direct comparison of simulations differing only
in force smoothing --- for the $n=0$ EdS 
model at the final in Fig. \ref{fig_compare_Gadget_smoothing}, 
and the $n=-1$ EdS model in Fig. \ref{fig_compare_N=256} ---
that the the power spectrum is significantly suppressed 
below its true value above $k \sim 0.1 \pi/\varepsilon$. 
From what can be inferred from the discussion given 
in \cite{smith}, it appears that the quoted results for the 
exponents have been derived from a Fourier space 
analysis only, and considering
as only relevant lower cut-off scale one of order
$\pi/\varepsilon$ \footnote{On page 11 of \cite{smith} it is 
stated that power is expected to be suppressed at 
$k \sim 1700 k_b$ which compares with  $\pi/\varepsilon=2000 k_b$.}.
It is very evident from Fig. \ref{fig_compare_Gadget_smoothing}
 (and, likewise from Fig. \ref{fig_compare_N=256}, for the 
case $n=-1$) that extrapolating a 
fit to $\Delta^2(k)$ much beyond about one tenth of $\pi/\varepsilon$ 
will lead to a lower fitted exponent, as reported by \cite{smith}. 
Thus we believe that the exponents reported by
\cite{smith} significantly below the stable clustering
prediction are a result of including in the fitted region 
at least some points at which the power is suppressed 
significantly by smoothing. 
Given our conclusions about finite size effects in
the cases $n=-1$ and $n=-2$, it appears probable
to us that the same effect explains the measurement
by \cite{smith} of exponents lower than those 
predicted by the stable clustering  hypothesis 
in these cases, and even lower than our average
measured exponents in these cases.
As we have discussed self-similarity itself 
should provide a ``protection" against such 
effects, and in particular a careful identification 
of the {\it lower cut-off} to self-similarity
and how it evolves in time. Such an
analysis has not apparently been performed 
in \cite{smith}, and in particular not in real
space in which the effects of smoothing
can be controlled for more easily.
 The importance of 
the determination of this lower cut-off, even
in real space, is illustrated well by the plots of 
$\xi$ for the EdS models ($\kappa=1$) in 
Fig. \ref{fig-fits-two-xis_n-2_n-1}:
there is a significant part of the measured 
correlation function well above $\varepsilon$, 
with an effective exponent well below that of 
the stable clustering region, which is not self-similar
and which must be excluded from our fit.
If this region beyond the ``bending" in
the correlation function is self-similar,
we would need to see the scale 
at which this bending take place
remain fixed in the self-similarity rescaling
plots. This is definitely something we have
not observed in our simulations of these models: in
our self-similarity plots the bending scale
is clearly always moving to smaller 
scales, and there is therefore always,
 as we have emphasized at 
the end of Section \ref{Non-linear self-similar clustering: two point correlation properties},
an identifiable region where the clustering measured at 
smaller scales is not self-similar, and therefore cannot
be taken to represent the physical clustering of the
self-similar model. We note finally that \cite{smith} 
report their derivation of exponents using the 
commonly used representation 
of $\Delta^2(k)$ as a function of the variable $\Delta_L^2(k_L)$
(following the ansatz of \cite{peacock}). We have 
checked our results also in this representation and
find results in agreement with those of our direct
analysis of $\xi(x)$ and $\Delta^2(k)$, i.e., we find,
restricting to the region in which self-similarity is
observed (to which we always detect a clear lower
cut-off) an exponent in the strongly non-linear region
in agreement with stable clustering.

\subsubsection{One dimension}

It is interesting to compare our results with
those of \cite{benhaiem+joyce+sicard_2013}, 
which performed the exactly analogous study of 
the one dimensional version of this class of models.
Compared to the three dimensional case, this model
has the interest of admitting ``exact" numerical 
integration (i.e. limited only by numerical roundoff)
and much greater spatial resolution than in 
three dimensions at very modest numerical cost. 
Thus, for example, in certain cases a strongly 
non-linear correlation function extending over 
as much as four orders in magnitude is obtained.
While the same qualitative difficulties are 
observed as in three dimensions --- the 
range of non-linear scales accessible in a 
simulation of given size shrinks rapidly
as $\gamma_{sc}$ (in its one dimensional
version) decreases --- this greater 
spatial resolution makes a crucial difference 
in the study: we are able to clearly identify
in one dimension a value of $\gamma_{sc}$,
roughly at $\gamma_{sc} \approx 0.15$, 
above which the stable clustering 
approximation works very well, and below
which the correlation function appears to
become independent of both $n$ and 
$\kappa$, i.e., there is apparently 
a truly universal region at sufficiently
small values of $\gamma_{sc}$. 
Further we note that the study of 
\cite{sl1,sl2,sl3} of the static case 
($\kappa=0$) in three dimensions,
for the case $n=2$ and $n=0$, 
finds evidence that in this limit
the non-linear correlation function,
albeit in a limited range of amplitude
where it can be measured,
is independent of initial conditions.
This latter result, and those in one 
dimension,  thus suggest that in
three dimensions there is probably
likewise a region in the $(n,\kappa)$
space where the strongly non-linear 
two point correlation function is indeed truly
universal, independent of both 
initial conditions and cosmology.
We have not been able to determine 
whether this region exists in this study,
but significantly larger simulations
might do so.

\subsection{Perspectives}
We finally mention some questions and directions for future work
with this class of models.  



\begin{itemize}
\item 
 {\it Analysis with other tools, notably halos}:
In this paper we have focussed solely on the characterization
of clustering with two point correlation statistics. Non-linear
clustering --- and in particular its compatibility with the
stable clustering approximation --- can evidently be probed 
with many other statistical tools: higher order correlation functions,
box counting statistics and associated fractal dimensions,
etc., and scale free models are particularly interesting in
that a careful separation of the self-similar part of the signal 
can be used to select the numerically resolved region.
In the cosmological context a widely used, albeit less
precisely defined statistical tool, is an analysis in terms of
``halos" through appropriately defined numerical
algorithms which select non-linear, and approximately
virialized, clumps and characterize the density field in 
terms of their density profiles. Apparent  ``universality",
of non-linear clustering,i.e., independence of both cosmology 
and initial conditions --- in particular in terms of  mass and phase 
space density profiles of halos --- have been extensively 
described in numerical simulations 
(\cite{navarro1, navarro2, taylor+navarro, moore_etal_1999})
and much discussed in the literature. The degree of, and
indeed origin of, such a putative universality remains an
open and important question. The set of models we have 
studied here provide a very well defined framework in which 
to address these questions, within the very constrained
setting of self-similar models. Indeed the
usual EdS model with power law initial conditions 
has been considered as an important
reference model for the study of halo 
properties' dependence on initial conditions
\cite{navarro2, knollmann_etal_2008}. In
principle a universality of halos profiles requires
a breakdown of stable clustering, as 
simple halo models are compatible with stable 
clustering predictions for correlation functions
only if their profiles depend also on initial conditions 
(see e.g. \cite{ma+fry_2000c, yano+gouda_2000}).


\item {\it Larger simulations}: It is evident that significantly larger 
simulations  of this class of models than those which we have reported 
here --- say with characteristics like those of the current
largest cosmological simulations (e.g. \cite{springel_05, deus})
may be able to resolve clearly the most interesting, and
indeed relevant, questions.  We have only been able to 
establish that the fully non-linear, and self-similar,
part of the correlation function which we can access
with our simulations can be described well in all 
cases by the predictions of stable clustering. 
Amongst the goals of such a larger study would be
the following. {\it Firstly}, for the cases 
($\gamma_{sc} \gtapprox 1$) where we have 
observed a region of power law clustering 
in line with the stable clustering prediction, 
to try locate a lower cut-off scale to its validity.
Indeed as we have discussed, 
at any given  $(n, \kappa)$, we would
expect there must be such a cut-off as the 
assumption that a structure is stable must 
break down at some finite time (due to merging).
Such a detection would require the identification
of a scale, {\it with a self-similar scaling},  
marking a break from the power law behaviour 
of the correlation, and which can be given
as some fraction of the non-linearity scale.
{\it Secondly}, in
the region of smaller $\gamma_{sc} \ltapprox 1$
where we have been unable to place any constraint,
to measure the correlation function to higher amplitudes
and assess whether it can be described in any
range by stable clustering; in principle we expect 
to see at some point a complete breakdown of this
approximation, and perhaps a tendency to a result
which is completely universal, i.e., independent of 
$n$ and $\kappa$.  {\it Thirdly}, to determine the
properties of halos in these models and understand
in detail their relation to two-point (and possibly 
higher order) correlation properties, with particular
attention to the adequacy of the approximation
of the distribution by smooth halos. 

\item {\it Application to ``realistic" cosmologies}:
The ultimate goal of this research is of course
to contribute to a better understanding of  the non-linear
regime of models with initial conditions
and cosmological evolutions like those
in ``realistic" models. In this respect we note that in 
principle a huge class of such models (e.g. any
homogeneous dark energy model)
should be characterized by a {\it trajectory}
in the space $(n, \kappa)$, through the
values of these quantities defined at
each time by the instantaneous value 
of $\kappa$ defined by the expansion rate,
and the  logarithmic slope of the power spectrum
at the scale of non-linearity.  Indeed 
numerous existing results in the literature on 
different (non self-similar) models can probably be
unified and understood in a simple framework
in this way: for example, the observation that
in both open universe models and those
with a cosmological constant, non-linear
clustering is amplified compared to that
in the EdS model (see e.g. \cite{peacock}),
and that stable clustering becomes a 
better approximation in an open universe
models (\cite{padmanabhan_etal_1996}).
Further, if a much fuller description of the self-similar
non-linear clustering in the $(n,\kappa)$
space can be obtained, and leads to
the identification of a truly ``universal 
region" in this space, this trajectory
would probably allow conclusions
to be drawn about any model with
respect to universality, and also the
relevance of stable clustering. In this respect 
we note that the region of  $\gamma_{sc} \ltapprox 1$,
where we have been unable to place any 
constraint on the non-linear regime, is 
a part of the $(n, \kappa)$ space which
is extremely  relevant to viable cosmological
models: in $\Lambda$CDM (or similar models with 
a dynamical dark energy component) the slope
of the power spectrum of initial 
fluctuations relevant to structure formation
varies in the range from $n=-1$ to $n=-3$, and 
the effective value of $\kappa$ is in the range 
between $1$ and  $\approx 2$ (see 
Fig. \ref{fig-kappa-lambdaCDM}).


We have not analysed the quasi-linear regime carefully,
but analysis in these models, and in particular as a function
of $\kappa$ at fixed $n$,  may provide stringent tests of
various phenomenological and/or theoretical proposal which
have been made to understand it (\cite{quasi-linear_padmanabhan})
and phenomenological fitting procedures (\cite{peacock}).
Closer to the linear regime, these models may also provide a 
powerful tool to test perturbative approaches beyond linear
order, e.g., using the ``renormalized perturbation theory" 
of \cite{crocce-scoccimarro}. Finally, using a description of a models
as a trajectory in the $(n, \kappa)$ space, 
it would be interesting to try to construct, using
the kind of approach developed in
in \cite{hamilton, peacock}, 
alternative semi-analytical models 
for non-linear clustering.

\item {\it Testing for discreteness effects}: 
As we have underlined, the self-similarity of 
evolution in such models is potentially a powerful
tool to study and control to what extent $N$ body
simulations can be confidently taken to really
reproduce, as required, the clustering in the 
continuum (Vlasov-Poisson) limit: indeed 
any deviation from self-similarity implies
a dependence on scales introduced by
the $N$ body method --- amongst which, notably, 
the particle discreteness scale $\Lambda$ and
force smoothing $\varepsilon$.  The importance
of understanding better the importance of
such effects, and {\it how they may depend notably 
on the model simulated}, is evident from the
difficulties observed in reproducing self-similarity
in the limit of smaller $\gamma_{sc}$, which 
is the regime extremely relevant to all
current realistic models. Further the confidence 
we have that stable clustering
should apply in some of them, could potentially also
be used as a strong test. We note that a few cases of the
usual EdS model have been exploited in this way
in studies focussing on the effects of discreteness
such as \cite{melott_etal_1997, splinter_1998, kuhlman},
and also on finite size effects \cite{orban_2012, orban_2013}.
In the spirit of these studies, it would be very interesting
also to use this class of models to test more directly
whether it is indeed reasonable to simulate in
the regime $\varepsilon \ll \Lambda$, which has
been a specific subject of debate in the literature
(see e.g.  \cite{discreteness3_mjbm} and references therein).
Despite the considerable
numerical challenge, it may be feasible to probe
a range of non-linear clustering even with 
$\varepsilon > \Lambda$, for larger values 
of $\gamma_{sc}$ and see if they match the
self-similar stable clustering behaviour observed
in the $\varepsilon \ll \Lambda$ simulations.

\item {\it Calibration of precision measures}: 
Along the lines of the previous point, a specific
point is control on the numerical accuracy of simulations 
by monitoring the energy evolution, of which
we have included some discussion in this paper.  
We will discuss  details of this technical issue further in a separate 
article.  More generally it would be interesting to explore 
whether in the framework of this class of models it might be
possible to calibrate such tools, i.e., provide approximate
rules on what deviations of different indicators may be
tolerated.

\item {\it Studies in lower dimensions}: 
The framework for this study was suggested to us initially
by an exploration of the same class of models in the simplified
context of one dimensional models. The fact that the results of
the three dimensional study reported here have turned out 
to be so strikingly in line in many respects with this simple
model, motivate in turn further study of this case, where further
modest numerical effort could probably help in particular
to better characterize what happens in the apparently
universal region. Further these studies suggest that it
may also be very instructive to consider the intermediate
case of two dimensions, closer to the real case but
with still a considerable gain compared to it in terms
of spatial resolution (see e.g. \cite{beacom_etal_1991}
and references therein). 

\end{itemize}

\section{Acknowledgements}
We thank Francesco Sylos Labini, Fran\c{c}ois Sicard
and Pascal Viot for many useful discussions in the context of collaboration 
on related projects. Numerical simulations have been performed at the cluster of the SIGAMM
hosted at ``Observatoire de C\^ote d'Azur'', Universit\'e de Nice --
Sophia Antipolis. This work was partly supported by the ANR
09-JCJC-009401 INTERLOP project and the CNPq PDS 158378/2012-1 grant.

\bibliographystyle{mn2e}

\appendix
\section{Integration scheme}
\label{app-gadget2}
In order to integrate Eq.~\eqref{3d-equations-3} we discretize 
using a leap-frog scheme, following a standard method used
in the simulation of Langevin dynamics (see e.g.  \cite{izaguirre2010multiscale}).
The latter introduces in the molecular dynamics a stochastic 
term and a fluid damping, and our case is obtained simply
by suppressing the former term. Each component of the equation
of motion for a particle of mass $m$ is discretized as
\begin{eqnarray}\label{mov1}
	\Delta x &=& v \,\Delta \tau  \\
	m \,\Delta v &=& F \left(x\right) \,\Delta \tau - m  \Gamma  v \, \Delta \tau  \label{mov2}
\end{eqnarray}
with the obvious notations.
As in the usual Leapfrog, position and velocity are updated separately at
different times separated by $\frac {\Delta \tau}{2}$, but in order 
to add the effect of the friction, another ``Kick-Drift-Kick" operation is added
to the core structure of the code. 
The code thus integrates Eqs.~\eqref{mov1} and \eqref{mov2} with the folloxwing steps:
\begin{eqnarray}
	v^{n+1/2}=e^{-\Gamma \frac {\Delta \tau}{2}} v^n + \left( \frac {1 - e^{\Gamma \frac {\Delta \tau} {2} }}{\Gamma} \right) m^{-1} F (x^{n}) \\
	x^{n+1}= x^{n} + \Delta \tau v^{n+1/2} \\
	v^{n+1}=e^{-\Gamma \frac {\Delta \tau}{2}} v^{n+1/2} + \left( \frac {1 - e^{\Gamma \frac {\Delta \tau} {2} }}{\Gamma} \right) m^{-1} F(x^{n+1}),	
\end{eqnarray}
where $v^{n}=v(\tau)$, $v^{n+1/2}=v(\tau+\Delta \tau /2)$, $v^{n+1}=v(\tau + \Delta \tau )$, and an analogous notation for the $x$ variable.

\end{document}